\documentclass{emulateapj}

\shorttitle{Observations of SN 2005cf} \shortauthors{Wang et~al.}

\def\gsim{\;\lower4pt\hbox{${\buildrel\displaystyle >\over\sim}$}\;}
\def\lsim{\;\lower4pt\hbox{${\buildrel\displaystyle <\over\sim}$}\;}
\def\grls{\;\lower4pt\hbox{${\buildrel\displaystyle >\over <}$}\;}

\begin{document}

\title{A Golden Standard Type Ia Supernova SN 2005cf: \\ Observations from the Ultraviolet
to the Near-Infrared Wavebands}
\author{
X.~F.~Wang\altaffilmark{1,2}, W.~D.~Li\altaffilmark{1},
A.~V.~Filippenko\altaffilmark{1}, R.~J.~Foley\altaffilmark{1,3,4},
R.~P.~Kirshner\altaffilmark{3}, M.~Modjaz\altaffilmark{3,1,5},
J.~Bloom\altaffilmark{1}, P.~J.~Brown\altaffilmark{6}, D.~Carter\altaffilmark{7},
A.~S.~Friedman\altaffilmark{3}, A.~Gal-Yam\altaffilmark{8}, M.~Ganeshalingam\altaffilmark{1},
M.~Hicken\altaffilmark{3}, K.~Krisciunas\altaffilmark{9},
P.~Milne\altaffilmark{10}, N.~B.~Suntzeff\altaffilmark{9},
W.~M.~Wood-Vasey\altaffilmark{3,11}, S.~B.~Cenko\altaffilmark{1,12}, P.~Challis\altaffilmark{3},
D.~B.~Fox\altaffilmark{6}, D.~Kirkman\altaffilmark{14},\\
J.~Z.~Li\altaffilmark{2}, T.~P.~Li\altaffilmark{2},
M.~A.~Malkan\altaffilmark{15}, D.~B.~Reitzel\altaffilmark{15},
R.~M.~Rich\altaffilmark{15}, F.~J.~D.~Serduke\altaffilmark{1},
R.~C.~Shang\altaffilmark{2},\\ J.~M.~Silverman\altaffilmark{1},
T.~N.~Steele\altaffilmark{1},
B.~J.~Swift\altaffilmark{1},
C.~Tao\altaffilmark{16}, D.~S.~Wong\altaffilmark{1}, and
S.~N.~Zhang\altaffilmark{2}}

\altaffiltext{1}{Department of Astronomy, University of California,
Berkeley, CA 94720-3411; wangxf@astro.berkeley.edu .}
\altaffiltext{2}{Physics Department and Tsinghua Center for
Astrophysics (THCA), Tsinghua University, Beijing, 100084, China;
wang\_xf@mail.tsinghua.edu.cn .}
\altaffiltext{3}{Harvard-Smithsonian Center for Astrophysics,
 60 Garden Street, Cambridge, MA, 02138.}
\altaffiltext{4}{Clay Fellow.}
\altaffiltext{5}{Miller Fellow.}
\altaffiltext{6}{Pennsylvania State University, Department of
Astronomy \& Astrophysics, University Park, PA 16802.}
\altaffiltext{7}{Astrophysics Research Institute, Liverpool John
Moores University, Tweleve Quays House, Egerton Wharf, Birkenhead
CH41 1LD, UK.}
\altaffiltext{8}{Benoziyo Center for Astrophysics,
Weizmann Institute of Science, 76100 Rhovot, Israel.}
\altaffiltext{9}{Department of Physics, Texas A\&M University,
College Station, Texas, 77843.}
\altaffiltext{10}{Steward Observatory, University of Arizona,
933 North Cherry Avenue, Tucson, AZ 85721.}

\altaffiltext{11}{Department of Physics, University of Pittsburgh,
100 Allen Hall, Pittsburgh, PA 15260.}

\altaffiltext{12}{Space Radiation Laboratory, MS 220-47, California
Institute of Technology, Pasadena, CA 91125}

\altaffiltext{13}{Department of Astronomy \& Astrophysics,
525 Davey Laboratory, Pennsyvania State University University park, PA 16802}

\altaffiltext{14}{CASS 0424, University of California, San Diego,
9500 Gilman Drive, La Jolla, CA 92093-0424.}

\altaffiltext{15}{Department of Physics and Astronomy,
University of California, Los Angeles, CA 90095-1547.}

\altaffiltext{16}{CPPM/CNRS Centre de Physique des Particules de
Marseille \& LAM/CNRS Laboratoire d'Astrophysique de Marseille
Universite de la Mediterranee, France.}

\begin{abstract}

We present extensive photometry at ultraviolet (UV), optical, and
near-infrared (NIR) wavelengths, as well as dense sampling of
optical spectra, for the normal type Ia supernova (SN Ia) 2005cf.
The optical photometry, performed at eight different telescopes,
shows a $1\sigma$ scatter of $\lesssim$0.03 mag after proper
corrections for the instrument responses.  From the well-sampled
light curves, we find that SN 2005cf reached a $B$-band maximum at
$13.63 \pm 0.02$ mag, with an observed luminosity decline rate
$\Delta m_{15}(B) = 1.05 \pm 0.03$ mag. The correlations
between the decline rate and various color indexes, recalibrated on
the basis of an expanded SN~Ia sample, yield a consistent estimate for
the host-galaxy reddening of SN 2005cf, e.g.,$E(B - V)_{\rm host}
= 0.09 \pm 0.03$ mag. The UV photometry was obtained with the
{\it Hubble Space Telescope} and the {\it Swift} Ultraviolet/Optical
Telescope, and the results match each other to within 0.1--0.2 mag.
The UV light curves show similar evolution to the broadband $U$, with
an exception in the 2000--2500~\AA\ spectral range (corresponding to
the F220W/uvm2 filters), where the light curve appears broader and much fainter
than that on either side (likely owing to the intrinsic spectral
evolution). Combining the UV data with the ground-based optical and
NIR data, we establish the generic UV-optical-NIR bolometric light
curve for SN 2005cf and derive the bolometric corrections in the
absence of UV and/or NIR data. The overall spectral evolution of SN
2005cf is similar to that of a normal SN~Ia, but with variety in the
strength and profile of the main feature lines.  The spectra at
early times displayed strong, high-velocity (HV) features in the
Ca~II H\&K doublet and NIR triplet, which were distinctly detached
from the photosphere ($v \approx 10,000$ km s$^{-1}$) at a velocity
ranging from 19,000 to 24,000 km s$^{-1}$. One interesting feature
is the flat-bottomed absorption observed near 6000~\AA\ in the
earliest spectrum, which rapidly evolved into a triangular shape and
then became a normal Si~II $\lambda$6355 absorption profile at about
one week before maximum brightness. This pre-maximum spectral
evolution is perhaps due to the blending of the Si~II $\lambda$6355
at photospheric velocity and another HV absorption component (e..g.,
Si~II shell at a velocity $\sim$18,000 km s$^{-1}$) in the outer
ejecta, and may be common in other normal SNe~Ia. The possible
origin of the HV absorption features is briefly discussed.

\end{abstract}

\keywords {supernovae: general -- supernovae: individual (SN
2005cf)}

\section{Introduction}

Type Ia supernovae (SNe~Ia) play important roles in diverse areas of
astrophysics, from chemical evolution of galaxies to observational
cosmology. They, together with the core-collapse SNe, are responsible
for most of the heavy elements in the universe. SNe~Ia have also
been used over the past decade as the most powerful tool probing the
expansion history of the universe. Owing to a relatively homogeneous
origin --- probably an accreting carbon-oxygen white dwarf (WD) with a mass close
to the Chandrasekhar limit ($\sim$1.4 M$_{\odot}$) in a binary system
(for a review see Hillebrandt \& Niemeyer 2000) --- most SNe~Ia show strikingly
similar spectral and photometric behavior (e.g., Branch \& Tammann
1992; Suntzeff 1996; Filippenko 1997).  In particular, the observed
peak luminosities of SNe~Ia have been shown to correlate with the
shapes of their light or color curves (e.g., Phillips 1993; Hamuy
et~al. 1996; Riess et~al. 1996; Perlmutter et~al. 1997; Wang et~al.
2003; Wang et~al. 2005; Guy et~al. 2005; Prieto et~al. 2006;
Jha et~al. 2007; Guy et~al. 2007; Colney et~al. 2008), leading to an uncertainty
of $\lesssim$ 10\% in distance measurements from SN~Ia.

Based on the observations of SNe~Ia at redshifts $z \approx 0.5$,
Riess et~al. (1998) and Perlmutter et~al. (1999) first reported the
discovery of an accelerating universe. The evidence for the
acceleration expansion from SNe~Ia improved markedly with follow-up
studies (Barris et~al. 2004; Tonry et~al. 2003; Knop et~al. 2003;
Riess et~al. 2004, 2007; Astier et~al. 2006; Wood-Vasey et~al.
2007a), suggesting that $\sim$70\% of the universe is composed of a
mysterious dark energy (for a review see, e.g., Filippenko 2005a).
Elucidating the nature of dark energy would require a large
sample of well-observed SNe~Ia at even higher redshifts (e.g., $z \gtrsim 1.0$),
and also relies on the improvement of the SN~Ia standardization (e.g., $\lesssim$ 0.01 mag).
Progress can be made by searching for additional luminosity-dependent parameters,
or by identifying a subsample of SNe~Ia with the lowest scatter of the luminosity.
This depends on the degree of our understanding of SN~Ia physics as well as on the
good controlling of various systematic effects such as the photometry itself, the
SN luminosity evolution, and the absorption by dust. Clarification of the above issues
demands a large sample of SNe~Ia with well-observed spectra and light curves,
from which we can get better constraints of their physical properties.


Quite a few detailed studies have been conducted of spectroscopically and/or
photometrically peculiar SNe~Ia such as SNe 1991T (Filippenko et~al.
1992a; Phillips et~al. 1992), 1991bg (Filippenko et~al. 1992b,
Leibundgut et~al. 1993), 2000cx (Li et~al. 2001; Thomas et~al.
2003; Candia et~al. 2003), 2002cx (Li et~al. 2003; Branch et~al.
2004; Jha et al. 2006a), and 2006gz (Hicken et~al. 2007). A comparable number
of relatively normal SNe~Ia have also been individually studied,
including SNe 1994D (Patat et~al. 1996), 1996X (Salvo et~al. 2001),
1998aq (Branch et~al. 2003; Riess et~al. 2005), 1998bu (Jha et~al.
1999), 1999ee (Stritzinger et~al. 2002; Hamuy et~al. 2002), 2001el
(Krisciunas et~al. 2003), 2002er (Pignata et~al. 2004), 2003cg
(Elisa-Rosa et~al. 2006), 2003du (Stanishev et~al. 2007),
2004eo (Pastorello et~al. 2007a), 2002bo (Benetti et~al. 2004),
and 2006X (Wang et~al. 2008a), though the last two may differ
from other typical SNe~Ia due to an unusually high
expansion velocity of their photospheres and a relatively flat evolution of
their $B$-band light curves starting from the early nebular phase
(Wang~et al. 2008a). However, a much larger sample of
SNe~Ia must be investigated in order to determine the dispersion among their
properties and refine possible systematic effects for precision cosmology.
In addition, the sample of ``golden standard'' SNe~Ia having extensive
observations from ultraviolet (UV) through
near-infrared (NIR) wavelengths is sparse. The UV properties could
provide clues to the diversity and evolution of the progenitor
system, as they are more sensitive to the metallicity of the ejecta
as well as the degree of mixing of the synthesized $^{56}$Ni
(H\"{o}flich et~al. 1998; Blinnikov \& Sorokina 2000), while the NIR
data are particularly suitable for the study of dust properties and
the determination of absorption corrections. The UV and NIR data are
also important in helping to determine, by means of the light
curves, the bolometric luminosity of SNe~Ia.

In this paper, we present extensive observations of the SN~Ia 2005cf
in UV, optical, and NIR bands, providing a ``golden standard'' with
which to compare other SNe~Ia. Pastorello et~al. (2007b; hereafter
P07) and Garavini et~al. (2007; hereafter G07) have previously
studied the optical properties of SN 2005cf, but our unique UV data
along with an excellent independent optical/NIR dataset allow us to
provide better constraints on the properties of SN 2005cf. We
compare our results with those of P07 and G07 where appropriate. Our
observations and data reduction are described in \S 2, while \S 3
presents the UV-optical-NIR ($uvoir$) light curves, the color
curves, and the reddening estimate. The spectral evolution is given
in \S 4. In \S 5 we construct the bolometric light curve of SN
2005cf. Our discussion and conclusions are given in \S 6.

\section{Observations and Data Reduction}

SN 2005cf was discovered at an unfiltered magnitude of 16.4 on 2005
May 28.36 (UT dates are used throughout this paper) by Pugh \& Li
(2005) during the Lick Observatory Supernova Search (LOSS) with the
0.76~m Katzman Automatic Imaging Telescope (KAIT; Filippenko et al.
2001; Filippenko 2005b), with J2000 coordinates $\alpha$ =
15$^{h}$21$^{m}$32$^{s}$.21 and $\delta$ =
$-07^{\circ}24^{'}47^{''}.5$. It exploded in the vicinity of the
tidal bridge connecting the S0 galaxy MCG$-$01-39-003 with the
nearby Sb galaxy MCG$-$01-39-002 (NGC 5917); see Figure 1. Assuming
MCG$-$01-39-003 is the host galaxy of SN 2005cf, we find that the
supernova was $15.7''$ west and $123.0''$ north of the galaxy
nucleus.

An optical spectrum taken on 2006 May 31.22 revealed that SN 2005cf
was a very young SN~Ia, at a phase of $\sim$10~d before maximum
brightness (Modjaz et~al. 2005). On this basis, we requested
frequent optical and NIR imaging, as well as optical spectroscopy;
we collected a total of 634 photometric datapoints and 39 optical
spectra. Moreover, {\it Hubble Space Telescope (HST)} UV and NIR
observations were soon triggered (proposal GO-10182; PI A. V.
Filippenko) with the Advanced Camera for Surveys (ACS) and the
Near-Infrared Camera and Multi-Object Spectrometer (NICMOS) at 12
different epochs. UV and optical photometry was also obtained with
the Ultraviolet/Optical Telescope (UVOT) on the space-based {\it
Swift} telescope.

\begin{figure}
\figurenum{1} \vspace{-0.2cm}
\centerline{\includegraphics[angle=0,width=110mm]{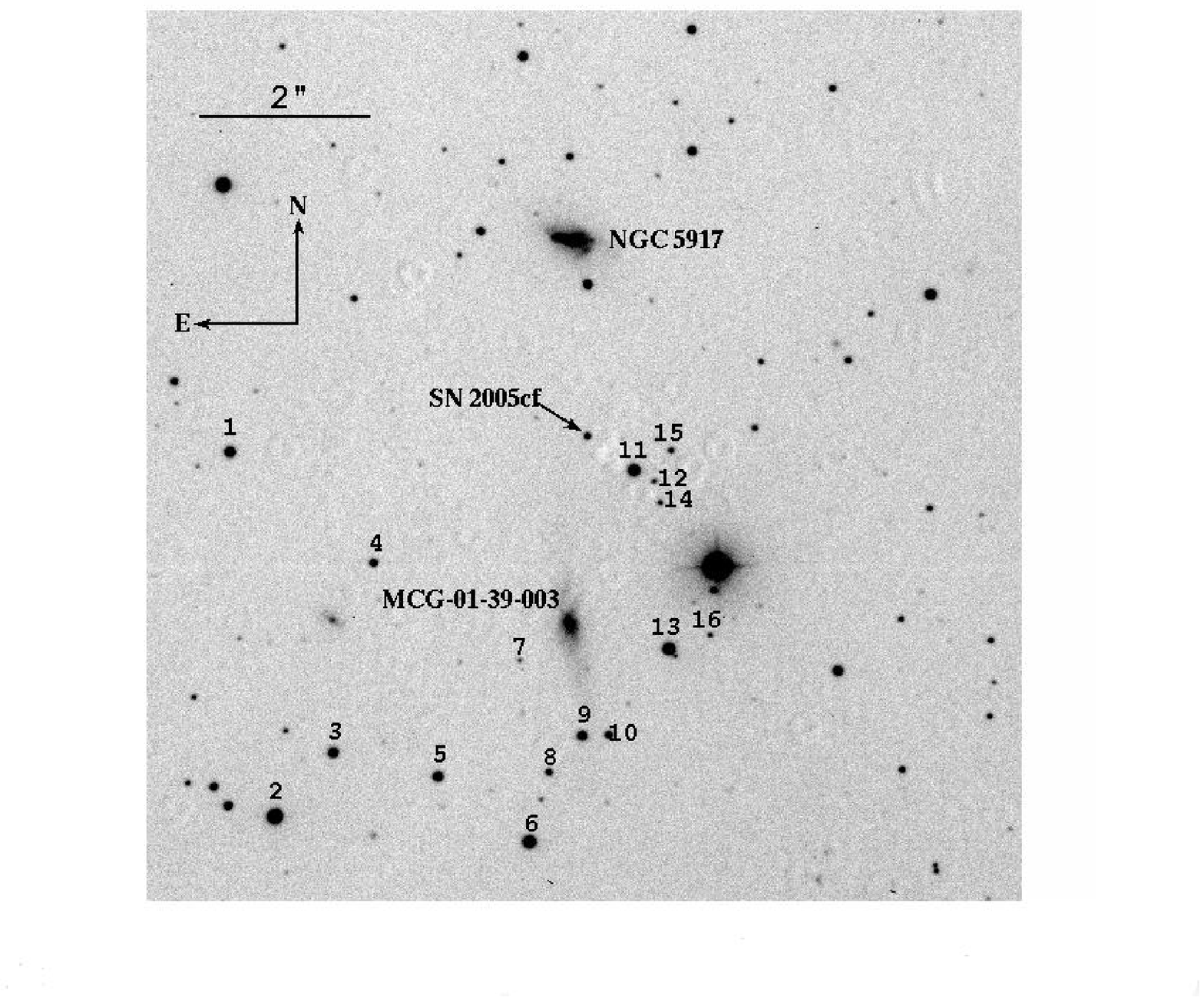}}
\vspace{-0.2cm} \caption{SN 2005cf in MCG$-$01-39-003. This is a
$V$-band image taken with the 0.8~m TNT on 2005 Sep. 21. The
supernova and 16 local reference stars are marked. North is up, and
east is to the left.} \label{fig-1} \vspace{-0.0cm}
\end{figure}

\subsection{Ground-Based Observations}

\subsubsection{Optical and NIR Photometry}

The ground-based optical photometry of SN 2005cf, spanning from 12~d
before to 3 months after the $B$-band maximum, was obtained with the
following telescopes: (1) the KAIT and the 1.0~m Nickel telescope at
Lick Observatory in California; (2) the 1.2~m telescope at the Fred
Lawrence Whipple Observatory (FLWO) of the Harvard-Smithsonian
Center for Astrophysics (CfA) in Arizona; (3) the 1.3~m and 0.9~m
telescopes at Cerro Tololo Inter-American Observatory (CTIO) in
Chile; (4) the 1.5~m telescope at Palomar Observatory in California
(Cenko et~al. 2006); (5) the 2.0~m Liverpool telescope at La~Palma in Spain; and (6) the
0.8~m THCA-NAOC Telescope (TNT) at Beijing Xinglong Observatory
(BAO) in China. Broad-band $BVRI$ photometry was obtained with all
the above telescopes, except for the FLWO 1.2~m and Liverpool 2.0~m
telescopes which followed SN 2005cf in $BV$ and Sloan $ri$ filters.
Observations made with KAIT, FLWO 1.2~m, the CTIO 0.9~m, and the
Nickel 1.0~m also sampled the $U$ band. Table 1 lists the average
color terms for all of the involved telescopes and filters. The NIR
($JHK_s$) photometry was obtained with the 1.3~m Peters Automated
Infrared Imaging Telescope (PAIRITEL; Bloom et~al. 2006) at FLWO.

As part of routine processing, all CCD images were corrected for
bias, flat fielded, and cleaned of cosmic rays. Since SN 2005cf is
isolated far away from the host-galaxy center, we omitted the usual
step of subtracting the galaxy template from the SN images; instead,
the foreground sky was determined locally and subtracted.
Instrumental magnitudes of the SN and the local standard stars
(labeled in Fig. 1) were measured by the point-spread function (PSF)
fitting method, performed using the IRAF\footnote{IRAF, the Image
Reduction and Analysis Facility, is distributed by the National
Optical Astronomy Observatory, which is operated by the Association
of Universities for Research in Astronomy (AURA), Inc. under
cooperative agreement with the National Science Foundation (NSF).}
DAOPHOT package (Stetson 1987).

The transformation from the instrumental magnitudes to the standard
Johnson $UBV$ (Johnson 1966) and Kron-Cousins $RI$ (Cousins 1981)
systems was established by observing, during a photometric night, a
series of Landolt (1992) standards covering a large range of
airmasses and colors. The average value of the photometric
zeropoints determined on 7 photometric nights was used to calibrate
the local standard stars in the field of SN 2005cf. Table 2 lists
the standard $UBVRI$ magnitudes and uncertainties of 16 comparison
stars which were used to convert the instrumental magnitudes of the
supernova to those of the standard system. Note, however, that our
$U$-band calibration may have an uncertainty larger than that quoted
in Table 2, as it was established on only a single photometric
night. The $UBV$ calibrations of SN 2005cf were also used by Li
et~al. (2006) to calibrate the {\it Swift} UVOT optical
observations.


  A comparison of 9 standard stars in common with P07 reveals that
some systematic differences exist between the datasets. With respect
to P07, our measurements are fainter by $0.138 \pm 0.029$ mag in
$U$, $0.023 \pm 0.012$ mag in $B$, $0.038 \pm 0.016$ mag in $V$,
$0.041 \pm 0.006$ mag in $R$, and $0.029 \pm 0.007$ mag in $I$. The
discrepancies are non-negligible and worrisome, especially in the
$U$ band. Such differences were also noticed by Stanishev et~al.
(2007) in studying photometry of the comparison stars of SN 2003du;
their measurements of the stars in common were found to be
systematically brighter than those given in Leonard et~al. (2005)
and Anupama et~al.(2005) by 0.04-0.06 mag in some wavebands. The
origin is unclear, and further studies are needed if systematic
errors are to be minimized (to $\lesssim 0.01$ mag) when using
SNe~Ia to measure cosmological distances.

To reasonably assemble the optical photometric data for SN 2005cf
obtained with different telescopes and place them on the
Johnson-Cousins $UVBRI$ system [e.g., from the Sloan $ri$ to the
broad-band $RI$], we applied additional magnitude corrections
($S$-corrections; Stritzinger et~al. 2002) to the photometry.  This
is because the color-term corrections only account for the
differences derived from normal stars, whereas the spectra of SNe
are quite dissimilar. Properly modeling the instrumental response is
essential for deriving reliable $S$-corrections. The normalized
instrumental response functions, obtained by multiplying the filter
transmission functions with the quantum efficiency of the CCD
detectors and the atmospheric transmission, are shown in Figure 2.
Details of the application of the $S$-corrections are given in the
Appendix.

\begin{figure} 
\figurenum{2} \vspace{-0.2cm}
\centerline{\includegraphics[angle=0,width=105mm,height=90mm]{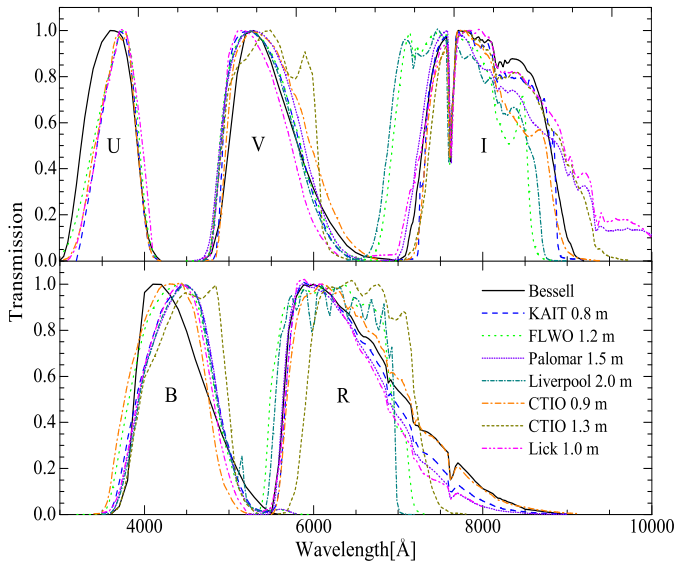}}
\vspace{-1.0cm} \caption{Comparison of instrumental responses of the
seven telescopes, normalized to the peak transmission, with the
standard Bessell Johnson/Kron-Cousins functions.} \label{fig-2}
\end{figure}        

Figure 3 shows the time evolution of the $S$-corrections and the
resulting color variances for different telescopes, computed with
the spectra of SN 2005cf presented in this paper and those published
by G07, as well as with some late-time spectra of SN 2003du
(Stanishev et~al. 2007). Note that all of the $U$-band filter
responses were cut off at 3300~\AA\ in the convolution due to the
spectral coverage. The resulting $S$-corrections are generally small
in $BVR$ bands, but can be noticeably large in the $U$ and $I$ bands
(e.g., $\sim$0.1--0.2 mag). It is worth noting that, without
applying such a systematic magnitude correction, the $B - V$ color
measured at the CTIO 0.9~m and Palomar 1.5~m (or Lick 1.0~m)
telescopes could differ by $\sim$0.1 mag in the early nebular phase
when the colors are usually used as the reddening indicators.
\begin{figure*} 
\vspace{-1.0cm}
\figurenum{3}
\centerline{\includegraphics[angle=0,width=150mm]{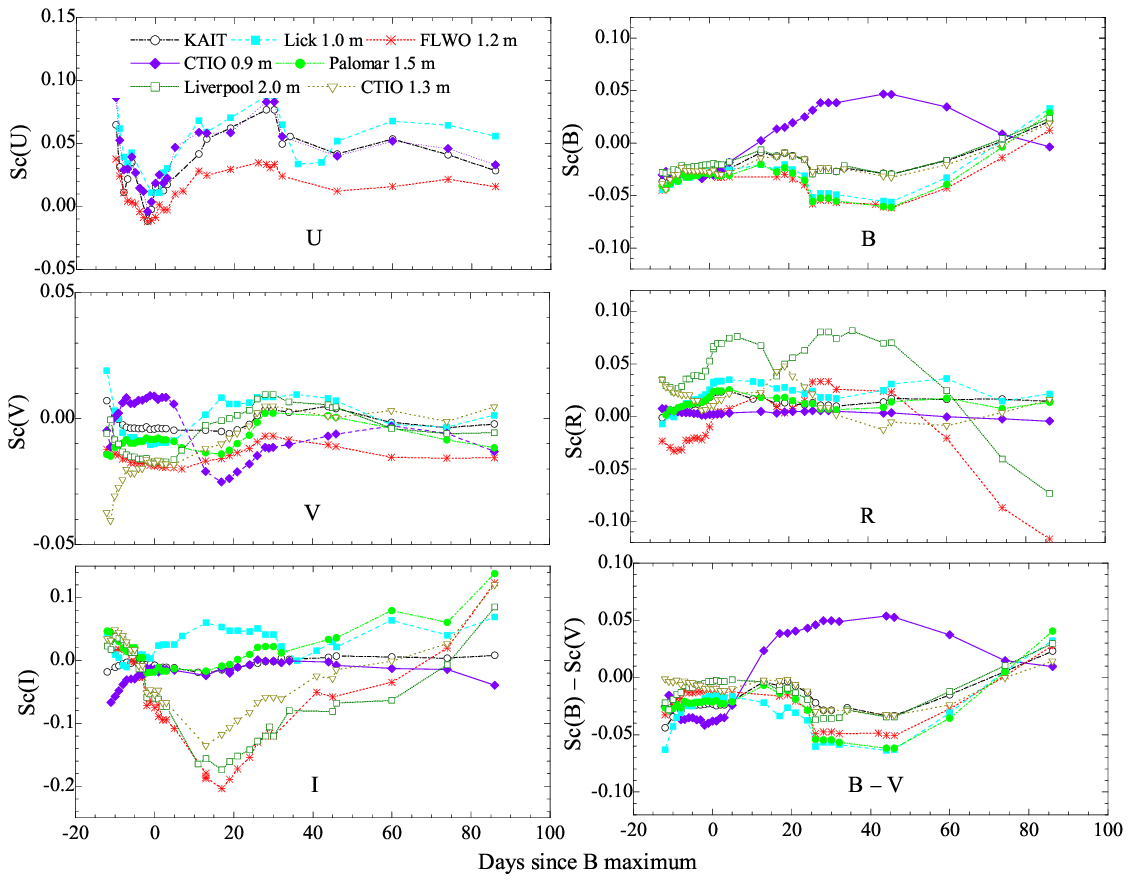}}
\vspace{-1.0cm}
 \caption{Time evolution of the $S$-corrections at the various telescopes.}
 \label{fig-3}
\end{figure*}

Applying the $S$-corrections to the photometry noticeably improved
the consistency of the datasets obtained with different telescopes.
This is demonstrated in Fig. 4, where the scatter around the
best-fit curve decreases from 0.06 mag to 0.03 mag in the $I$ band.
Improvements are also achieved for the other bands, with the
photometric scatter reduced to within 0.02--0.03 mag. Such a
normalization in the photometry could be potentially important when
comparing the properties of SNe measured with different systems.
This is because a shift of a few percent in the $B - V$ color might
systematically bias the extinction correction, hence producing an
error of $\sim$10\% (a factor of 3--4 larger) in the derived
luminosity of the SN.

We further applied $K$-corrections to our photometry using the same
set of spectra as for the $S$-corrections. We note that the $U$-band
$K$-correction could reach $\sim$0.06 mag at the earliest phases,
perhaps due to a rapid evolution of the spectral shape. The large
$K$- and $S$-corrections required in the $U$ band, which were
usually unavailable due to the insufficient spectral coverage, might
partially account for the large scatter seen in the $U$-band light
curve of SN 2005cf.

The final calibrated $UBVRI$ magnitudes, after performing the $K$-
and $S$-corrections, are presented in Table 3. The error bars (in
parentheses) are dominated by the uncertainty in the calibration of
the comparison stars.

Since the NIR observations of SN 2005cf were conducted with
PAIRITEL, the instrument that defines the 2MASS photometric system,
we use the 2MASS point-source catalog (Cutri et~al. 2003) to
calibrate the supernova. The calibrated $JHK_s$ magnitudes of SN
2005cf are given in Table 4, which were corrected for the
$K$-corrections (Columns (7)-(9)) computed from the NIR spectra
of SN 1999ee. No $S$-corrections were applied to the NIR photometry
because of the similarity of the transmission curves between the 2MASS
system and the Persson et~al. (1998) system. The filter transmission
curves of these two systems are shown in Figure 5.

\begin{figure}
\figurenum{4} \vspace{-0.2cm}
\centerline{\includegraphics[angle=0,width=100mm,height=90mm]{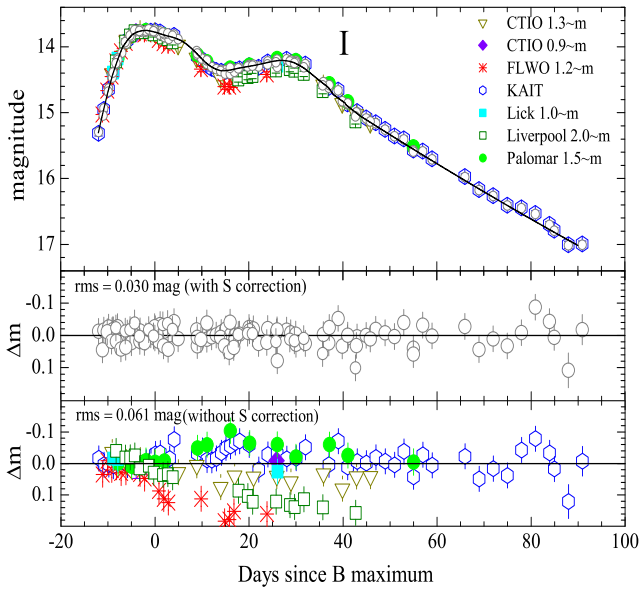}}
\vspace{-0.8cm} \caption{The $I$-band light curve of SN 2005cf with
and without applying the $S$-corrections. The solid line in the
upper panel represents the best fit to the $S$-corrected light
curve.} \label{fig-4}
\end{figure}

\subsubsection{Optical spectroscopy}

Low-resolution spectra of SN 2005cf were obtained with the Kast
double spectrograph (Miller \& Stone 1993) on the 3.0~m Shane
telescope at Lick Observatory and the FAST spectrograph (Fabricant
et~al. 1998) on the Tillinghast 1.5~m telescope at FLWO. Two
late-time spectra were also obtained at the W. M. Keck Observatory:
one with the Low Resolution Imaging Spectrometer (LRIS; Oke et~al.
1995) mounted on the 10~m Keck I telescope, and the other with the
Deep Extragalactic Imaging Multi-Object Spectrograph (DEIMOS; Faber
et al. 2003) mounted on the 10~m Keck II telescope. A journal of the
spectroscopic observations is given in Table 5.

All spectra were reduced using standard IRAF routines (e.g., Foley et~al. 2003).
For the Lick/Kast observations, flatfields for the red-side spectra
were taken at the position of the object to reduce NIR fringing effects.
For two Lick/Kast spectra taken on 2005 June 10 and June 11, there was
condensation on the red-side camera, producing non-standard and
variable absorption features. To compensate for this effect, we
created a two-dimensional surface map of a flat field image. We
smoothed the surface map to remove any fringing in the flat. We then
divided our images by this surface to remove the absorption features
from our images. Although this process produced significantly
improved spectra, there are still some persistent systematic features
in those spectra. The Keck/DEIMOS data were reduced using a modified
version of the DEEP pipeline and our own routines as described by Foley et~al (2007).

Flux calibration of the spectra was performed by means of spectrophotometric
standard stars observed at similar airmass on the same night as the SN.
Using our own IDL routines, the extracted, wavelength calibrated spectra
were corrected for continuum atmospheric extinction using
mean extinction curves for FLWO and Lick Observatory; moreover, telluric
lines were removed from the data. For all the spectra observed at Lick and FLWO,
the slit was always aligned along the parallactic angle to avoid chromatic
effects in the data (Filippenko 1982).

\subsection{HST UV and NIR Observations}

\subsubsection{ACS UV Photometry}
Imaging of SN 2005cf was carried out with the {\it HST} ACS High
Resolution Camera (HRC) during Cycle 13. The field of view of this
CCD-based instrument is about $29''\times 25''$ with a scale of
$0.028'' \times 0.025''$ pixel$^{-1}$. The observations were made in
the F220W, F250W, and F330W bands, with exposure times of 1040~s,
800~s, and 360~s, respectively. The exposure time was split into
several equal segments that were used in the data reduction process
to reject cosmic-ray events. The data produced by the STScI
reduction pipeline had bias and dark-current frames subtracted and
were divided by a flatfield image. In addition, known hot pixels and
other defects were masked. Individual exposures were combined using
the MultiDrizzle task within STSDAS to reject cosmic rays and
perform geometric-distortion corrections.

To get the optimal signal-to-noise ratio (S/N), we performed
aperture photometry in all of the drizzled images using an aperture
radius of 4 pixels ($\sim0.1''$).  The background level was
determined from the median counts in an annulus of radius 100--130
pixels. The measured magnitudes were further corrected to an
infinite-radius aperture and placed on the Vega magnitude system
(Sirianni et~al. 2005).  The final {\it HST} ACS UV magnitudes of SN
2005cf are listed in Table 6. Uncertainties were calculated by
combining in quadrature the Poisson photon errors, the readout-noise
errors from the pixels within the aperture, and errors in the
aperture correction.

\begin{figure}
\vspace{-0.7cm} \figurenum{5}
\centerline{\includegraphics[angle=0,width=100mm]{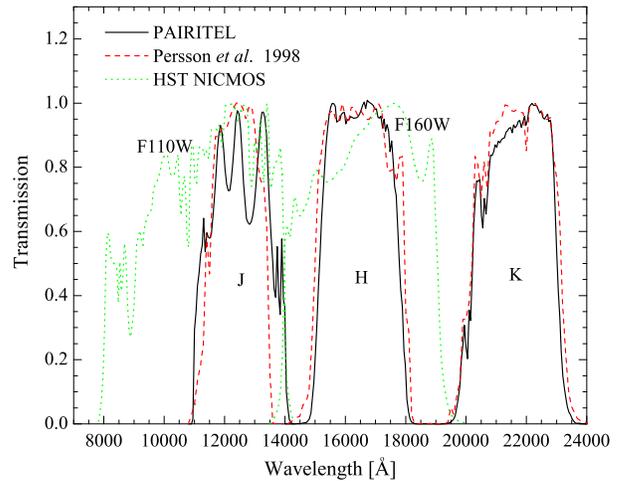}}
\vspace{-0.5cm}
 \caption{Comparison of near-infrared transmission curves
 of PAIRITEL (ex-2MASS) and {\it HST} NICMOS3 with that of the
 Persson et~al. (1998) system.}
 \label{fig-5}
\end{figure}
\subsubsection{NICMOS NIR Photometry}

The infrared observations were obtained with the {\it HST} NICMOS3,
which has a scale of $0.20''$ pixel$^{-1}$ and a field of view of
$51'' \times 51''$. Images were acquired through the F110W and F160W
filters (see Fig. 5 for the transmission curves) at similar epochs
as the optical ones with the ACS. The data were preprocessed using
the STSDAS package CALNICA and the latest reference files provided
by STScI. Unlike the ACS, the NICMOS calibrated data are given in
count rate (DN s$^{-1}$, where DN are data-number counts).

Aperture photometry was performed on the calibrated NICMOS3 images.
Counts were summed within an aperture of 5.5 pixel radius ($1.10''$,
the size of the aperture used for the standard-star measurements)
centered on the source. To correct for a nominal infinite aperture,
the measured count rates in F110W and F160W were, respectively,
multiplied by 1.056 and 1.087. The total count rates were then
converted into flux using the recently determined photometric scale
factors, $1.59 \times 10^{-6}$ Jy s DN$^{-1}$ at 1.1~$\mu$m and
$1.93 \times 10^{-6}$ Jy s DN$^{-1}$ at 1.6~$\mu$m. Corresponding
zeropoints were calculated on the Vega system, assuming the
zero-magnitude flux densities of 1886 and 1086 Jy and the effective
wavelengths of 1.12~$\mu$m and 1.60~$\mu$m for F110W and F160W,
respectively. All of the above parameters involved in the
calculations were taken from the website for {\it HST} NICMOS
photometry\footnote{http://www.stsci.edu/hst/nicmos/performance/photometry/}.

As the NICMOS3 filters do not match ground-based $JHK$ well
(especially the F110W filter, which is much broader and bluer), it
is necessary to apply $S$-corrections as well as color-term
corrections to place the photometry on the ground-based $JHK$
system.  A total of six stars (2 white dwarfs, 2 solar analogs, and
2 red stars) have been observed through both the NICMOS NIR system
(F110W, F160W, and F222M filters) and the ground-based $JHK$ system
(Persson et~al. 1998). These stars, along with the Sun, Vega, and
Sirius with model IR spectra from the Kurucz web site\footnote{
http://kurucz.harvard.edu/}, allow us to establish a rough
transformation between the NICMOS3 and the ground-based NIR systems
for normal stars. In computing the NIR $S$-corrections we used data
for SN 1999ee (Hamuy et~al. 2002), which has relatively good
temporal coverage of the NIR spectra and a value of $\Delta m_{15}$
similar to that of SN 2005cf.

Table 7 gives the color-, $K$- and $S$-corrected NIR photometric results
from the {\it HST} NICMOS3 images, consistent with the ground-based
measurements within the error bars. The $S$- and color-corrections
that were added to normalize the photometry to the Persson et~al. (1998)
system are also listed.

\subsection{Swift UVOT Optical/UV Observations}

SN 2005cf was also observed with the Ultraviolet/Optical Telescope
(UVOT; Roming et~al. 2005) onboard the {\it Swift} observatory
(Gehrels et~al.  2004), covering from 8~d before to 42~d after the
$B$-band maximum.  The photometric observations were performed in
three UV filters (uvw1, uvm2, and uvw2) and three broadband optical
filters ($U$, $B$, and $V$). As shown in Figure 6, the instrumental
response curves of the UVOT uvm2 and uvw1 filters are, respectively,
very similar to those of the {\it HST} ACS F220W and F250W filters.
The UVOT optical filters are close to the standard Johnson system in
$B$ and $V$ but exhibit noticeable difference in $U$.

\begin{figure}[htp]
\figurenum{6} \vspace{-0.5cm}
\centerline{\includegraphics[angle=0,width=105mm] {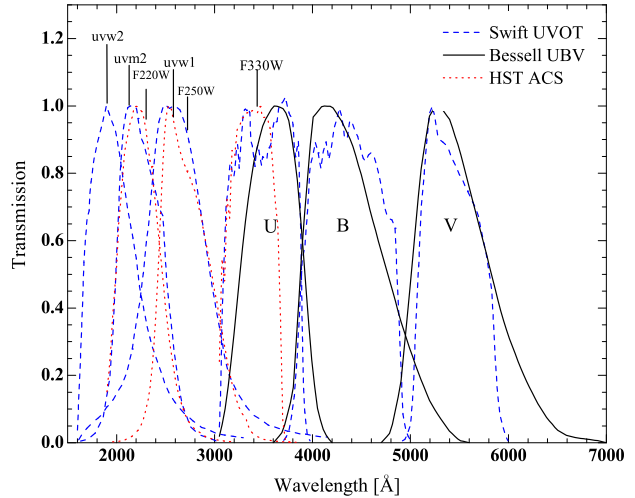}}
\vspace{-0.3cm}
 \caption{Comparison of transmission curves of {\it Swift} UVOT filters
 with {\it HST} ACS in the UV and with the Johnson system in $UBV$.}
 \label{fig-6}
\end{figure}

Images of SN 2005cf were retrieved from the {\it Swift} Data Center,
and were reprocessed utilizing an improved plate scale for the uvw2
images and corrections to the exposure times in the image headers
(Brown et~al. 2008). To maximize the S/N, we performed aperture
photometry using an optimal aperture of $3.0''$ radius suggested by
Li et~al. (2006), after first subtracting the host-galaxy light
using a template image. Since the UVOT is a photon-counting detector
and suffers from coincidence losses when observing bright sources,
the observed counts of SN 2005cf were corrected for such losses
using the empirical relation described by Poole et~al. (2008).
Finally, aperture corrections were applied to bring the measurements
from an aperture of $3.0''$ to the $5.0''$ aperture with which the
photometric zeropoints are calibrated on the Vega magnitude system.

As the instrumental response curves of the UVOT optical filters do
not follow exactly those of the Johnson $UBV$ system (see also Fig.
6), some color terms are expected. We calculated the synthetic color
terms by convolving the instrumental response of the UVOT in the
optical with the 93 spectrophotometric Landolt standard stars
published by Stritzinger et~al. (2005). These are $-$0.104,
$-$0.012, and $-$0.030 in a linear correlation of the parameter
pairs ($U$, $U-B$), ($B$, $B-V$), and ($V$, $B-V$), respectively.

The synthetic color terms are small in the $B$ and $V$ bands but
relatively large in the $U$ band. Our determinations are consistent
with those measured by Poole et~al. (2008) who preferred a
polynomial expression. The $S$-corrections of the UVOT optical
filters were derived using spectra of SN 2005cf; they are given in
Table 8 (columns 10--12) and are non-negligible in the $U$ band
(e.g., 0.1--0.2 mag). Columns 4--9 in Table 8 list the final UVOT
UV/optical magnitudes. The magnitudes in optical have been
corrected for the color- and $S$-corrections.

\section{Light Curves of SN 2005cf}


Figure 7 shows the $uvoir$ light curves of SN 2005cf. The
$S$- and $K$-corrections have been applied to the light curves in all of the
optical and NIR bands. No S- or K-corrections were applied to
the UV data. The optical light curves were sampled nearly daily during the period
$t \approx -12$ to +90~d, making SN 2005cf one of the best-observed SNe~Ia.
The morphology of the light curves resembles that of a normal SN~Ia,
having a shoulder in the $R$ band and pronounced secondary-maximum features
in the NIR bands. The NIR light curves of SN 2005cf reached their peaks
slightly earlier than the $B$-band curve. This is also the case for the
UV light curves, which are found to have narrower peaks, with the exceptions
of the {\it HST} ACS F220W and the {\it Swift} UVOT uvm2 filters. Detailed
analysis of the light curves is described in the following
subsections.

\begin{figure}
\vspace{-1.0cm}
\figurenum{7}
\centerline{\includegraphics[angle=0,width=110mm]{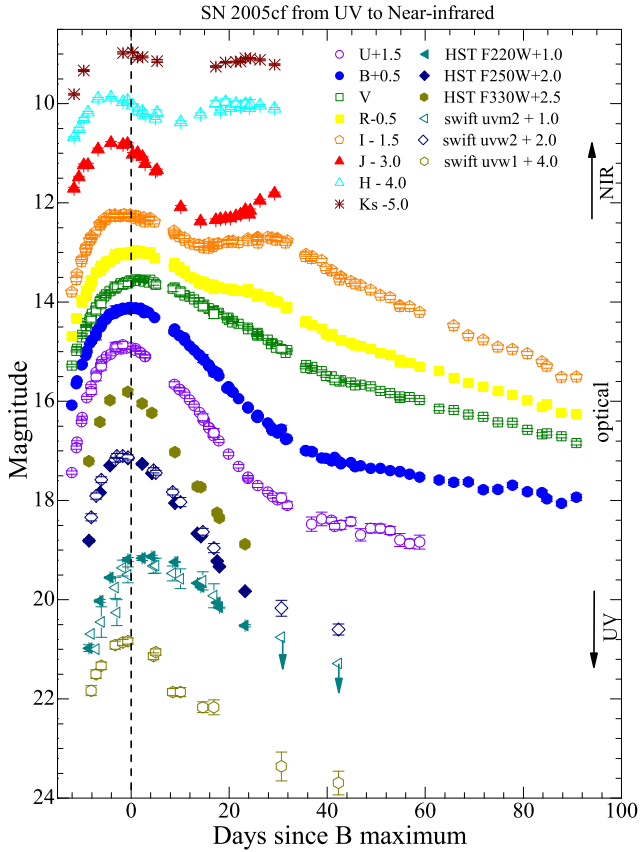}}
\vspace{-1.8cm} \hspace{5.5cm} \caption{The UV, optical, and NIR
light curves of SN 2005cf. The UV light curves were obtained from
the photometry with {\it HST} ACS and the {\it Swift} UVOT; the
$UBVRI$ optical photometry was collected at eight ground-based
telescopes as well as the {\it Swift} UVOT; and the $JHK_s$ data
were taken with the 1.3~m PAIRITEL and with {\it HST} NICMOS3.}
 \label{fig-7}
\end{figure}

\subsection{Optical Light Curves}

A polynomial fit to the $B$-band light curve around maximum
brightness yields $B_{\rm max} = 13.63 \pm 0.02$ mag on JD
2,453,533.66 $\pm$ 0.28 (2005 June 12.16). The maximum epoch $t_{\rm
max}(B)$ is consistent with the estimate by P07 while $B_{\rm max}$
itself is fainter than that of P07 by 0.09 mag.  The derived value
of $t_{\rm max}(B)$ indicates that our observations started from
$-$11.9~d and extended to +90.3~d with respect to the $B$ maximum.
Likewise, the $V$-band light curve reached a peak magnitude of
$13.55 \pm 0.02$ on JD 2,453,535.54 $\pm$ 0.33, about 1.9~d after
$t_{\rm max}(B)$. A comparable measurement was obtained by P07 who
gives $V_{max} = 13.53 \pm 0.02$ mag. The fitting parameters for the
maxima in the other bands are presented in Table 9, together with
the initial decline rates within the first 15~d after their maxima
(see below).

From the $B$- and $V$-band light curves, we derive an observed
$\Delta m_{15}(B) = 1.05 \pm 0.03$ mag and $B_{\rm max} - V_{\rm
max} = 0.08 \pm 0.03$ mag.\footnote{The true $\Delta m_{15}(B)$ for
SN 2005cf is $1.07 \pm 0.03$ mag, taking into account the reddening
effect on the light-curve shape (Phillips et~al. 1999).}  These
values are slightly smaller and redder than those given by P07
[$\Delta m_{15} = 1.11 \pm 0.05$ and $B_{\rm max} - V_{\rm max} =
0.01 \pm 0.03$]. We also measured the $B-V$ color at 12~d after the
$B$ maximum (Wang et~al. 2005), obtaining $0.47 \pm 0.04$ mag. After
removal of the Galactic reddening, these color indexes are slightly
redder than the intrinsic value (see also descriptions in \S 3.3),
suggesting some line-of-sight reddening toward SN 2005cf.

Based on the extremely well-sampled photometry in the optical, we
constructed the light-curve templates of SN 2005cf by using a
smoothing spline function. To obtain better sampling in the $U$
band, the late-phase data from P07 were also included in the fit. We
tabulate the template light curves in Table 11.  In Figure 8, we
compare the best-fit $UBVRI$ light-curve templates of SN 2005cf with
observations from the {\it Swift} UVOT and P07. We also compare them
with other well-observed, nearby SNe~Ia having similar $\Delta
m_{15}$ values, including SNe 2001el ($\Delta m_{15} = 1.15$ mag;
Krisciunas et~al. 2003), 2002bo ($\Delta m_{15} = 1.15$ mag; Benetti
et~al. 2004; Krisciunas et~al.  2004), 2002dj ($\Delta m_{15} =
1.08$ mag; Pignata et~al. 2008), 2003du ($\Delta m_{15} = 1.02$ mag;
Stanishev et~al. 2007), and 2004S ($\Delta m_{15} = 1.10$ mag;
Krisciunas et~al. 2007; Chornock \& Filippenko 2008).
These 6 objects include all SNe~Ia with $1 < \Delta m_{15} < 1.15$
and at least 15 epochs per band of published $UBVRIJHK$ data.
The $BVRI$ data for SNe 2002bo, 2002dj, and 2004S obtained with KAIT
(Ganeshalingam et~al., in prep.) are overplotted. The photometric data for the
above comparison SNe~Ia were $S$-corrected. To be consistent with SN
2005cf, the $K$-corrections computed with the spectra of the
comparison SNe and/or the spectra of SN 2005cf were also applied to
their photometry. All of the light curves of the comparison SNe~Ia
have been normalized to the epoch of their $UBVRI$ maxima and
shifted in their peaks to match the templates of SN 2005cf.

\begin{figure*}
\vspace{-0.5cm} \figurenum{8}
\centerline{\includegraphics[angle=0,width=150mm]{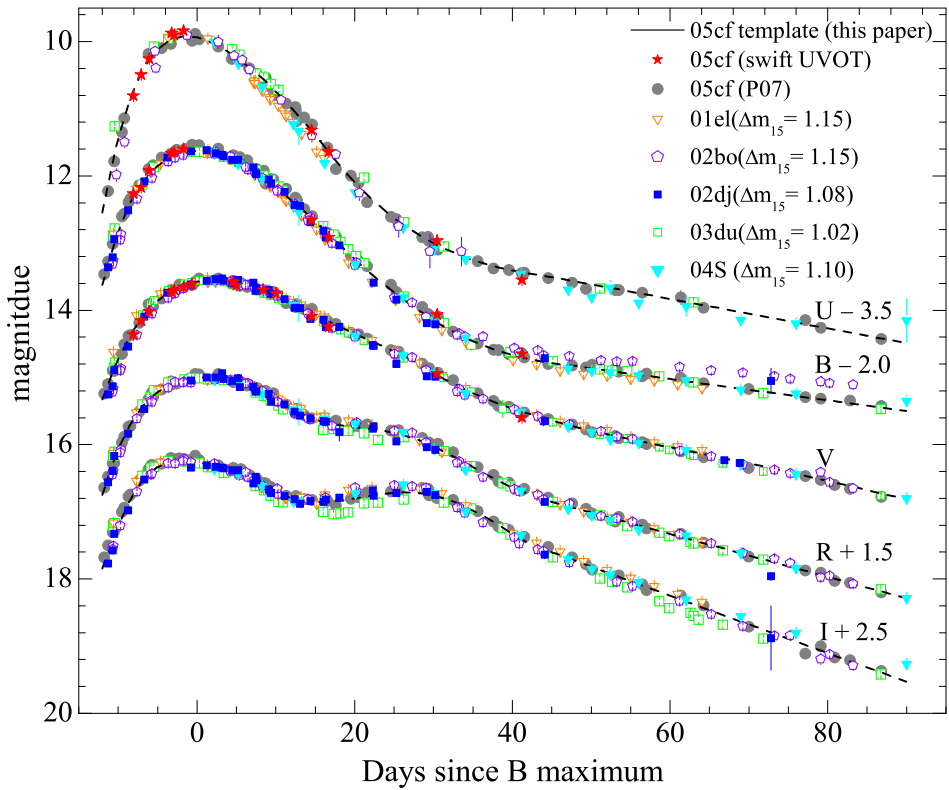}}
\vspace{-0.8cm} \caption{Comparison of $UBVRI$ light curves of SN
2005cf with those published by Pastorello et~al. (2007) and other
well-observed SNe Ia: SNe 2001el, 2002bo, 2002dj, 2003du, and 2004S.
See text for the references.} \label{fig-8}
\end{figure*}

The overall comparison with the photometry of P07 reveals some
systematic differences, especially at the earlier phases, as shown
in Figure 9.  The magnitudes in P07 are found to be brighter than
ours on average over time by $0.047 \pm 0.005$ mag in $B$, $0.009
\pm 0.004$ mag in $V$, $0.040 \pm 0.004$ mag in $R$, and $0.041 \pm
0.005$ mag in $I$, while they are fainter by $0.050 \pm 0.008$ mag
in $U$ (see the filled circles in Fig. 9). This is primarily due to
calibration differences, especially in the $BRI$ bands.  It is
puzzling, however, that a remarkable difference of $\sim$0.14 mag in
the $U$ band for the comparison stars (see \S 2.1.1) is not also
seen in the $U$-band SN photometry. Additional, uncorrected
systematic effects are probably present.

With the constructed light-curve templates of SN 2005cf, one may
also examine whether the {\it Swift} UVOT optical observations are
consistent with the ground-based data. The mean values of the
computed residuals between these two measurements (see the star
symbols in Fig. 9) are $0.015 \pm 0.024$ mag in $U$, $-0.045 \pm
0.019$ mag in $B$, and $0.009 \pm 0.011$ mag in $V$.  There appears
to be a small systematic shift in $B$, but the consistency in the
$U$ and $V$ bands is satisfactory. This demonstrates the great
improvements recently achieved in {\it Swift} UVOT calibrations (Li
et~al. 2006; Poole et~al. 2008).

Although the light curves of the comparison SNe~Ia are similar to
each other near maximum brightness, they diverge at earlier (rising)
phases.  SNe 2001el and 2003du displayed a slower rise rate in each
of the $UBVRI$ bands. In contrast, the brightness of SNe 2002bo and
2002dj rose at a faster pace than that of SN 2005cf (see also
Pignata et~al. 2008).  This indicates that the rise time might
slightly vary among SNe~Ia having similar $\Delta m_{15}$ values,
though a larger sample having early-epoch observations is needed to
verify this trend.

Differences in the light curves also emerge at late phases,
especially in the $B$ band. We measured a late-time decline rate
$\beta = 1.62 \pm 0.05$ mag (100~d)$^{-1}$ in $B$ for SN 2005cf
during the interval $t = 40$--90~d, which is comparable to that of
SNe 2001el, 2003du, and 2004S at similar phase.  While SNe 2002bo
and 2002dj showed much slower decay rates, e.g., $\beta \approx 1.2$ mag
(100~d)$^{-1}$, and they happened to be events with high
expansion velocity of the photosphere. These fast-expanding SNe may generally have
flatter photometric evolution in $B$ at late times, similar to that
of SN 2006X (Wang et~al. 2008a).\footnote{Measurements of the late-time
decline rate in the $U$ band are usually difficult due to the lack of reliable
photometry at this phase. High-quality data are needed to test
whether the $U$-band data follow the same trend as shown in $B$.}

\begin{figure}
\vspace{-0.3cm} \figurenum{9}
\centerline{\includegraphics[angle=0,width=100mm]{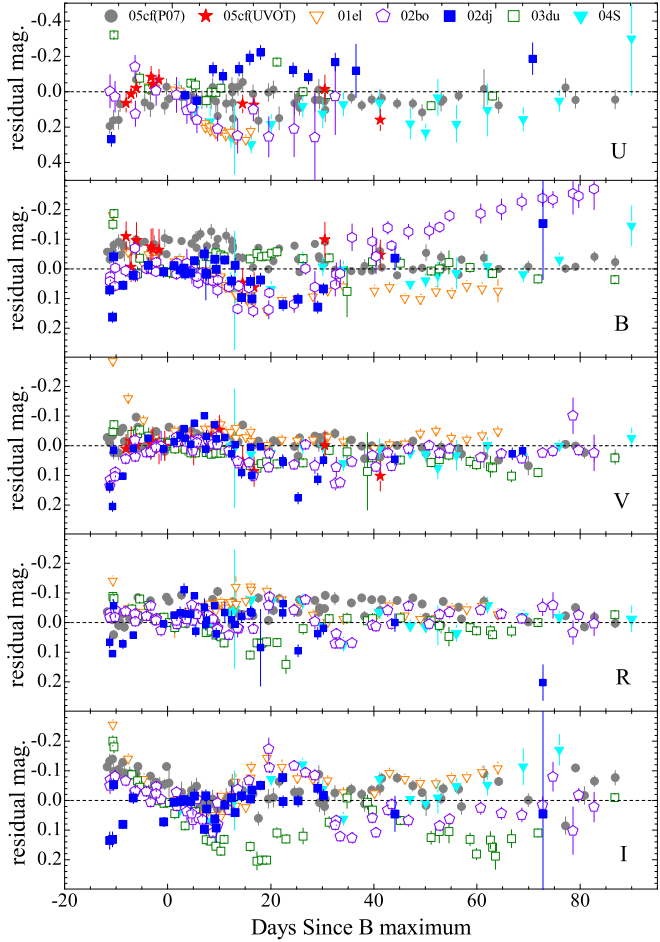}}
\vspace{-0.3cm} \caption{The residual magnitudes of the comparison
SNe~Ia with respect to the light-curve templates derived from our
observations of SN 2005cf.} \label{fig-9}
\end{figure}

In contrast to the $V$- and $R$-band evolution, the light curves in
the $U$ and $I$ bands appear more heterogeneous. The largest
dispersion is in the premaximum $U$-band evolution and in the
$I$-band secondary-maximum phase. Systematic effects due to filter
mismatches might not be fully removed by the $S$-corrections because
of the incomplete wavelength and/or temporal coverage for the SN
spectra. Nevertheless, part of the dispersion is likely
to be intrinsic. Significant spectral variations at the wavelengths of
the broad-band $U$ were also found in SNe~Ia at redshift z$\sim$0.5
(Ellis et~al. 2008). The $U$-band brightness has also been proposed to
depend sensitively on the metallicity of the progenitor
(e.g., H\"{o}flich et~al. 1998; Lentz et al. 2000; Sauer et al. 2008), which
may vary from one supernova to another; while emission in the $I$ band is
heavily affected by the Ca~II NIR triplet absorption, which is found
to vary substantially among SNe~Ia at early phases (see Fig. 17 in
\S 4.1).

\subsection{The UV Light Curves}

UV observations are important for understanding SN~Ia physics as
well as possible differences among the progenitors in a range of
environments.  Due to the requirement of space-based observations,
however, UV data for SNe~Ia have been sparse; see Panagia (2003) and
Foley, Filippenko, \& Jha (2008) for summaries of {\it International
Ultraviolet Explorer (IUE)} and {\it HST} spectra. In {\it HST}
Cycle 13, extensive UV observations of SNe~Ia were conducted
(program GO-10182; PI Filippenko); unfortunately, the Space
Telescope Imaging Spectrometer (STIS) failed just before the program
began, so the UV prism in ACS was used instead, yielding spectra far
inferior to those anticipated with STIS. Recently, {\it Swift} has
obtained UV photometry of SNe~Ia (Brown et al. 2008; Milne et al.
2008, in~prep.), as well as some low-quality UV ugrism spectra
(Bufano et~al., in prep.).

Figure 10 shows the UV light curves of SN 2005cf obtained with the
{\it Swift} UVOT and the uvw1, uvw2, and uvm2 filters, as well as
with the {\it HST} ACS and the F220W, F250W, and F330W filters.
These two UV data sets span the periods from $t = -8.1$~d to $t =
41.8$~d and from $t = -8.6$~d to $t = 23.4$~d, respectively.
Overlaid are the $U$-band templates of SN 2005cf, shifted to match
the peak of the UV bands. With respect to $U$, the F330W-band light
curve has a noticeably narrower peak. The UVOT $uvw1$ light curve
closely resembles that in the {\it HST} F250W band, with the former
being slightly brighter by $-0.09 \pm 0.02$ mag.  Despite the
similarity of the filter responses, the UVOT $uvm2$ light curve
appears to be systematically dimmer than that of {\it HST} F220W by
$\sim$0.2 mag at comparable phases.  This is perhaps due to the
calibration uncertainty or reduction errors, but more data are
obviously needed to clarify this difference.

\begin{figure}
\vspace{-0.5cm} \figurenum{10}
\centerline{\includegraphics[angle=0,width=100mm]{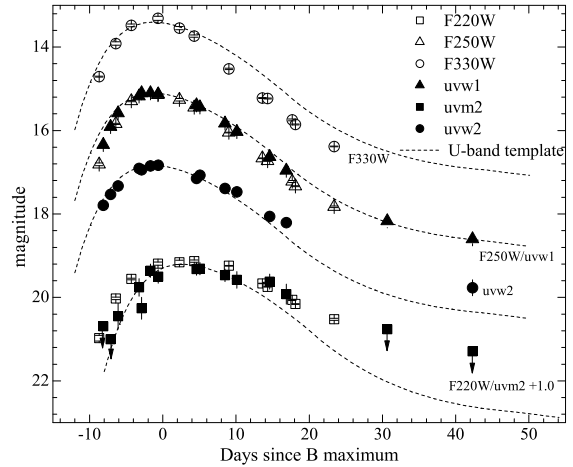}}
\vspace{-0.3cm} \caption{UV light curves of SN 2005cf, obtained with
the {\it HST} ACS (F220W, F250W, and F330W) and the {\it Swift} UVOT
($uvw1$, $uvm2$, $uvw2$). The first two and the last two data points
in the $uvm2$ band are 3-$\sigma$ upper limits of the detection.
Overlaid are the $U$-band template of SN 2005cf (dashed line).}
\label{fig-10}
\end{figure}

The temporal behavior of the UV photometry of SN 2005cf is similar
to the optical behavior, but with different epochs for the maximum
brightness. As in the $U$ band, most of the UV light curves reached
their maximum slightly before the optical, with the exception of the
uvm2/F220W-band photometry, which peaks $\sim$3--4~d after the $B$
maximum. Another interesting feature is that the light-curve peak in
these two filters appears to be much fainter and broader than that
of the other UV filters on both sides. A similar feature was seen in
SN 2005am (Brown et~al. 2005), a normal SN~Ia resembling SN 1992A
(Kirshner et al. 1993). Brown et~al. (2005) suggested that the
faintness in $uvm2$ could be explained in terms of a bump of the
extinction curve near 2200~\AA\ (Cardelli et~al. 1989). Assuming a
total reddening of $E(B - V) \approx 0.2$ mag (see \S 3.6), however,
we find that the change of $R_{V}$ could only result in a larger
extinction in $uvm2$/F220W by $\sim$0.5 mag. This is far below the
flux drop of $\sim$1.5 mag with respect to the neighboring bands, as
seen in SN 2005cf.
The red leak in the UV filters might cause the abnormal behavior in
the UV light curves by including some optical light, due to the red tail.
However, the report from the most recent check of the HST ACS CCDs shows that
such an effect is small in the $HST$ UV filters
\footnote{http://www.stsci.edu/hst/acs/documents/handbooks/cycle17}.
Moreover, the off-band contamination in the F220W filter is found to be
larger than that in F250W. This shows that the UV leak may not
be responsible for the faint brightness in F220W or uvm2 filters.

As in the optical, the light-curve parameters in the UV were
obtained by using a polynomial fit to the observations (see Table
9).  One can see that the luminosity in the F330W band is the
brightest, but it declines at the fastest pace after the initial
peak. The $uvm2$-band luminosity is the faintest, and has a
post-maximum decay rate much slower than the other UV filters.

\subsection{The NIR Light Curves}

The NIR photometry of SN 2005cf was obtained with PAIRITEL
(Wood-Vasey et~al. 2007b) and the {\it HST} NICMOS3, spanning
from $-11.4$~d to +29.1~d with respect to $t_{\rm max}(B)$.
Due to significant differences between these two photometric systems,
we applied both color- and $S$-corrections to the {\it HST} photometry
to normalize it to the PAIRITEL system.
As shown in Figure 11, the corrected F110W- and F160W-band
magnitudes (smaller filled circles) are consistent with the
ground-based results.

\begin{figure}
\vspace{-0.7cm} \figurenum{11}
\centerline{\includegraphics[angle=0,width=110mm]{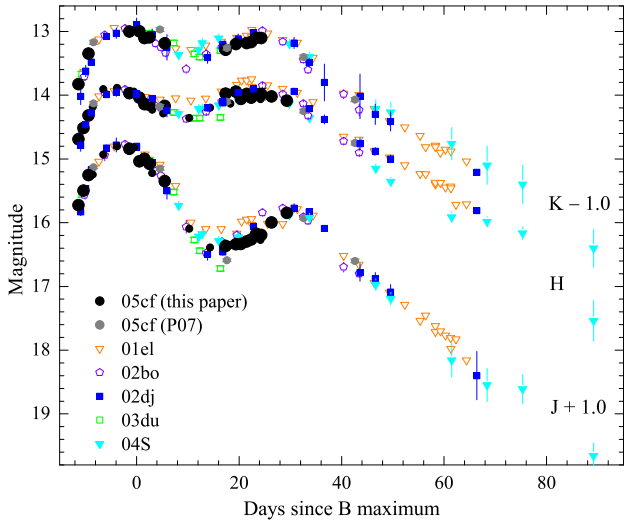}}
\vspace{-0.3cm} \caption{Comparison of the NIR light curves of SN
2005cf with those of SNe 2001el, 2002bo, 2002dj, 2003du, and 2004S.
All light curves are shifted in time and magnitude to fit the peak
values of SN 2005cf. The data sources are cited in the text.}
\label{fig-11}
\end{figure}

In Figure 11, the NIR data of the comparison SNe~Ia are overlaid.
Although the light curves of all these SNe are similar near maximum,
they diverge after the peak phase. As in the optical, SNe 2002dj and
2002bo rise at a faster pace than SNe 2005cf and 2003du in the NIR.
Scatter is also present near the secondary-maximum phase; the
fast-expanding events tend to exhibit more prominent peaks. The
secondary-peak feature in the $J$ band appears somewhat less
pronounced in SN 2001el.

The peak magnitudes in the NIR, estimated by fitting the data with
the Krisciunas et al. (2004a) templates, are reported in Table 9. In
contrast with the rapid decline in $J$ after the primary maximum,
the $H$- and $K$-band light curves show much slower decay at
comparable phases.

\subsection{The Spectral Energy Distribution Evolution}

The evolution of the spectral energy distribution (SED) can be best
studied through spectroscopy; however, only the optical spectra of
SN 2005cf were involved in our study. As an alternative, a rough SED
can be constructed from the observed fluxes in various passbands at
the same or similar epochs. Since we have photometry in the UV,
optical, and NIR bands, covering the 1600--24,000~\AA\ region, we
can study the SED evolution of SN 2005cf by means of photometry.
Because of the numerous instruments involved in the observations of
SN 2005cf, it was not always possible to observe all bands at
exactly the same time, and our definition of the ``same epoch''
refers to a reference date $\pm$1~d. The observed apparent
magnitudes in each passband were converted to fluxes using the
reddening derived in \S 3.6.

\begin{figure}
\vspace{-0.0cm} \figurenum{12}
\centerline{\includegraphics[angle=0,width=110mm]{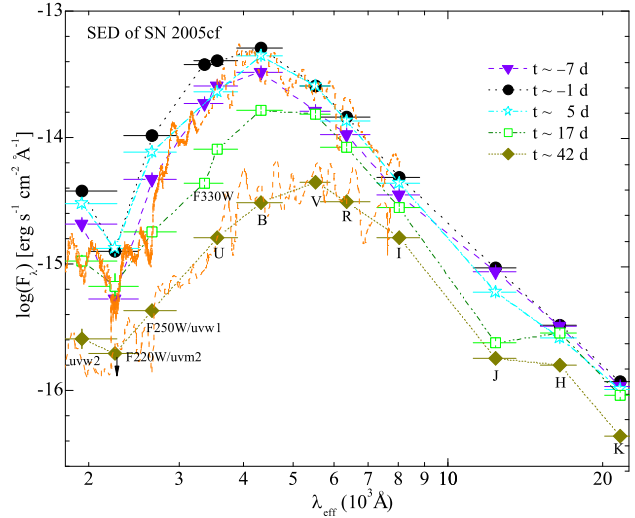}}
\vspace{-7.0cm} \caption{The spectral energy distribution evolution
of SN 2005cf at $t = -7$, $-1$, +5, +18, and +42~d after the
$B$-band maximum. The photometric points are shown with error bars
(vertical ones for uncertainties and horizontal ones for the FWHM of
the filters). Overlayed are the {\it HST} FOS spectra of SN 1992A
(dotted curves) from Kirshner et~al. (1993), obtained at $t \approx
+5$~d and +45~d, respectively. Note that the error bar of the flux
is in most cases smaller than the size of the symbol.}
\label{fig-12}
\end{figure}

Figure 12 shows SEDs at five selected epochs, $t \approx -7$, $-1$,
+5, +17, and +42~d after the $B$-band maximum. The SN 2005cf SED
went through dramatic changes in going from epochs near $B$ maximum
to the nebular phase. The SEDs at $t \approx -7$, $-1$ and +5~d are
very similar, peaking in $B$ but showing a flux deficit in
F220W/uvm2 ($\sim$2000--2500~\AA). At these early phases, the
emission from the supernova dominated in the blue. By $t \approx
17$~d the SED showed a significant deficit at short wavelengths
compared with earlier epochs, whereas the NIR remained fairly
constant (with the exception of the $J$ band). The flux peak shifted
to the $V$ band by $t \approx 42$~d, and the deficit in F220W/uvm2
might become less noticeable, though the measurement from the
$swift$ UVOT is around the detection limit.

To better understand the evolution of the 2000--2500~\AA\ region, we
overplot the {\it HST} FOS spectra of SN 1992A (Kirshner et~al.
1993) in Figure 12. One sees that the F220W/uvm2-band flux deficit
in SN 2005cf at $t \approx 5$~d is consistent with the strong
absorption trough present in the UV spectrum of SN 1992A at similar
epochs, although the UV brightness of the latter seems relatively
fainter with respect to their optical emission. This absorption feature
could arise from multiplets of the iron-peak elements at ejection velocities
above $\sim$16,000 km s$^{-1}$, such as Fe~II $\lambda\lambda$2346, 2357, 2365, and 2395,
according to the synthetic-spectrum fit by Kirshner et~al. (1993).
The complex of iron-peak element line blending might also explain
the less prominent, absorption-like feature in the SED blueward of
2500~\AA\ at days 42 and 45. As SN 1992A suffers negligible
reddening from interstellar material in both the Milky Way and the
host galaxy (e.g., Wang et~al. 2006; Jha et~al. 2007), the observed
absorption feature around 2300~\AA\ is not caused by novel dust
extinction. We suggest that the flux deficit in the F220W/uvm2
filter is likely an indication of the spectral evolution common to
normal SNe~Ia.

\subsection{The Color Curves}

Figure 13 shows the optical color curves of SN 2005cf ($U-B$, $B-V$,
$V-R$, and $V-I$), corrected for the reddening derived in \S 3.6.
Also overplotted are the color curves of the Type Ia SNe 2001el,
2002bo, 2002dj, 2003du, and 2004S, all corrected for reddening in
both the Milky Way and the host galaxies.

After a rapid decline at early phases, the $U-B$ color of SN 2005cf
reached a minimum at $t \approx -5$~d and then became progressively
redder in a linear fashion until $t \approx +23$~d, when the color
curves entered a plateau phase (Fig. 13a). SN 2003du exhibited
similar behavior.  The overall color of SN 2005cf is redder than
that of SNe 2001el and 2003du but bluer than that of SN 2004S. The
scatter in $U-B$ at maximum can reach $\pm$0.2 mag between SNe~Ia
having similar values of $\Delta m_{15}$, suggesting that a large
uncertainty might be introduced when using this color index to
estimate the reddening of SNe~Ia.

\begin{figure*}[htbp]
\vspace{-1.0cm}
\figurenum{13}
\centerline{\includegraphics[angle=0,width=160mm]{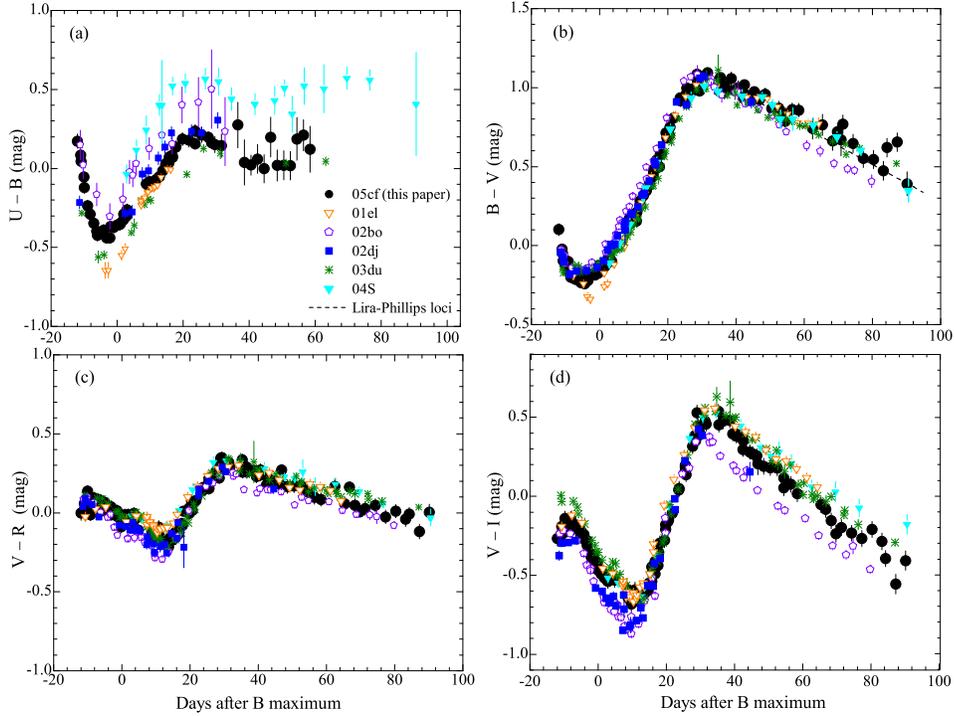}}
\vspace{-1.5cm} \caption{$U-B$, $B-V$, $V-R$, and $V-I$ color curves
of SN 2005cf compared with those of SNe 2001el, 2002bo, 2002dj,
2003du, and 2004S. All of the comparison SNe have been dereddened.
The dash-dotted line shows the unreddened Lira-Phillips loci. The
data sources are cited in the text.} \label{fig-13} \vspace{-0.0cm}
\end{figure*}

The $B-V$ colors of the selected SNe (Fig. 13b) show similar
evolution, except for SN 2001el near maximum and SN 2002bo at early
nebular phases.  We note that SN 2001el reached the bluest color
slightly later, and it is also bluer than other comparison SNe~Ia at
this point, as with $U-B$.  The $B - V$ color of SN 2002bo after $t =
+40$~d evolves at a faster pace than the Lira-Phillips relation
because of the flatter photometric evolution in the $B$ band. This
serves as a reminder that the Lira-Phillips relation does not hold
for all SNe~Ia, specifically the fast-expanding events (Wang et~al.
2008a) and the SN 2000cx-like objects (Li et~al. 2001).

The $V-R$ and $V-I$ color curves of SN 2005cf (Fig. 13c and 13d)
exhibit a behavior that is quite similar to that of the normal
SNe~Ia. By contrasting with other normal SNe, SNe 2002bo and 2002dj
are very blue in $V-I$ (and possibly in $V - R$), and they also
reach their blue peak in $V-I$ about 4~d earlier.

Figure 14 shows the observed $V-JHK$ colors of SN 2005cf, together
with the comparison SNe~Ia corrected for reddening. The $V-J$ and
$V-H$ colors are redder than the average values of the comparison
SNe~Ia by $\sim$0.2 mag, but the $V-K$ color shows little
difference. As with $V-I$, SNe 2002bo and 2002dj are bluer in all
of the $V-JHK$ colors. We notice, however, that most of the color
difference would disappear, were a smaller total-to-selective
absorption ratio applied to their extinction corrections ($R_{V} =
2.0$ rather than 3.1). This suggests that either the NIR
luminosities of the fast-expanding events were relatively faint with
respect to the optical luminosities, or a lower $R_{V}$ value is
required for their dust extinction (Wang et~al., in prep.).

\begin{figure}
\figurenum{14}
\centerline{\includegraphics[angle=0,width=70mm]{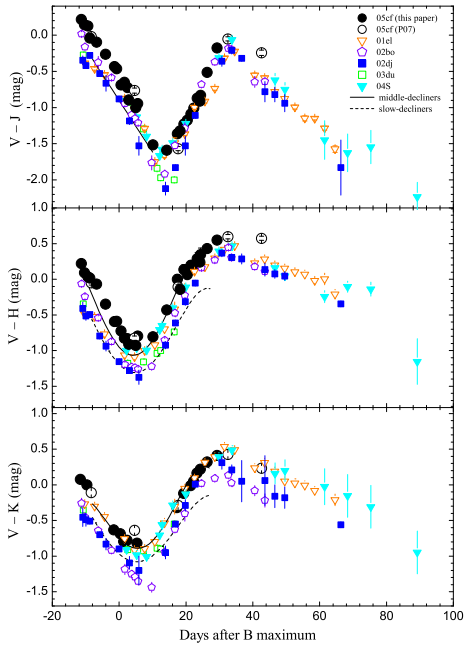}}
\vspace{-0.2cm} \caption{The $V - JHK$ color curves of SN 2005cf,
together with those of SNe 2001el, 2002bo, 2002dj, 2003du, and
2004S. The dashed lines represent the mean loci of the unreddened
SNe~Ia with $\Delta m_{15} = 0.8$--1.0 mag; the solid lines denote
those with $\Delta m_{15} = 1.0$--1.3 mag (Krisciunas et~al.
2004a).} \label{fig-14} \vspace{-0.0cm}
\end{figure}
The overall color evolution of SN 2005cf closely resembles the
selected normal SNe~Ia with similar $\Delta m_{15}$. We notice that
SNe 2002bo and 2002dj exhibit a distinguished bluer $V-IJHK$
colors with $R_{V}$ =3.1.

\subsection{Interstellar Extinction}

The Galactic extinction toward SN 2005cf is $A^{\rm{Gal}}_{V}=0.32$
mag (Schlegel, Finkbeiner, \& Davis 1998), corresponding to a color
excess of $E(B - V) = 0.097$ mag adopting the standard reddening law
of Cardelli, Clayton, \& Mathis (1989). In this section, we use
several empirical methods to derive the host-galaxy reddening of SN
2005cf. All methods assume that SN 2005cf has intrinsic colors
similar to those of normal SNe~Ia, with either similar evolution in
some colors or comparable $\Delta m_{15}$ values.

Phillips et~al. (1999) proposed correlations between the light-curve
width parameter $\Delta m_{15}$ and the intrinsic $B_{\rm max} -
V_{\rm max}$ or $V_{\rm max} - I_{\rm max}$ values (or $C_{\rm
max}$). The $B - V$ color at early nebular phases (e.g., 30~d
$\lesssim t \lesssim$ 90~d] was found to evolve in a similar fashion
for most SNe~Ia (dubbed the ``Lira-Phillips relation''; Phillips
et~al. 1999), allowing one to statistically separate the reddening
from the intrinsic color component. In addition, Wang et~al. (2005)
suggested using the $B - V$ color at 12~d after $B$ maximum [$(B -
V)_{12}$, or $C_{12}$] as a reddening indicator because the
intrinsic value of this post-maximum color was found to be a tight
function of $\Delta m_{15}$.  Based on the Lira-Phillips relation,
Jha et~al. (2007) also proposed to use the color at $t = 35$~d to
measure the host-galaxy reddening of SNe~Ia.

\begin{figure}
\figurenum{15} \vspace{-0.8cm}
\centerline{\includegraphics[angle=0,width=80mm]{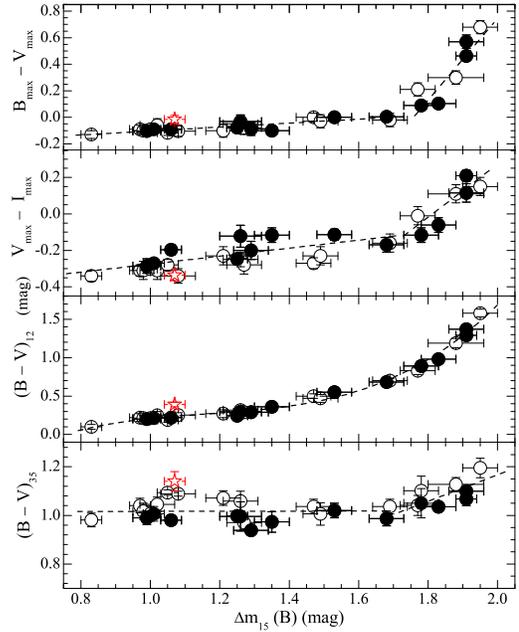}}
\vspace{-0.7cm} \caption{Correlation between the decline rate
$\Delta m_{15}$ and the observed colors for low-reddening SNe~Ia.
Open circles represent the sample collected from the literature
(Hamuy et~al. 1996; Riess et~al. 1999; Riess et~al. 2005; Jha et~al.
2006b; Garnavich et~al. 2004; Patat et~al. 1996; Salvo et~al. 2001;
Krisciunas et~al. 2004a); filled circles show those from the KAIT
photometry (Ganeshalingam et~al. in prep.). The observed colors of
SN 2005cf, corrected for the Galactic reddening, were also
overplotted (star symbol).} \label{fig-15}
\end{figure}

As shown in Figure 15, the empirical relations between the observed
colors and $\Delta m_{15}$ were recalibrated using 28 well-observed,
low-reddening SNe~Ia\footnote{SNe 1992A, 1992al, 1992bc, 1992bl,
1992bo, 1993H, 1993O, 1994D, 1994S, 1996X, 1998aq, 1998bp, 1998de,
1999by, 1999ej, 2000ca, 2000dk, 2000dr, 2001ba, 2002dl, 2002ha,
2002fk, 2003gs, 2003hv, 2003du, 2004at, 2005el, and 2005ki.}, of
which 15 are from the literature and 13 are from the new KAIT
photometric database (Ganeshalingam et~al., in prep.). The selected
SN sample closely follows the Lira-Phillips relation [e.g., with a
mean slope of $-0.011 \pm 0.001$ mag d$^{-1}$ and with the measured
$E(B - V)_{\rm host} \lesssim 0.05$ mag]. With the above sample, the
coefficients for the $C_{\rm max} - \Delta m_{15}$ relation were
determined and are reported in Table 11.

Our determinations are generally comparable to the earlier results
by Phillips et~al. (1999), but with slightly steeper slope in
$B_{\rm max} - V_{\rm max}$ and bluer unreddened zeropoints. Note
that the ($B_{\rm max} - V_{\rm max})-\Delta m_{15}$ correlation
break down at large $\Delta m_{15}$ values; a much steeper slope is
required for very fast decliners. The $V_{\rm max} - I_{\rm max}$
color does not show a great correlation with the decline rate, with
a root-mean square (rms) scatter of $\sim$0.06 mag. Applying the
$C_{\rm max} - \Delta m_{15}$ relation to SN 2005cf, we obtain $E(B
- V)_{\rm max} = 0.07 \pm 0.04$ mag and $E(V - I)_{\rm max} = - 0.08
\pm 0.06$ mag. From the $B - V$ color of SN 2005cf in the nebular
phase, we measured $E(B - V)_{\rm tail} = 0.15 \pm 0.04$ mag.

The $C_{12} - \Delta m_{15}$ relation was also reexamined with the
new sample.  The zeropoint of the intrinsic color at the nominal
decline rate is bluer than in previous reports (Wang et~al. 2006) by
0.08 mag. This difference is perhaps due to the stricter criteria of
selecting the training sample and also indicates the difficulty of
separating reddening from the intrinsic color. This post-maximum
color index gives $E(B - V)_{12} = 0.12 \pm 0.04$ mag.

In comparison with the sample observed with other systems, the KAIT
sample of 13 SNe~Ia seems to be slightly bluer by $\sim$0.05 mag at
$t = 35$~d (see the bottom panel in Fig. 15), while this difference
does not hold for the other color indices shown in the same plot.
Such a discrepancy may arise from systematic effects, such as the
$S$-corrections, which were found to show noticeable divergence in
the nebular phase (see Fig. 3). In addition, we point out that the
$(B - V)_{35}$ color may not be a constant for the fast decliners; a
larger $\Delta m_{15}$ corresponds to a redder $(B - V)_{35}$. With
the new unreddened loci, we measured $E(B - V)_{35} = 0.13 \pm 0.05$
mag for SN 2005cf.

Krisciunas et~al. (2000, 2004a) have shown that the intrinsic $V -
JHK$ colors of SNe~Ia are uniform and can be used as reddening
indicators. Based on Krisciunas unreddened loci (see Fig. 14), for
SN 2005cf we obtain $E(V - J) = 0.28 \pm 0.08$ mag, $E(V - H) = 0.25
\pm 0.06$ mag, and $E(V - K) = 0.16 \pm 0.06$ mag. This corresponds
to an E$(B - V)$ reddening of 0.13$\pm$0.08 mag, 0.10$\pm$0.06 mag,
and 0.06$\pm$0.06 mag, respectively.

The color excesses of SN 2005cf derived from various empirical
methods are summarized in Table 12. They are consistent with each
other within the uncertainties, except the $V - I$ color which shows
larger scatter and may be not a reliable reddening color index. The
mean value gives $E(B - V)_{\rm host} = 0.09 \pm 0.03$ mag,
indicating that SN 2005cf suffers a small but non-negligible
reddening in the host galaxy.

\section{Optical Spectra}

There are a total of 38 optical spectra of SN 2005cf obtained at
Lick Observatory and FLWO, spanning from $t = -12.0$ to $t =
+83.0$~d with respect to the $B$ maximum.  The complete spectral
evolution is displayed in Figure 16, and two late-time nebular
spectra are presented in Figure 18. The earliest spectra show very
broad and highly blueshifted absorption features at 3700~\AA\ (Ca~II
H\&K), 6020~\AA\ (Si~II $\lambda$6355), and 7900~\AA\ (Ca~II NIR
triplet). In particular, a flat-bottomed feature is distinctly seen
in Si~II $\lambda$6355. The spectral evolution near maximum
generally follows that of a normal SN~Ia, with the distinctive
``W''-shaped S~II lines near 5400~\AA\ and the blended lines of
Fe~II and Si~II near 4500~\AA. We discuss in detail the spectral
evolution of SN 2005cf in the following subsections.

\begin{figure}
\figurenum{16}
\vspace{-1.2cm}
\centerline{\includegraphics[angle=0,width=105mm,height=160mm]{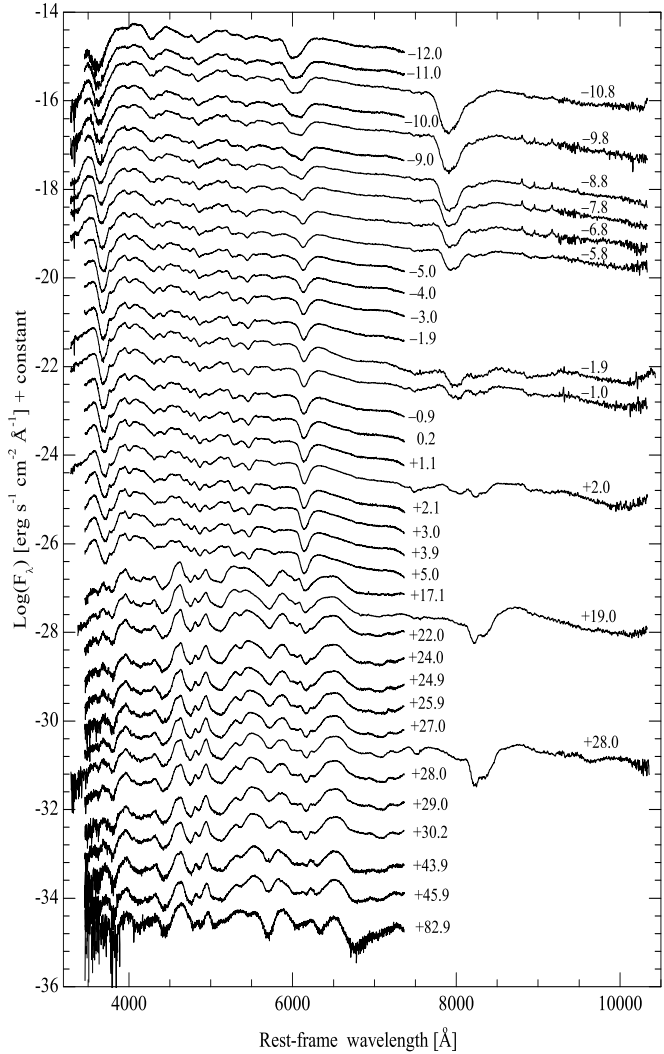}}
\vspace{-0.5cm} \caption{Optical spectral evolution of SN 2005cf.
The spectra have been corrected for the redshift of the host galaxy
($v_{\rm hel}$ = 1967 km s$^{-1}$) but not reddening, and they have
been shifted vertically by arbitrary amounts for clarity. The
numbers on the right-hand side mark the epochs of the spectra in
days after $B$ maximum.} \label{fig-16}
\end{figure}

\subsection{Temporal Evolution of the Spectra}

In Figure 17, we compare the spectra of SN 2005cf with those of
SNe~Ia having similar $\Delta m_{15}$ at four different epochs ($t
\approx -11$~d, $-7$~d, 0~d, and 18~d past $B$ maximum). All spectra
have been corrected for reddening and redshift. For SN 2001el,
$R_{V} = 2.1$ is assumed according to the analysis by Krisciunas
et~al. (2007). For the other SNe, the host-galaxy reddening is
measured using the empirical correlations presented in Table 11 and
the extinctions are corrected using the standard value $R_{V} =
3.1$. The line identifications adopted here are taken from Branch
et~al. (2005, 2006).

\begin{figure*}
\figurenum{17} \vspace{-0.8cm}
\centerline{\includegraphics[angle=0,width=190mm]{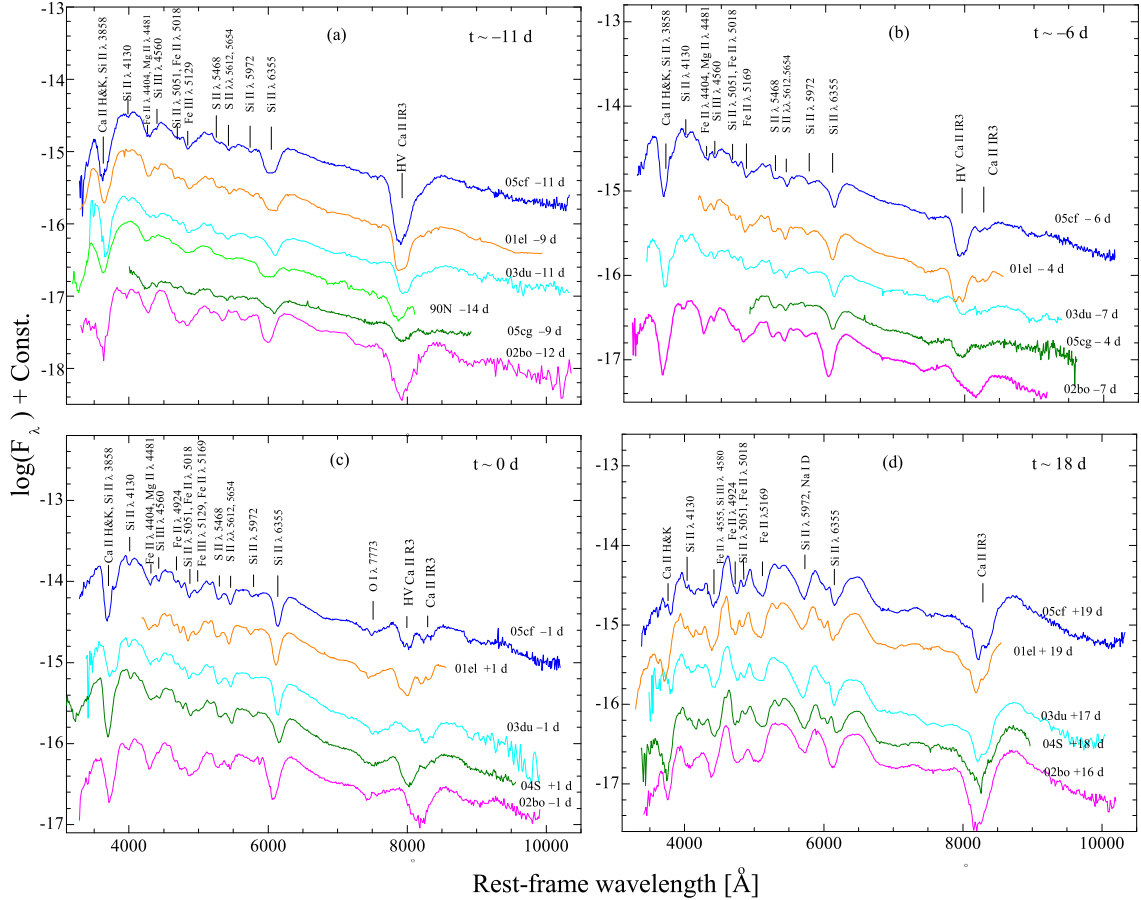}}
\vspace{-1.0cm} \caption{The spectrum of SN 2005cf at $-$11~d,
$-$6~d, 0~d, and +18~d after $B$ maximum, overplotted with the
comparable-phase spectra of SNe 1990N (Leibundgut et~al. 1991),
1994D (Filippenko et~al. 1997), 2001el (Wang et~al. 2003; Mattila
et~al. 2005), 2002bo (Benetti et~al. 2004), 2003du (Stanishev et~al.
2007), 2004S (Krisciunas et~al. 2007), and 2005cg (Quimby et~al.
2006). All spectra shown here have been corrected for the reddening
and redshift of the host galaxy.  For clarity of display, the
spectra were arbitrarily shifted in the vertical direction.}
\label{fig-17}
\end{figure*}

Figure 17a shows the comparison of the spectra at $t \approx -11$~d.
As in the comparison SNe~Ia, the absorption near 3700~\AA\ due to
the blending of Ca~II H\&K and Si~II $\lambda$3858 is prominent in
the earliest spectrum of SN 2005cf. The Si~II $\lambda$4130
absorption feature appears common in the early spectra, except in
the spectrum of SN 1990N. In the 4000--4500~\AA\ wavelength range,
all of the SNe show a strong absorption feature at $\sim$4300~\AA,
probably owing to a blend of Mg~II $\lambda$4481 and Fe~III
$\lambda$4404. The weak Si~III $\lambda\lambda$4553, 4568 blend can
be identified in the comparison SNe, though its absence in SN 2002bo
may indicate a cooler temperature in the photosphere.  The
``W"-shaped feature of the two S~II lines appears at this phase.
The lines of Si~II $\lambda$6355 and the Ca~II NIR triplet of SN
2005cf are very broad and deep, comparable to those in SNe 2001el
and 2002bo. One interesting feature is the flat-bottomed profile of
Si~II $\lambda$6355, which was previously only observed in SN 1990N
at $t = -14$~d and in SN 2001el at $t = -9$~d. In contrast, SNe
2003du and 2005cg displayed a triangular-shape feature at similar
phases.

Figure 17b shows the comparison at $t \approx -7$~d. A second
minimum begins to develop on the red side of the Ca~II H\&K
absorption, as in SN 2003du.  The weak features (e.g., Si~II
$\lambda$4130, Si~III $\lambda$4560, and the S~II ``W") strengthen
with time. The flat-bottomed feature associated with Si~II
$\lambda$6355 is barely visible in the SN 2005cf spectrum, with only
a small notch on the blue side of the absorption minimum. The Ca~II
NIR triplet now weakens and shows a double minimum on the red side
of the main absorption.

In Figure 17c, we compare the spectrum of SN 2005cf near maximum
with those of SNe 2001el, 2002bo, 2003du, and 2004S. At $t \approx
-1$~d, the spectrum of SN 2005cf has evolved while maintaining most
of its characteristics shown at earlier epochs. The second
absorption minimum in Ca~II H\&K now becomes noticeable in SNe
2005cf and 2003du, while it is still barely observed in SN 2004S.
(The spectrum of SN 2001el did not cover this wavelength range.) The
Si~II $\lambda$6355 absorption in SN 2005cf now appears quite
similar to that of the other normal SNe~Ia. The O~I $\lambda$7773
line strengthens in all cases in our sample. The two absorption
components of the Ca~II NIR triplet evolve rapidly, with the blue
component becoming weak and the red one gaining strength in the
minimum, similar to SNe 2001el and 2004S.  SN 2002bo still shows
broader and deeper absorptions of Si~II $\lambda$6355 and the Ca~II
NIR triplet with less substructure. We measured the line-strength
ratio of Si~II $\lambda\lambda$5958, 5979 to Si~II $\lambda$6355,
known as $R$(Si~II) (Nugent et al. 1995), to be $0.28 \pm 0.04$ for
SN 2005cf near maximum, in good agreement with the measurement
reported by G07.

In Figure 17d, we compare the spectra at $t \approx 18$~d. SN 2005cf
exhibits spectral evolution quite similar to that of the other
SNe~Ia. The Fe~II and Si~II features are fairly developed in the
range 4700--5000~\AA. The stronger Fe~II lines dominate around
5000\AA, and Na~I~D appears in the region overlapping with Si~II
$\lambda$5972. The Si~II $\lambda$6355 trough becomes affected by
Fe~II $\lambda\lambda$6238, 6248 and Fe~II $\lambda\lambda$6456,
6518. Although the Ca~II NIR triplet shows the most diverse features
at the earlier epochs, they develop into a rather smooth absorption
profile by 2 weeks after $B$ maximum.

Comparison of the spectra of SNe~Ia having similar $\Delta m_{15}$
values reveals that they show the most diversity at the earliest
epochs, with significantly different strengths and profiles of the
main features (e.g., Si~II $\lambda$6355 and the Ca~II NIR triplet).
In general, the overall spectral evolution of SN 2005cf at early
phases closely resembles that of SN 2001el.

\begin{figure}
\figurenum{18} \vspace{-0.3cm}
\centerline{\includegraphics[angle=0,width=100mm]{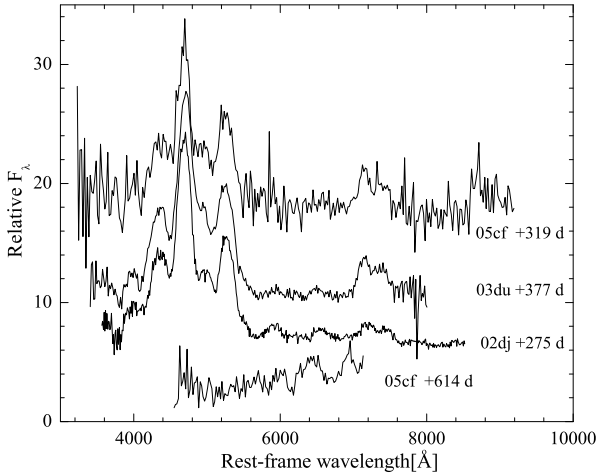}}
\vspace{-0.2cm} \caption{Late-time nebular spectra of SN 2005cf,
slightly smoothed. The nebular spectra of SN 2003du (Stanishev
et~al. 2007) and SN 2002dj (Pignata et~al. 2008) at about the same
phase as the earlier SN 2005cf spectrum are shown for comparison.
Both reddening and redshift corrections have been applied to the
spectra.} \label{fig-18}
\end{figure}

Two late-time nebular Keck spectra, obtained with LRIS on day +319
and with DEIMOS on day +614, are shown in Figure 18. They do not
exhibit any detectable signature of a low-velocity hydrogen
emission, though the S/N is low, especially in the case of the day
+614 spectrum. (See also the deep observations of SN 2005cf
performed by Leonard 2007.) The spectrum on day +319 is dominated by
the forbidden lines of singly and doubly ionized Fe and Co lines;
its overall shape is very similar to that of SN 2003du at a similar
phase. The nebular spectrum of SN 2002dj at this phase is also quite
normal, without showing extra flux (or an evidence for presence of
an echo) at shorter wavelengths as in the extreme HV event SN 2006X
(Wang et~al. 2008b). We note, however, that SN 2002dj is a slightly
fast-expanding object with less significant reddening (Pignata
et~al. 2008).

\subsection{The Photospheric Expansion Velocity}

\begin{figure}
\figurenum{19} \vspace{-0.6cm}
\centerline{\includegraphics[angle=0,width=105mm]{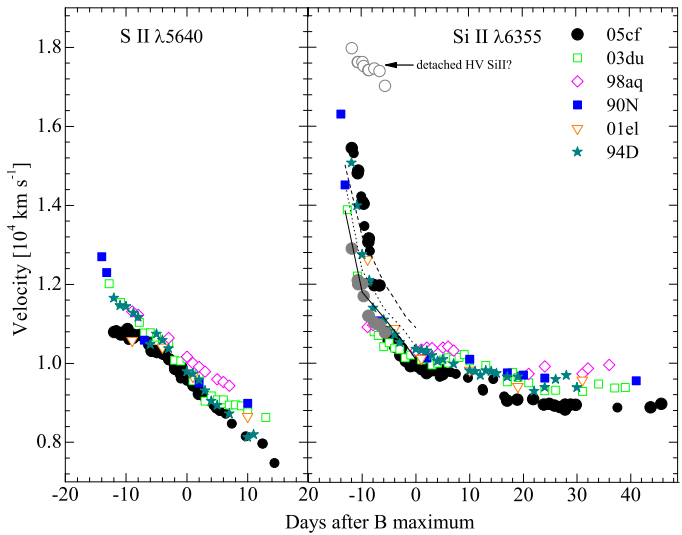}}
\vspace{-0.5cm} \caption{Evolution of the expansion velocity of SN
2005cf as measured from the minimum of Si~II $\lambda$6355 and S~II
$\lambda$5640, compared with the values of SNe 1990N, 1994D, 1998aq,
2001el, and 2003du (see text for the references). The grey dots show
the low-velocity component decomposed from the Si~II $\lambda$6355
absorption of SN 2005cf by using the double-Gaussian model (see \S
4.3.1 and Fig.20); the open circles represent the detached
high-velocity component. Overplotted are velocities predicted by the
Lentz et~al. (2000) model for cases of $\times$10 (dashed line),
$\times$3 (dotted line), and $\times$1/3 (solid line) solar C + O
layer metallicity.} \label{fig-19}
\end{figure}

In this paper, we examine the photospheric expansion velocity
$v_{\rm exp}$ from the velocity evolution of the Si~II $\lambda$6355
and S~II $\lambda$5640 lines. The derived $v_{\rm exp}$ values of SN
2005cf from Si~II $\lambda$6355 and S~II $\lambda$5640 as a function
of time are shown in Figure 19, together with those of the
comparison SNe~Ia. The measurements from the spectra published by
G07 are also overplotted (small filled circles). All velocities have
been corrected for the redshifts of the host galaxies.

At the earliest phases, the photospheric expansion velocity implied
from Si~II $\lambda$6355 for SN 2005cf is higher than that for SNe
1990N (Leibundgut et~al. 1992), 1998aq (Branch et~al. 2003), and
2003du (Stanishev et~al. 2007) by $\sim$2000 km s$^{-1}$, and comparable
to that for SNe 1994D (Filippenko 1997, Patat et~al. 1996) and 2001el
(Wang et~al. 2003; Mattila et~al. 2005). This expansion velocity declines
very rapidly within the first two weeks before $B$ maximum and then
maintains a plateau phase for about a month.  Such evolution may be
related to the fact that the Si~II absorption region is close to the
photosphere at earlier phases but becomes more detached at later
times (Patat et~al. 1996). In contrast, the velocity yielded from
the S~II $\lambda$5640 line is slightly lower than that of the other
SNe~Ia and shows a flat evolution from $t \approx -12$~d to $-$5~d.
SN 2001el may show a similar plateau feature, though the data are
sparse.

Following Benetti et~al. (2005), we calculate the velocity gradient
$\dot{v}$ of Si~II $\lambda$6355 for SN 2005cf as $31 \pm 5$ km
s$^{-1}$ d$^{-1}$ during the period $t \approx 0$--30~d, which puts
SN 2005cf in the group of normal SNe~Ia having low velocity
gradients (LVGs).


\subsection{The High-Velocity Features}

In addition to the evolution of the photospheric expansion, the
well-sampled spectra of SN 2005cf (especially those at earlier
phases) provide a good opportunity to study the high-velocity (HV)
features that were only seen in the earliest spectra.  The HV
material is usually located in the outermost layers of the ejecta
where SNe~Ia show the highest degree of heterogeneity.

\subsubsection{The High-Velocity Si~II}

Figure 20 (see the left panel) shows the pre-maximum evolution of
the velocity-space distribution of Si~II $\lambda$6355 in the
spectra of SN 2005cf. In the $-$12~d spectrum, this absorption
profile is broad and asymmetric with a stronger minimum on the blue
side. At $t = -11$~d, the Si~II line exhibits a flat-bottomed
feature with comparable absorption minima on both red and blue
sides. As the spectrum evolves, the red-side minimum gradually
dominates. By $t = -6$~d, the Si~II absorption feature gradually
develops into a single minimum with a typical velocity of
$\sim$10,500 km s$^{-1}$, though the line profile may still be
affected by the blue component until around maximum light (G07).

The observed line profiles of Si~II $\lambda$6355 at earlier phases
can be well fit by a double-Gaussian function with separate central
wavelengths, probably suggesting the presence of another HV
absorption component. The HV component could be a thin pure Si shell
or a mixed layer of Si~II and C~II $\lambda$6580 (Fisher et~al.
1997; Mazzali et~al. 2001, 2005). Assuming that the blue-side
absorption component is primarily produced by the HV Si~II detached
from the photosphere, the mean absorption-minimum velocity is
estimated to be $17,500 \pm 500$ km s$^{-1}$ during the period from
$t = -12$~d to $t = -7$~d (see also the open circles in Fig.19).
This velocity is much higher than the value inferred from the
absorption minimum on the red side.

The presence of HV Si~II in SN 2005cf was also proposed by G07 by
modeling the observed spectra with the use of the parameterized code
for supernova synthetic spectroscopy, SYNOW (Fisher 2000). Their
analysis suggested that the HV feature of Si~II $\lambda$6355 in SN
2005cf is detached at 19,500 km s$^{-1}$.  Given the contamination
by such a HV Si~II feature, the photospheric velocity measured
directly from the overall line profile was overestimated.  The
recomputed $v_{\rm exp}$ values from the Si~II line (see the gray
dots in Fig. 19), after removing the HV component, closely match the
velocity evolution predicted from the 1/3 solar metallicity models
by Lentz et~al (2000).


\begin{figure}
\figurenum{20} \vspace{-0.6cm}
\centerline{\includegraphics[angle=0,width=105mm]{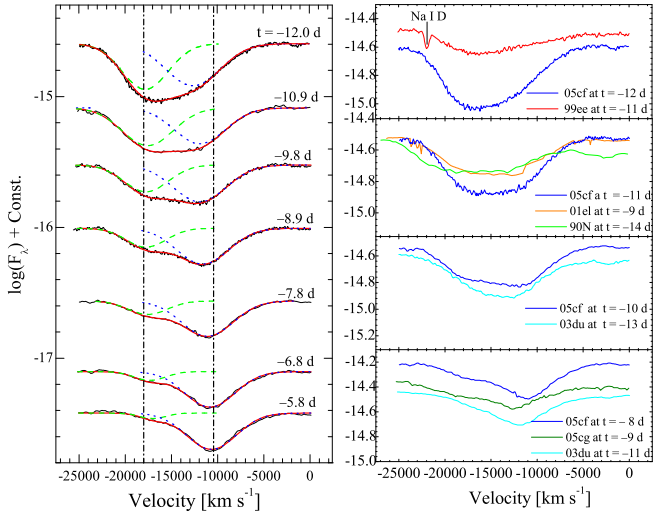}}
\vspace{-0.2cm}
\caption{{\it Left Panel}: The evolution in velocity space of Si~II
$\lambda$6355 in SN 2005cf compared with the double-Gaussian fit.
Dashed green lines show the velocity distribution on the blue side,
and dotted blue lines show the component on the red side. The red
solid lines represent the best-fit curve to the observed profile.
{\it Right Panels}: Comparison of pre-maximum evolution of the Si~II
$\lambda$6355 profile of SN 2005cf with that of SNe 1990N, 1999ee,
2001el, 2003du, and 2005cg at four selected epochs (see text for the
references).} \label{fig-20}
\end{figure}

In the right panels of Figure 20, we also compare the Si~II
absorption features in the earliest spectra.  The presence of the HV
feature of Si~II has also been suggested in SNe 1990N, 1999ee,
2001el, and SN 2005cg (Mazzali et~al. 2001; Mattila et~al. 2005;
Quimby et~al. 2006), but none of them could be securely established
due to the sparse coverage of the pre-maximum spectra. Of the above
SNe~Ia, the flat-bottomed profile of the Si~II line is seen only in
SN 1990N at $t = -14$~d and SN 2001el at $t = -9$~d. Inspection of
the $t = -13$~d spectrum of SN 2003du reveals a Si~II absorption
reminiscent of the feature seen in SN 2005cf at $t = -10$~d. A
triangular-shaped profile is present in the spectra of all the other
events. It is therefore likely that such a ``peculiar" line profile
is just a snapshot of the common evolutionary pattern (see similar
arguments by Stanishev et~al. 2007).




In addition to the absorption by the HV material, alternative
interpretations have also been proposed for the formation of the
broad profile of Si~II absorption. Mattila et~al. (2005) suggest
that the flat-bottomed line shape in SN 2001el can be produced by
pure scattering within a thin region moving at the continuum
photospheric velocity; it disappears as the photosphere recedes and
the scattering region widens.  The ejecta slowly extending to the HV
part, typical for delayed-detonation models (Khokhlov 1991), may
also account for the triangular feature of the Si~II profile in SN
2005cg (Quimby et~al. 2006). However, neither of these two models
can explain the asymmetric line profile with a stronger HV
component, as observed in SN 2005cf at $t = -12$~d and in SN 1999ee
at $t\approx-11$~d (Hamuy et~al. 2002)(see the top right panel of
Fig. 20).

\subsubsection{The High-Velocity Ca~II}

Figure 21 presents the detailed evolution of the Ca~II NIR triplet
and Ca~II H\&K of SN 2005cf. In comparison with the Si~II line, the
HV features in the Ca~II NIR triplet are more frequently observed in
SNe~Ia (e.g., Mazzali et~al. 2005), as they are more pronounced and
may have a longer duration. The HV component dominates in the Ca~II
NIR lines at earlier phases, but it gradually loses its strengths
with time. In the Ca~II NIR triplet, the HV components are more
separated from the photospheric components than in the Si~II line.
At $t = -11$~d, the HV component shows an expansion velocity at
about 22,000 km s$^{-1}$. By $t = +2$~d, the velocity is around
19,000 km s$^{-1}$ and the absorption minimum is beginning to be
dominated by the photospheric component at $\sim$10,000 km s$^{-1}$.

\begin{figure}
\figurenum{21} \vspace{-0.5cm}
\centerline{\includegraphics[angle=0,width=90mm]{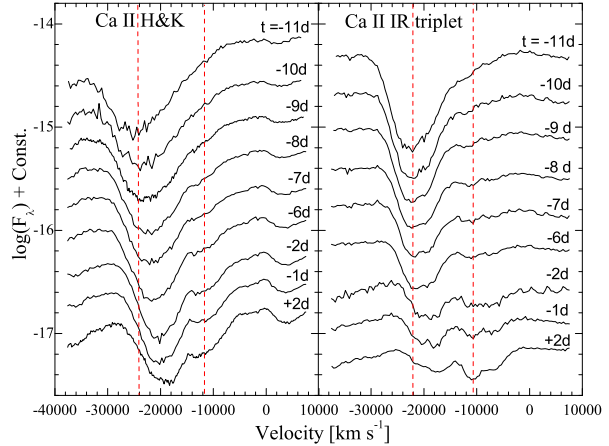}}
\vspace{-0.2cm}
\caption{Premaximum evolution in velocity space for the Ca~II H\&K and NIR triplet of
SN 2005cf. Vertical dashed lines indicate, respectively, the highest velocity in the
earliest phase ($\sim$ 23000 km s$^{-1}$) and the photospheric velocity
(at $\sim$ 11000 km s$^{-1}$).} \label{fig-21}
\vspace{0.3cm}
\end{figure}

The Ca~II H\&K lines may show similar HV features, but they overlap
with the Si~II $\lambda$3858 line at earlier phases. Due to the
severe line blending, it is difficult to disentangle the HV
component and quantify its strength. Assuming a double-Gaussian
model, we measure the velocity of the HV component at $\sim$24,500
km s$^{-1}$ and 20,000 km s$^{-1}$ in the $t = -11$~d and $-1$~d
spectra, respectively, similar to those measured for the Ca~II NIR
triplet. The higher velocity of the HV component in the Ca~II lines
with respect to the Si~II line perhaps suggests a different
abundance distribution in the ejecta.


\subsubsection{Origin for the HV features}

Currently, the origin of the HV features is still debated. In
principle, their formation can be due to an abundance and/or a
density enhancement in the ejecta at outer layers. According to
Mazzali et~al. (2005), the single abundance enhancement may not
account for the observed HV features, as the nuclear burning cannot
produce enough Si and Ca required in the outer region. In contrast,
an enhancement in the local density could lead to a good
reproduction of the HV spectral evolution, as demonstrated by the
three-dimensional (3D) explosion models (e.g., Ropke et~al. 2006).
One possible scenario for the density enhancement is the interaction
of the outermost ejecta with the circumstellar matter (CSM) produced
in the vicinity of the SN (e.g., Gerardy et~al. 2005). On the other
hand, an aspherical ejecta model could also cause the HV density
enhancement. Variations of the strength of the HV features may be
explained by different viewing angles if they result from aspherical
structures like a torus or clumps (Tanaka et~al. 2006).

Inspection of the early-time spectra presented in Figure 17 reveals
that the strength of the HV features from Si~II and Ca~II may be
correlated in a given SN, though their strength varies from SN to
SN. This holds true for SN 2005cf and all the comparison objects,
perhaps favoring the origin of the observed HV features from either
CSM interaction or aspherical structure of the ejecta produced by
the explosion itself. Note that the detached HV features discussed
here seem different from those observed in SN 2006X or even in SN
2002bo; however, the same configuration with a much larger
photospheric velocity may explain the difference (Tanaka et~al.
2008).

\section{The Distance and Luminosity of SN 2005cf}

The extensive photometric observations of SN 2005cf, from the UV to
the NIR bands, enable us to construct the $uvoir$ ``bolometric"
light curve within 0.2--2.4~$\mu$m. For this calculation, we used
the normalized passband transmission curves given by Bessell (1990).
The integrated flux in each filter was approximated by the mean flux
multiplied by the effective width of the passband, and was corrected
for the reddening. Since the filter transmission curves do not
continuously cover the spectrum and some also overlap, we corrected
for these gaps and overlaps by adjusting the effective wavelengths
of the filters in the UV, optical, and NIR passbands.

\subsection{The Distance to SN 2005cf}

The distance to SN 2005cf is important for deriving the bolometric
luminosity. Direct measurements were unavailable in the literature,
so we used several methods to estimate the distance.  According to
Wang et~al. (2006), a nominal standard SN~Ia with Cepheid-based
calibration has an absolute magnitude of $-19.27 \pm 0.05$ mag in
the $V$ band. Combining this value with the fully corrected apparent
magnitude of SN 2005cf, we obtain a distance modulus $\mu = 32.29
\pm 0.12$ mag. We also determine the distance using the latest
version of the MLCS2k2 fitting technique (Jha et~al. 2007), which
yields $\mu = 32.43 \pm 0.13$ mag.

Krisciunas et~al. (2004b) propose that SNe~Ia are more uniform in the
NIR bands than in the optical; their NIR peak luminosities are found
to be nearly independent of the light-curve shape. The absolute NIR
peak magnitudes for SNe~Ia with $\Delta m_{15} < 1.7$ are reported
as $-$18.61 in $J$, $-$18.28 in $H$, and $-$18.44 mag in $K$,
respectively. Assuming the same NIR magnitudes for SN 2005cf, we
derive a mean distance modulus $\mu = 32.20 \pm 0.10$ mag.

Averaging the above three distances, we obtain a weighted mean of
$\mu = 32.31 \pm 0.11$ mag for SN 2005cf. Note that this estimate
may still suffer from an additional uncertainty of 0.12--0.15 mag,
due to the intrinsic luminosity dispersion of SNe~Ia in both optical
and NIR bands (Krisciunas et~al. 2004; Wang et~al. 2006; Wood-Vasey
et~al. 2007b).

\subsection{The Missing Flux Below the Optical Window}

The filled circles in Figure 22 show the temporal evolution of the
ratio of the NIR-band emission (9000--24,000~\AA) to the optical
(3200--9000~\AA), $F_{\rm IR}/F_{\rm opt}$. The flux ratios obtained
from SNe 2001el and 2004S are overplotted. The dashed curve in the
plot represents the best fit to the data of SN 2005cf.  In the fit,
the NIR contribution at $t > 40$~d is assumed to be the same as that
of SN 2001el and SN 2004S. This assumption is reasonable because the
$F_{\rm IR}/F_{\rm opt}$ ratio of SN 2005cf agrees well with the
corresponding values of these two SNe~Ia from the very beginning to
$t \approx +40$~d. Initially, the ratio shows a sharp decline with a
minimum $\sim$4~d after the $B$ maximum. Then it rises rapidly in a
linear fashion and reaches a peak ($\sim$20\%) at $t \approx +30$~d,
when the secondary maximum appears in the NIR. At nebular phases,
the NIR contribution gradually declines and becomes $<$10\% at $t
\approx +80$~d, similar to that found by Suntzeff in studying SN
1992A (1996).

The contribution of the UV flux in the 1600--3200~\AA\ range to the
optical is shown with the open circles in Fig. 22. The dotted line
indicates the best fit to the observed data points, assuming that
the UV flux remains constant at $t > +30$~d.  A flat contribution at
late times was evidenced by the UV data of SN 2001el from the {\it
HST} archive (see Fig. 14 in Stanishev et~al. 2007).  The UV
contribution is found to be generally a few percent of the optical
in SN 2005cf, with a peak ($\sim$10\%) at around the $B$ maximum.
Three weeks after the maximum, the ratio remains at a level of
3--4\%.

\begin{figure}
\figurenum{22} \vspace{-0.4cm}
\centerline{\includegraphics[angle=0,width=90mm]{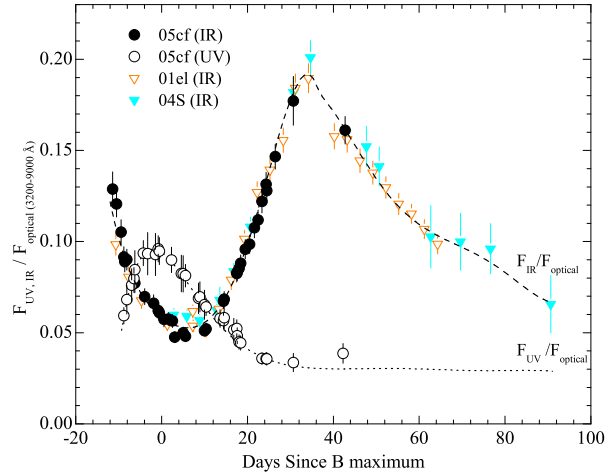}}
\vspace{-0.5cm}
\caption{The ratio of the UV and NIR fluxes to the optical for SN 2005cf.
Overplotted are the NIR flux ratios of SNe 2001el and 2004S.}
\label{fig-22}
\end{figure}

It is also interesting to estimate the ratios of the optical fluxes
to the bolometric fluxes, as most SNe~Ia were not observed in the UV
or the NIR bands. These ratios are useful for correcting the
observed optical luminosity to the bolometric luminosity. Figure 23
shows the correction factors (defined as the missing fraction of the
observed flux relative to the generic $uvoir$ bolometric luminosity)
obtained from SN 2005cf; one can see that the bolometric luminosity
is dominated by the optical fluxes. The missing flux beyond the
optical window is about 20\%, with a slightly larger fraction at $t
\approx +30$~d due to the appearance of the secondary maximum in the
NIR. The corrections, based on the fluxes in the $UBVRI$, $BVRI$,
$BVI$, $BV$, and $V$ bands, are also shown in the plot. We note that
the bolometric correction for the $V$ filter exhibits the least
variation with time: the overall variation is less than 4\% before
$t \approx +30$~d, and at later phases the correction stays nearly
constant. This validates previous assumptions that the $V$-band
photometry can well represent the bolometric light curve, in
particular at later phases, at a constant fraction of about 20\%.

\begin{figure}
\figurenum{23} \vspace{0.3cm}
\centerline{\includegraphics[angle=0,width=100mm]{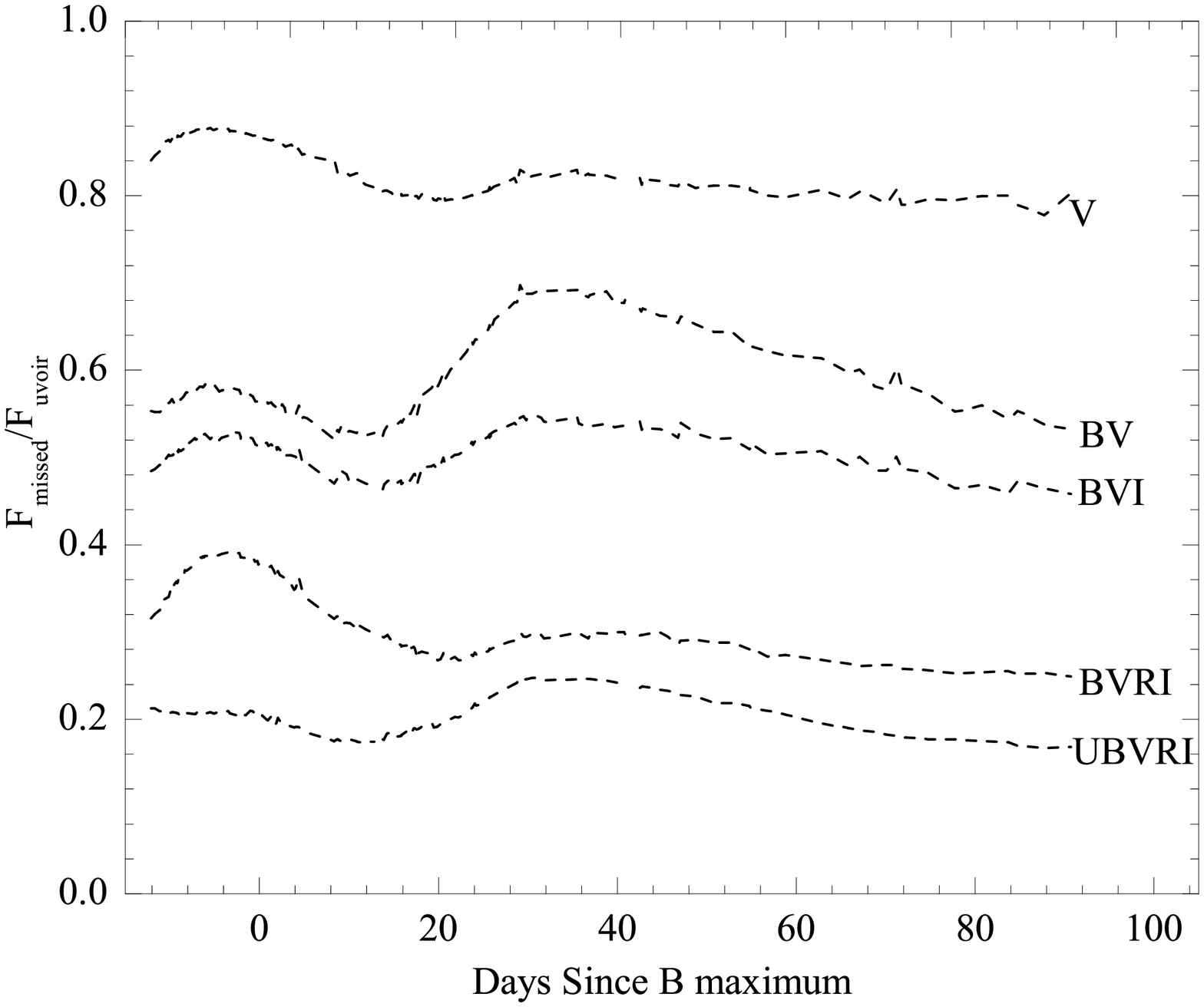}}
\vspace{-1.0cm} \caption{Correction factors for missing passbands.
The correction factors are obtained by comparing the fluxes in the
passbands with the total $uvoir$ flux.} \label{fig-23}
\end{figure}

\subsection{The Bolometric Light Curve}

Figure 24 shows the ``$uvoir$'' bolometric light curves of SN 2005cf
and several other SNe~Ia. The UV emission of all the comparison SNe was
corrected on the basis of SN 2005cf. Similar corrections were
applied to their light curves when the NIR observations are missing.
The distances to the comparison SNe~Ia were derived using the methods
adopted in \S 5.1. The $uvoir$ bolometric light curves of our SN~Ia are
overall very similar in shape, with the exceptions of SN 2002bo and SN 2002dj,
which seem to rise at a faster pace.  The maximum bolometric
luminosity of SN 2005cf is estimated to be $(1.54 \pm 0.20) \times
10^{43}$ erg s$^{-1}$ around the $B$-band maximum, which is similar
to that of SNe 2001el and 2003du but less than that of SNe 2002bo
and 2002dj. However, differences in the bolometric luminosity at
maximum may be due to errors in the absorption corrections and/or
the distance modulus.


\begin{figure}
\figurenum{24}
\centerline{\includegraphics[angle=0,width=100mm]{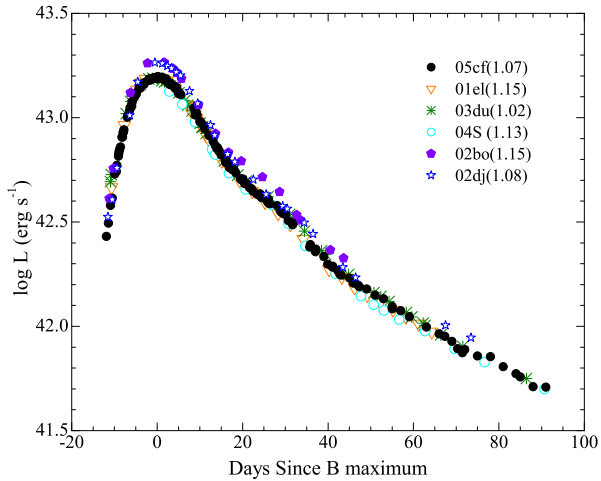}}
\vspace{-1.0cm} \caption{The $uvoir$ bolometric light curve of SN
2005cf. Overplotted are the corresponding light curves of the
comparison SNe~Ia. The number in parentheses represent the $\Delta
m_{15}$ values for the SNe~Ia.}
 \label{fig-24}
\end{figure}

With the derived bolometric luminosity, we can estimate the
synthesized $^{56}$Ni mass --- one of the primary physical
parameters determining the peak luminosity, the light curve width,
and the spectroscopic evolution of SNe~Ia (e.g., Kasen et al. 2006).
Assuming the Arnett law (Arnett 1982; Arnett et~al. 1985; Branch
1992), the maximum luminosity produced by the radioactive $^{56}$Ni
can be written as (Stritzinger \& Leibundgut 2005)
\begin{equation}
L_{\rm max} = (6.45e^{\frac{-t_{r}}{(8.8~{\rm d})}} +
 1.45e^{\frac{-t_{r}}{(111.3~{\rm d})}})(\frac{M_{\rm Ni}}{M_{\odot}})
\times 10^{43} \,\rm{erg\ s^{-1}},
\end{equation}

where $t_{r}$ is the rise time of the bolometric light curve, and
$M_{Ni}$ is the $^{56}$Ni mass (in units of solar masses,
M$_{\odot}$). With the photometric data in the $R$ band and our
earliest unfiltered data from the database of the KAIT survey, we
estimate the rise time to the $B$ maximum as $18.4 \pm 0.5$~d.
Inserting this value and the maximum bolometric luminosity into the
above equation, we derive a nickel mass of $0.77 \pm 0.11$
M$_{\odot}$ for SN 2005cf. This is within the reasonable range of
$^{56}$Ni masses for a normal SN~Ia. The quoted error bar includes
uncertainties in the rise time and in the peak luminosity.

The lower bolometric luminosity and smaller nickel mass obtained by
P07 is primarily due to the neglected host-galaxy extinction for SN
2005cf in their analysis. Table 13 lists all of the important
parameters for SN 2005cf and its host galaxy, as we derived in the
previous sections.

\section{Discussion and Conclusions}

In this paper we present extensive optical, UV, and NIR photometry
as well as optical spectroscopy of SN 2005cf. In particular, the
photometric observations in the optical bands were extremely
well-sampled with numerous telescopes. To minimize systematic
deviations from the standard system, we carefully computed the
$S$-corrections and applied them to SN 2005cf. Photometry without
such corrections may potentially lead to a noticeably inconsistent
measurement of the reddening, which makes precise extinction
corrections difficult.

The {\it Swift} UVOT optical photometry is found to be consistent
with the ground-based observations to within $\lesssim$0.05 mag,
after applying the $S$-corrections. The {\it Swift} UVOT UV uvm2 and
uvw2 photometry is relatively close to that of {\it HST} ACS F220W
and F250W, respectively, with a small shift of up to 0.1--0.2 mag.


Our observations show that SN 2005cf is a normal SN~Ia with a
typical luminosity decline rate $\Delta m_{15}$(true) = $1.07 \pm
0.03$ mag. Based on the $V - JHK$ colors and the refurbished
Color-$\Delta m_{15}$ relations, we estimated the host-galaxy
reddening of $E(B - V)_{\rm host} = 0.09 \pm 0.03$ mag, which is
small but non-negligible. The NIR light curves closely resemble
those of normal SNe~Ia, with a prominent secondary maximum. One
distinguishing feature is the UV luminosity in the 2000--2500~\AA
range which appears much fainter and peaks later than that of the
neighboring bands. This is likely caused by intrinsic spectral
features in the UV, an explanation favored by the evolution of the
SED.

The comprehensive data, from UV to NIR bands, allow us to establish
the $uvoir$ bolometric light curve of SN 2005cf in the
1600--24,000~\AA\ range. The maximum bolometric luminosity is found
to be $(1.54 \pm 0.20) \times 10^{43}$ erg s$^{-1}$, corresponding
to a synthesized nickel mass of $0.77 \pm 0.11$~M$_{\odot}$. The
bolometric luminosity is dominated by the optical emission; the UV
contribution is found to be a few percent of the optical, and the
peak value ($\sim$10\%) occurs at around maximum light. The NIR flux
contribution shows more complicated temporal evolution: the ratio
decreases from the beginning of the explosion and reaches a minimum
of $\sim$5\% at $t \approx +5$~d; it then rises up to a peak value
of $\sim$20\% at $t \approx 35$~d, and finally declines in a linear
fashion.

In general, the optical spectra of SN 2005cf are similar to those of
normal SNe~Ia. Strong HV features are distinctly present in the
Ca~II NIR triplet and the Ca~II H\&K lines; these are detached from
the photosphere at a velocity $\sim$19,000-24,000 km s$^{-1}$. A HV
Ca~II feature is commonly seen in other SNe~Ia, while the
flat-bottomed shape of the Si~II $\lambda$6355 line is relatively rare
perhaps due to the paucity of the very early spectra. The excellent
temporal coverage of the spectra of SN 2005cf reveals that either
the flat-bottomed feature or the triangular-shaped feature
associated with Si~II $\lambda$6355 in some SNe~Ia might be due to
contamination by another HV absorption component, such as the pure
Si~II shell.

Although HV absorptions of Ca~II and Si~II are observed in many
SNe~Ia, they diverge in strength and duration. For instance, the HV
features are very strong in SN 2005cf and SN 2001el, remaining until
a few days after $B$ maximum, whereas they are relatively weak in SN
2003du and SN 2005cg and becomes marginally detectable around $B$
maximum. Given a common origin of the HV absorption from the
aspherical ejecta (e.g., a torus or clumps), the diversity in
strength of the HV features could be interpreted by a line-of-sight
effect (see also Tanaka et~al. 2006). Aspherical structure of the
ejecta is favored by the evidence that the degree of polarization
for the line features in SN 2001el is much higher than that for the
continuum (Wang et~al. 2003; Kasen et~al. 2003; see also Chornock \&
Filippenko 2008 for SN 2004S). It is thus interesting to examine
whether the strength of the HV line features correlates, at least
for a subset of SNe~Ia, with that of the line polarization.
Unfortunately, similar high-quality polarimetric measurements at
early phases are sparse for SNe~Ia (but see Wang et~al. 2007).
Obviously, polarization spectra obtained from the very beginning
would allow us to penetrate the layered geometrical structure of the
SN ejecta as well as the immediate environment surrounding the
explosion site.

The well-observed, multi-wavelength data presented in this paper
makes SN 2005cf a rare "golden standard" sample of normal type Ia supernova,
which could be used as a testbed either for the theoretical models of SN Ia, or
for the studies of the systematic errors in SN~Ia cosmology.


\acknowledgments
We thank Ryan Chronock and Stefanie Blondin for useful discussions.
Some of the data presented herein were obtained at the W. M. Keck
Observatory, which is operated as a scientific partnership among the
California Institute of Technology, the University of California,
and the National Aeronautics and Space Administration (NASA). The
Observatory was made possible by the generous financial support of
the W. M. Keck Foundation. We thank the the Lick Observatory,
Palomar Observatory, NAOC, and CTIO staffs for their assistance with
the observations. The research of A.V.F.'s supernova group at UC
Berkeley is supported by NSF grant AST--0607485, the TABASGO
Foundation, Gary and Cynthia Bengier, and Department of Energy grant
DE-FG02-08ER41563. Additional support was provided by NASA grant
GO--10182 from the Space Telescope Science Institute (STScI), which is operated by
AURA, Inc., under NASA contract NAS5-26555. We are also grateful to
the National Natural Science Foundation of China (NSFC grant
10673007), the 973 Key Program of China (2009CB824800), and the
Basic Research Funding at Tsinghua University (JCqn2005036). Supernova
research at Harvard University is supported by NSF grant AST06-06772.
M.M. is supported by a fellowship from the Miller Institute for Basic Research
The work of AG is supported by grants from the Israeli Science
Foundation  and the EU Marie Curie IRG program, and by the
Benoziyo Center for Astrophysics, a research grant from the
Peter and Patricia Gruber Awards, and the William Z. and Eda Bess
Novick New Scientists Fund at the Weizmann Institute. KAIT and its
ongoing operation were made possible by donations from Sun Microsystems,
Inc., the Hewlett-Packard Company, AutoScope Corporation, Lick
Observatory, the NSF, the University of California, the Sylvia
\& Jim Katzman Foundation, and the TABASGO
Foundation. The PAIRITEL project is operated by the Smithsonia
Astrophysical Observatory (SAO) and was made possible by a grant
from the Harvard University Milton Fund, a camera loan from the
University of Virginia and continued support of the SAO and UC
Berkeley. PAIRITEL is further supported by the NASA/$Swift$ Guest
Investigator grant NNG06GH50G. The CTIO 1.3~m telescope is operated
by the Smart and Moderate Aperture Research Telescope System
(SMARTS) Consortium. We are particularly grateful for the scheduling
flexibility of SMARTS. The Liverpool Telescope is operated on the island of La Palma by
Liverpool John Moores University in the Spanish Observatorio del Roque de
los Muchachos of the Instituto de Astrofisica de Canarias with financial
support from the UK Science and Technology Facilities Council. We made
use of the NASA/IPAC Extragalactic
Database (NED), which is operated by the Jet Propulsion Laboratory,
California Institute of Technology, under contract with NASA.

\clearpage
\begin{deluxetable*}{lcrrrr}
\tablecolumns{6} \tablewidth{0pc}
\tablecaption{Instrumental Color Terms for Different Telescopes}
\tablehead{ \colhead{Telescope} & \colhead{$U$} & \colhead{$B$} &
\colhead{$V$} & \colhead{$R$} & \colhead{$I$} } \startdata
KAIT 0.76~m  & $-$0.085(017) &$-$0.043(011) & 0.035(007)  &  0.070(012) &  $-$0.010(006)  \\
FLWO 1.2~m & $-$0.037(003) &$-$0.080(003) & 0.039(002)   &  0.207(010) &     0.109(008)  \\
CTIO 1.3~m & \nodata  &$-$0.035(001) & 0.049(005)  &  0.029(009) &     0.070(006)  \\
CTIO 0.9~m & $-$0.100(004)&0.096(011)&$-$0.016(001)&  0.006(001) &  $-$0.006(001)  \\
Palomar 1.5~m &\nodata&$-$0.100(011) &0.020(007)   &  0.070(012) &     0.040(006)  \\
Lick 1.0~m & $-$0.080(010)&$-$0.080(011) &0.060(007)   &  0.100(012)& $-$0.035(006)  \\
Liverpool 2.0~m&\nodata&$-$0.045(004)&0.054(005)   &  0.200(008) &     0.100(004)  \\
TNT 0.8~m   &  \nodata      &$-$0.132(004) & 0.080(004)  &  0.106(006) &  $-$0.037(003)  \\
\enddata
\tablenotetext{}{Note: Uncertainties, in units of 0.001 mag, are
$1\sigma$.}
\end{deluxetable*}

\begin{deluxetable*}{lllllr}
\tablecolumns{6} \tablewidth{0pc}
\tablecaption{Magnitudes of Photometric Standards in the SN 2005cf
Field\tablenotemark{a}} \tablehead{ \colhead{Star} & \colhead{$U$} &
\colhead{$B$} & \colhead{$V$} & \colhead{$R$} & \colhead{$I$} }
\startdata
1 &  \nodata   & 15.265(010) & 14.380(007) &13.877(006)&13.493(093) \\
2 & 13.650(030)& 13.486(012) & 12.799(013) &12.417(076)&12.035(017) \\
3 & 16.370(030)& 15.625(008) & 14.676(012) &14.086(002)&13.560(008) \\
4 & 16.582(030)& 16.466(008) & 15.756(007) &15.349(008)&14.947(014)\\
5 & 15.311(030)& 15.328(011) & 14.820(011) &14.519(005)&14.199(002) \\
6 & 14.014(030)& 14.059(027) & 13.604(014) &13.329(005)&13.037(019) \\
7 & 18.297(030)& 18.434(017) & 17.786(001) &17.518(020)&17.156(025) \\
8 & 18.926(030)& 17.810(020) & 16.264(007) &15.236(007)&14.048(007) \\
9 & 15.831(030)& 15.660(023) & 14.986(010) &14.597(007)&14.248(023) \\
10& 18.294(030)& 17.124(016) & 15.947(011) &15.155(012)&14.488(018) \\
11& 14.956(030)& 14.747(008) & 14.022(005) &13.594(001)&13.201(017) \\
12& 19.353(030)& 18.338(012) & 17.327(017) &16.708(009)&16.205(025) \\
13& 15.340(030)& 14.760(006) & 13.883(006) &13.337(005)&12.823(015) \\
14& 18.558(030)& 18.139(025) & 17.393(027) &16.925(030)&16.451(019) \\
15& 18.297(030)& 17.591(011) & 16.715(007) &16.187(005)&15.756(011) \\
16& 17.933(030)& 17.981(020) & 17.450(015) &17.107(039)&16.724(074) \\
\enddata
\tablenotetext{a}{See Fig.~1 for a chart of SN 2005cf and the
comparison stars.} \tablenotetext{}{Note: Uncertainties, in units of
0.001 mag, are $1\sigma$.}
\end{deluxetable*}

\clearpage

\LongTables
\begin{deluxetable}{lcccccccc}
\tablewidth{0pt} \tabletypesize{\scriptsize} \tablecaption{The $K$-
and $S$-corrected Optical Photometry of SN 2005cf.}
\tablehead{
 \colhead{UT Date} &
 \colhead{JD$-$2,450,000} &
 \colhead{Phase\tablenotemark{a}}&
 \colhead{$U$} &
 \colhead{$B$} &
 \colhead{$V$} &
 \colhead{$R$} &
 \colhead{$I$} &
 \colhead{Instrument\tablenotemark{b}}
 }
\startdata
2005    May 31  &   3521.75 &   -11.91  &   15.862(0.038)    &   15.600(0.022)    &   15.346(0.022)    &   15.189(0.032)    &   15.360(0.041)    &   2   \\
2005    May 31  &   3521.77 &   -11.89  &   15.850(0.038)    &   15.579(0.029)    &   15.293(0.024)    &   15.206(0.041)    &   15.312(0.041)    &   1   \\
2005    Jun 1   &   3522.74 &   -10.92  &   15.360(0.035)    &   15.151(0.022)    &   15.000(0.022)    &   14.838(0.032)    &   15.062(0.041)    &   2   \\
2005    Jun 1   &   3522.87 &   -10.79  &   15.254(0.035)    &   15.096(0.022)    &   14.955(0.022)    &   14.846(0.032)    &   14.953(0.041)    &   1   \\
2005    Jun 1   &   3523.15 &   -10.51  &   \nodata          &   15.054(0.041)    &   14.954(0.032)    &   14.793(0.032)    &   14.896(0.041)    &   3   \\
2005    Jun 2   &   3523.77 &   -9.89   &   14.851(0.035)    &   14.774(0.022)    &   14.721(0.022)    &   14.474(0.032)    &   14.701(0.041)    &   2   \\
2005    Jun 2   &   3523.87 &   -9.79   &   14.768(0.035)    &   14.765(0.022)    &   14.670(0.022)    &   14.536(0.032)    &   14.651(0.041)    &   1   \\
2005    Jun 2   &   3524.13 &   -9.53   &   \nodata          &   14.751(0.022)    &   14.654(0.022)    &   14.488(0.032)    &   14.573(0.041)    &   3   \\
2005    Jun 3   &   3524.63 &   -9.03   &   \nodata          &   14.587(0.022)    &   14.497(0.022)    &   14.411(0.032)    &   14.464(0.041)    &   4   \\
2005    Jun 3   &   3524.68 &   -8.98   &   \nodata          &   14.543(0.022)    &   14.502(0.022)    &   14.291(0.032)    &   14.487(0.041)    &   2   \\
2005    Jun 3   &   3524.79 &   -8.87   &   \nodata          &   14.512(0.022)    &   14.489(0.022)    &   14.326(0.032)    &   14.409(0.041)    &   5   \\
2005    Jun 3   &   3524.85 &   -8.81   &   14.380(0.035)    &   14.494(0.022)    &   14.438(0.022)    &   14.298(0.032)    &   14.407(0.041)    &   1   \\
2005    Jun 3   &   3525.42 &   -8.24   &   \nodata          &   14.343(0.024)    &   14.344(0.022)    &   14.190(0.032)    &   14.269(0.041)    &   6   \\
2005    Jun 4   &   3525.69 &   -7.97   &   14.154(0.035)    &   14.304(0.022)    &   14.293(0.022)    &   14.079(0.032)    &   14.259(0.041)    &   2   \\
2005    Jun 4   &   3525.76 &   -7.90   &   \nodata          &   14.293(0.022)    &   14.299(0.022)    &   14.128(0.032)    &   14.199(0.041)    &   5   \\
2005    Jun 4   &   3525.87 &   -7.79   &   \nodata          &   14.274(0.022)    &   14.247(0.022)    &   14.095(0.032)    &   14.217(0.041)    &   7   \\
2005    Jun 5   &   3526.68 &   -6.98   &   \nodata          &   14.127(0.022)    &   14.122(0.022)    &   13.916(0.032)    &   14.108(0.041)    &   2   \\
2005    Jun 5   &   3526.75 &   -6.91   &   13.883(0.035)    &   14.101(0.022)    &   14.118(0.022)    &   \nodata          &   \nodata          &   5   \\
2005    Jun 5   &   3527.44 &   -6.22   &   \nodata          &   14.007(0.022)    &   14.003(0.022)    &   13.875(0.032)    &   13.948(0.041)    &   6   \\
2005    Jun 6   &   3527.64 &   -6.02   &   \nodata          &   14.014(0.022)    &   13.992(0.022)    &   13.884(0.032)    &   13.948(0.041)    &   4   \\
2005    Jun 6   &   3527.69 &   -5.97   &   13.734(0.035)    &   13.979(0.022)    &   13.981(0.022)    &   13.785(0.032)    &   13.970(0.041)    &   2   \\
2005    Jun 6   &   3527.85 &   -5.81   &   13.655(0.035)    &   13.950(0.022)    &   13.942(0.022)    &   13.796(0.032)    &   13.915(0.041)    &   1   \\
2005    Jun 6   &   3528.43 &   -5.23   &   \nodata          &   13.863(0.022)    &   13.896(0.022)    &   13.781(0.032)    &   13.845(0.041)    &   6   \\
2005    Jun 7   &   3528.75 &   -4.91   &   \nodata          &   13.861(0.022)    &   13.852(0.022)    &   13.708(0.032)    &   13.863(0.041)    &   7   \\
2005    Jun 7   &   3528.84 &   -4.82   &   13.564(0.035)    &   13.843(0.022)    &   13.843(0.022)    &   13.697(0.032)    &   13.828(0.041)    &   1   \\
2005    Jun 8   &   3529.43 &   -4.23   &   \nodata          &   13.755(0.025)    &   13.795(0.022)    &   13.704(0.032)    &   13.765(0.041)    &   6   \\
2005    Jun 8   &   3529.71 &   -3.95   &   13.527(0.035)    &   13.768(0.022)    &   13.783(0.022)    &   13.683(0.032)    &   13.778(0.041)    &   8   \\
2005    Jun 8   &   3530.42 &   -3.24   &   \nodata          &   13.716(0.022)    &   13.736(0.022)    &   13.629(0.032)    &   13.774(0.041)    &   6   \\
2005    Jun 8   &   3530.59 &   -3.07   &   \nodata          &   13.741(0.022)    &   13.712(0.022)    &   13.644(0.032)    &   13.755(0.041)    &   4   \\
2005    Jun 9   &   3530.68 &   -2.98   &   13.433(0.035)    &   13.718(0.022)    &   13.713(0.022)    &   13.558(0.032)    &   13.786(0.041)    &   2   \\
2005    Jun 10  &   3531.67 &   -1.99   &   13.395(0.035)    &   13.677(0.022)    &   13.655(0.022)    &   13.523(0.032)    &   13.783(0.041)    &   2   \\
2005    Jun 10  &   3531.79 &   -1.87   &   \nodata          &   13.666(0.022)    &   13.630(0.022)    &   13.551(0.032)    &   13.767(0.041)    &   7   \\
2005    Jun 10  &   3531.83 &   -1.83   &   13.422(0.035)    &   13.664(0.022)    &   13.639(0.025)    &   13.577(0.032)    &   13.767(0.041)    &   1   \\
2005    Jun 10  &   3532.42 &   -1.24   &   \nodata          &   13.654(0.022)    &   13.638(0.022)    &   13.550(0.032)    &   13.745(0.041)    &   6   \\
2005    Jun 11  &   3532.87 &   -0.79   &   13.403(0.035)    &   13.643(0.022)    &   13.600(0.022)    &   13.548(0.032)    &   13.766(0.041)    &   1   \\
2005    Jun 11  &   3533.42 &   -0.24   &   \nodata          &   13.621(0.021)    &   13.595(0.022)    &   13.559(0.032)    &   13.772(0.041)    &   6   \\
2005    Jun 12  &   3533.66 &   0       &   \nodata          &   13.650(0.022)    &   13.592(0.022)    &   13.578(0.032)    &   13.748(0.041)    &   4   \\
2005    Jun 12  &   3533.72 &   0.06    &   \nodata          &   13.623(0.022)    &   13.570(0.022)    &   13.515(0.032)    &   13.785(0.041)    &   7   \\
2005    Jun 12  &   3533.84 &   0.18    &   13.393(0.035)    &   13.619(0.022)    &   13.568(0.022)    &   13.517(0.032)    &   13.761(0.041)    &   1   \\
2005    Jun 13  &   3534.73 &   1.07    &   13.444(0.035)    &   13.637(0.022)    &   13.574(0.022)    &   13.485(0.032)    &   13.821(0.041)    &   2   \\
2005    Jun 13  &   3534.84 &   1.18    &   13.412(0.035)    &   13.622(0.022)    &   13.549(0.022)    &   13.525(0.032)    &   13.773(0.041)    &   1   \\
2005    Jun 13  &   3535.43 &   1.77    &   \nodata          &   13.649(0.022)    &   13.581(0.022)    &   13.540(0.032)    &   13.787(0.041)    &   6   \\
2005    Jun 14  &   3535.72 &   2.06    &   \nodata          &   13.641(0.022)    &   13.564(0.022)    &   13.505(0.032)    &   13.812(0.041)    &   7   \\
2005    Jun 14  &   3535.74 &   2.08    &   13.489(0.035)    &   13.654(0.022)    &   13.574(0.022)    &   13.496(0.032)    &   13.858(0.041)    &   2   \\
2005    Jun 14  &   3535.83 &   2.17    &   13.487(0.035)    &   13.631(0.022)    &   13.556(0.032)    &   13.509(0.032)    &   \nodata          &   1   \\
2005    Jun 14  &   3536.44 &   2.78    &   \nodata          &   13.719(0.021)    &   13.591(0.022)    &   13.539(0.032)    &   13.800(0.041)    &   6   \\
2005    Jun 15  &   3536.70 &   3.04    &   13.564(0.035)    &   13.702(0.022)    &   13.581(0.022)    &   13.519(0.032)    &   13.884(0.041)    &   2   \\
2005    Jun 15  &   3536.83 &   3.17    &   13.563(0.035)    &   13.681(0.022)    &   13.558(0.022)    &   13.545(0.032)    &   13.840(0.041)    &   1   \\
2005    Jun 15  &   3537.47 &   3.81    &   \nodata          &   13.717(0.022)    &   13.616(0.022)    &   13.551(0.032)    &   13.833(0.041)    &   6   \\
2005    Jun 16  &   3537.82 &   4.16    &   \nodata          &   \nodata          &   13.554(0.022)    &   13.527(0.032)    &   13.814(0.041)    &   1   \\
2005    Jun 17  &   3538.68 &   5.02    &   \nodata          &   13.815(0.022)    &   13.624(0.022)    &   13.638(0.032)    &   13.884(0.041)    &   4   \\
2005    Jun 21  &   3542.61 &   8.95    &   \nodata          &   14.093(0.045)    &   13.733(0.022)    &   13.813(0.032)    &   14.059(0.041)    &   4   \\
2005    Jun 21  &   3542.72 &   9.06    &   14.117(0.054)    &   14.090(0.022)    &   13.740(0.022)    &   13.789(0.032)    &   14.117(0.041)    &   1   \\
2005    Jun 21  &   3542.76 &   9.10    &   \nodata          &   14.040(0.035)    &   13.709(0.033)    &   13.742(0.032)    &   14.093(0.041)    &   7   \\
2005    Jun 21  &   3543.06 &   9.40    &   \nodata          &   \nodata          &   13.717(0.031)    &   13.795(0.032)    &   14.143(0.053)    &   3   \\
2005    Jun 22  &   3543.68 &   10.02   &   14.216(0.035)    &   14.171(0.022)    &   13.797(0.022)    &   13.814(0.032)    &   14.171(0.041)    &   2   \\
2005    Jun 22  &   3543.82 &   10.16   &   14.219(0.035)    &   14.182(0.022)    &   13.808(0.022)    &   13.868(0.032)    &   14.168(0.041)    &   1   \\
2005    Jun 23  &   3544.77 &   11.11   &   \nodata          &   14.217(0.021)    &   13.867(0.027)    &   13.888(0.032)    &   14.197(0.041)    &   7   \\
2005    Jun 23  &   3544.80 &   11.14   &   14.299(0.035)    &   14.278(0.023)    &   13.866(0.032)    &   13.941(0.035)    &   14.248(0.041)    &   1   \\
2005    Jun 24  &   3545.82 &   12.16   &   14.429(0.035)    &   14.377(0.022)    &   13.894(0.029)    &   14.013(0.032)    &   14.297(0.041)    &   1   \\
2005    Jun 25  &   3546.75 &   13.09   &   14.539(0.035)    &   14.465(0.026)    &   13.946(0.034)    &   14.054(0.034)    &   14.328(0.041)    &   1   \\
2005    Jun 26  &   3547.67 &   14.01   &   \nodata          &   14.553(0.022)    &   14.015(0.022)    &   14.141(0.032)    &   14.336(0.041)    &   4   \\
2005    Jun 26  &   3547.82 &   14.16   &   14.682(0.035)    &   14.570(0.022)    &   14.002(0.022)    &   14.103(0.032)    &   14.343(0.041)    &   1   \\
2005    Jun 27  &   3548.66 &   15.00   &   14.841(0.037)    &   14.647(0.022)    &   14.071(0.022)    &   14.117(0.032)    &   14.400(0.041)    &   2   \\
2005    Jun 27  &   3548.79 &   15.13   &   14.828(0.036)    &   14.670(0.037)    &   14.053(0.027)    &   14.154(0.032)    &   14.347(0.041)    &   1   \\
2005    Jun 28  &   3549.71 &   16.05   &   14.99 (0.035)    &   14.778(0.022)    &   14.134(0.022)    &   14.161(0.032)    &   14.402(0.041)    &   2   \\
2005    Jun 28  &   3549.74 &   16.08   &   \nodata          &   14.755(0.022)    &   14.118(0.023)    &   14.168(0.032)    &   14.309(0.041)    &   7   \\
2005    Jun 28  &   3549.79 &   16.13   &   14.983(0.035)    &   14.781(0.022)    &   14.119(0.029)    &   14.200(0.039)    &   14.349(0.041)    &   1   \\
2005    Jun 29  &   3550.66 &   17.00   &   15.141(0.035)    &   14.879(0.022)    &   14.178(0.022)    &   14.161(0.032)    &   14.371(0.041)    &   2   \\
2005    Jun 29  &   3550.67 &   17.01   &   \nodata          &   14.911(0.022)    &   14.195(0.022)    &   14.251(0.032)    &   14.340(0.041)    &   4   \\
2005    Jun 29  &   3550.79 &   17.13   &   15.082(0.035)    &   14.879(0.021)    &   14.162(0.022)    &   14.196(0.032)    &   14.332(0.041)    &   1   \\
2005    Jun 29  &   3551.49 &   17.83   &   \nodata          &   14.913(0.032)    &   14.222(0.022)    &   14.217(0.032)    &   14.317(0.041)    &   6   \\
2005    Jun 30  &   3551.73 &   18.07   &   15.251(0.035)    &   14.984(0.025)    &   14.210(0.026)    &   14.216(0.032)    &   14.314(0.041)    &   1   \\
2005    Jun 30  &   3552.07 &   18.41   &   \nodata          &   15.045(0.032)    &   14.279(0.022)    &   14.206(0.032)    &   14.316(0.041)    &   3   \\
2005    Jul 1   &   3553.44 &   19.78   &   \nodata          &   15.229(0.031)    &   14.323(0.022)    &   14.244(0.032)    &   14.290(0.041)    &   6   \\
2005    Jul 2   &   3553.73 &   20.07   &   15.513(0.039)    &   15.193(0.021)    &   14.308(0.022)    &   14.242(0.032)    &   14.270(0.041)    &   1   \\
2005    Jul 2   &   3553.83 &   20.17   &   \nodata          &   15.224(0.022)    &   14.351(0.022)    &   14.250(0.032)    &   14.279(0.041)    &   7   \\
2005    Jul 2   &   3554.45 &   20.79   &   \nodata          &   15.298(0.021)    &   14.405(0.022)    &   14.260(0.032)    &   14.280(0.041)    &   6   \\
2005    Jul 3   &   3554.66 &   21.00   &   \nodata          &   15.308(0.022)    &   14.369(0.022)    &   14.280(0.032)    &   14.263(0.041)    &   4   \\
2005    Jul 4   &   3555.77 &   22.11   &   15.761(0.038)    &   15.439(0.025)    &   14.430(0.026)    &   14.288(0.031)    &   14.294(0.041)    &   1   \\
2005    Jul 6   &   3557.68 &   24.02   &   15.977(0.035)    &   15.604(0.022)    &   14.535(0.022)    &   14.261(0.032)    &   14.260(0.041)    &   2   \\
2005    Jul 6   &   3557.73 &   24.07   &   16.008(0.034)    &   15.621(0.026)    &   14.506(0.024)    &   14.302(0.032)    &   14.217(0.041)    &   1   \\
2005    Jul 7   &   3559.48 &   25.82   &   \nodata          &   15.791(0.022)    &   14.636(0.022)    &   14.381(0.032)    &   14.230(0.041)    &   6   \\
2005    Jul 8   &   3559.59 &   25.93   &   \nodata          &   15.801(0.022)    &   14.663(0.022)    &   14.383(0.032)    &   14.223(0.041)    &   8   \\
2005    Jul 8   &   3559.61 &   25.95   &   \nodata          &   15.811(0.022)    &   14.645(0.022)    &   14.401(0.032)    &   14.201(0.041)    &   4   \\
2005    Jul 8   &   3559.70 &   26.04   &   16.149(0.044)    &   15.778(0.022)    &   14.626(0.021)    &   14.345(0.032)    &   14.197(0.041)    &   1   \\
2005    Jul 8   &   3559.73 &   26.07   &   \nodata          &   15.765(0.021)    &   14.610(0.027)    &   14.357(0.032)    &   14.192(0.041)    &   7   \\
2005    Jul 8   &   3559.76 &   26.10   &   \nodata          &   15.799(0.022)    &   14.662(0.022)    &   14.391(0.032)    &   14.304(0.041)    &   5   \\
2005    Jul 10  &   3561.73 &   28.07   &   16.280(0.035)    &   15.925(0.022)    &   14.738(0.024)    &   14.417(0.032)    &   14.186(0.041)    &   1   \\
2005    Jul 10  &   3562.42 &   28.76   &   \nodata          &   16.010(0.022)    &   14.829(0.022)    &   14.470(0.032)    &   14.220(0.041)    &   6   \\
2005    Jul 11  &   3562.63 &   28.97   &   \nodata          &   16.071(0.022)    &   14.813(0.022)    &   14.497(0.032)    &   14.203(0.041)    &   4   \\
2005    Jul 11  &   3562.72 &   29.06   &   16.373(0.035)    &   16.057(0.022)    &   14.817(0.022)    &   14.496(0.032)    &   \nodata          &   5   \\
2005    Jul 11  &   3563.03 &   29.37   &   \nodata          &   16.101(0.025)    &   14.811(0.026)    &   14.482(0.032)    &   14.196(0.041)    &   3   \\
2005    Jul 11  &   3563.42 &   29.76   &   \nodata          &   16.081(0.023)    &   14.904(0.022)    &   14.497(0.032)    &   14.245(0.041)    &   6   \\
2005    Jul 12  &   3563.70 &   30.04   &   \nodata          &   16.101(0.025)    &   14.859(0.022)    &   14.513(0.032)    &   14.232(0.041)    &   7   \\
2005    Jul 12  &   3563.73 &   30.07   &   16.433(0.045)    &   16.127(0.024)    &   14.854(0.023)    &   14.515(0.032)    &   14.214(0.041)    &   1   \\
2005    Jul 13  &   3565.39 &   31.73   &   \nodata          &   16.256(0.022)    &   15.004(0.022)    &   14.632(0.032)    &   14.311(0.041)    &   6   \\
2005    Jul 14  &   3565.71 &   32.05   &   16.544(0.071)    &   16.244(0.025)    &   14.943(0.022)    &   14.609(0.032)    &   14.263(0.041)    &   1   \\
2005    Jul 17  &   3569.40 &   35.74   &   \nodata          &   16.479(0.031)    &   15.302(0.022)    &   14.914(0.032)    &   14.587(0.041)    &   6   \\
2005    Jul 18  &   3569.54 &   35.88   &   \nodata          &   16.483(0.022)    &   15.244(0.022)    &   14.902(0.032)    &   14.536(0.041)    &   4   \\
2005    Jul 19  &   3570.70 &   37.04   &   16.926(0.141)    &   16.501(0.033)    &   15.323(0.027)    &   14.986(0.035)    &   14.624(0.041)    &   1   \\
2005    Jul 19  &   3570.80 &   37.14   &   \nodata          &   16.508(0.054)    &   15.285(0.027)    &   14.979(0.032)    &   14.583(0.041)    &   7   \\
2005    Jul 21  &   3572.70 &   39.04   &   16.818(0.138)    &   16.632(0.037)    &   15.368(0.022)    &   15.052(0.032)    &   14.665(0.041)    &   1   \\
2005    Jul 22  &   3573.60 &   39.94   &   \nodata          &   16.638(0.026)    &   15.473(0.022)    &   15.166(0.032)    &   14.829(0.041)    &   4   \\
2005    Jul 23  &   3574.70 &   41.04   &   16.845(0.059)    &   16.674(0.023)    &   15.482(0.022)    &   15.195(0.032)    &   14.878(0.041)    &   1   \\
2005    Jul 23  &   3574.76 &   41.10   &   \nodata          &   16.690(0.036)    &   15.480(0.024)    &   15.161(0.032)    &   14.865(0.041)    &   7   \\
2005    Jul 24  &   3576.40 &   42.74   &   \nodata          &   16.710(0.022)    &   15.580(0.022)    &   \nodata          &   15.061(0.041)    &   6   \\
2005    Jul 25  &   3576.61 &   42.95   &   \nodata          &   16.744(0.022)    &   15.554(0.022)    &   15.287(0.032)    &   15.002(0.041)    &   4   \\
2005    Jul 25  &   3576.73 &   43.07   &   16.945(0.088)    &   16.735(0.025)    &   15.574(0.022)    &   15.292(0.032)    &   14.997(0.041)    &   1   \\
2005    Jul 27  &   3578.69 &   45.03   &   16.873(0.086)    &   16.723(0.022)    &   15.601(0.022)    &   15.348(0.032)    &   15.096(0.041)    &   1   \\
2005    Jul 28  &   3579.55 &   45.89   &   \nodata          &   16.805(0.022)    &   15.661(0.022)    &   15.396(0.032)    &   15.140(0.041)    &   4   \\
2005    Jul 29  &   3580.69 &   47.03   &   17.141(0.123)    &   16.789(0.026)    &   15.649(0.022)    &   15.415(0.032)    &   15.169(0.041)    &   1   \\
2005    Jul 29  &   3581.00 &   47.34   &   \nodata          &   16.828(0.055)    &   15.691(0.022)    &   15.451(0.032)    &   15.236(0.065)    &   3   \\
2005    Jul 31  &   3582.69 &   49.03   &   17.011(0.095)    &   16.839(0.025)    &   15.683(0.032)    &   15.466(0.032)    &   15.270(0.041)    &   1   \\
2005    Aug 2   &   3584.69 &   51.03   &   17.014(0.088)    &   16.841(0.028)    &   15.775(0.032)    &   15.550(0.032)    &   15.373(0.041)    &   1   \\
2005    Aug 4   &   3586.69 &   53.03   &   17.056(0.084)    &   16.885(0.033)    &   15.820(0.032)    &   15.591(0.032)    &   15.419(0.041)    &   1   \\
2005    Aug 6   &   3588.69 &   55.03   &   17.248(0.132)    &   16.906(0.026)    &   15.924(0.032)    &   15.735(0.032)    &   15.589(0.041)    &   1   \\
2005    Aug 6   &   3588.71 &   55.05   &   \nodata          &   16.905(0.033)    &   15.886(0.032)    &   15.684(0.032)    &   15.606(0.042)    &   7   \\
2005    Aug 8   &   3590.69 &   57.03   &   17.323(0.094)    &   16.957(0.027)    &   15.901(0.032)    &   15.746(0.032)    &   15.605(0.041)    &   1   \\
2005    Aug 10  &   3592.69 &   59.03   &   17.291(0.142)    &   17.013(0.031)    &   15.963(0.032)    &   15.819(0.032)    &   15.727(0.041)    &   1   \\
2005    Aug 14  &   3596.68 &   63.02   &   \nodata          &   17.070(0.055)    &   16.140(0.037)    &   15.910(0.032)    &    \nodata         &   1   \\
2005    Aug 17  &   3599.68 &   66.02   &   \nodata          &   17.121(0.059)    &   16.164(0.032)    &   16.047(0.032)    &   16.002(0.041)    &   1   \\
2005    Aug 18  &   3601.01 &   67.35   &   \nodata          &   \nodata          &   16.240(0.032)    &   16.094(0.025)    &   16.074(0.041)    &   3   \\
2005    Aug 19  &   3602.67 &   69.01   &   \nodata          &   17.115(0.051)    &   16.259(0.032)    &   16.150(0.032)    &   16.201(0.041)    &   1   \\
2005    Aug 20  &   3604.01 &   70.35   &   \nodata          &   17.225(0.038)    &   16.309(0.032)    &   16.201(0.027)    &   16.297(0.041)    &   3   \\
2005    Aug 21  &   3605.11 &   71.35   &   \nodata          &   17.263(0.078)    &   16.350(0.064)    &   \nodata          &   \nodata          &   3   \\
2005    Aug 22  &   3605.67 &   72.01   &   \nodata          &   17.270(0.051)    &   16.310(0.032)    &   16.219(0.032)    &   16.295(0.041)    &   1   \\
2005    Aug 25  &   3608.67 &   75.01   &   \nodata          &   17.261(0.051)    &   16.421(0.032)    &   16.308(0.032)    &   16.443(0.041)    &   1   \\
2005    Aug 28  &   3611.67 &   78.01   &   \nodata          &   17.175(0.051)    &   16.429(0.032)    &   16.385(0.032)    &   16.487(0.041)    &   1   \\
2005    Aug 31  &   3614.66 &   81.00   &   \nodata          &   17.309(0.051)    &   16.573(0.032)    &   16.489(0.032)    &   16.575(0.041)    &   1   \\
2005    Sept 3  &   3617.66 &   84.00   &   \nodata          &   17.325(0.051)    &   16.663(0.034)    &   16.625(0.032)    &   16.744(0.041)    &   1   \\
2005    Sept 4  &   3618.65 &   84.99   &   \nodata          &   17.453(0.051)    &   16.646(0.032)    &   16.572(0.032)    &   16.836(0.041)    &   1   \\
2005    Sept 7  &   3621.66 &   88.00   &   \nodata          &   17.545(0.051)    &   16.706(0.031)    &   16.739(0.034)    &   17.061(0.054)    &   1   \\
2005    Sept 10 &   3624.65 &   90.99   &   \nodata          &   17.422(0.064)    &   16.844(0.039)    &   16.756(0.032)    &   17.054(0.047)    &   1   \\
\enddata
\tablenotetext{a}{Relative to the epoch of $B$-band maximum (JD =
2,453,533.66).} \tablenotetext{b}{1 = KAIT 0.76~m; 2 = FLWO 1.2~m; 3
= TNT 0.8~m; 4 = CTIO 1.3~m; 5 = Lick 1.0~m; 6 = Liverpool 2.0~m; 7
= Palomar 1.5~m; 8 = CTIO 0.9~m} \tablenotetext{*}{Note:
Uncertainties, in units of 0.001 mag, are $1\sigma$.}
\end{deluxetable}

\clearpage

\begin{deluxetable*}{llrcccrrr}
\tablecolumns{9} \tablewidth{0pc} \tabletypesize{\scriptsize}
\tablecaption{$JHK_s$ Magnitudes of SN 2005cf from PAIRITEL}
\tablehead{ \colhead{UT Date}&
\colhead{JD - 2,450,000} &
\colhead{Phase\tablenotemark{a}}&
\colhead{$J$} &
\colhead{$H$} &
\colhead{$Ks$} &
\colhead{$K_{J}$}&
\colhead{$K_{H}$}&
\colhead{$K_{Ks}$}
} \startdata
 31/05/05 & 3522.24 & -11.42 & 14.724(0.023) &  14.689(0.019) &  14.825(0.021) & 0.015&0.010&0.016 \\
 01/06/05& 3523.25 & -10.41 & 14.495(0.015)  &  14.514(0.031) &  \nodata      & 0.015&0.010&\nodata\\
 02/06/05 & 3524.24   & -9.42  & 14.254(0.028) &  14.316(0.031)& 14.347(0.021)& 0.015&0.010&0.016
   \\
 10/06/05& 3532.23   & -1.43  & 13.841(0.039) &  13.969(0.023) & 14.002(0.023)& 0.011&0.006&0.026
   \\
 12/06/05& 3534.23   &  0.57  & 14.035(0.044) &  14.040(0.036) & 13.995(0.015)& 0.010&0.004&0.030
   \\
 13/06/05& 3535.22   &  1.56  & 13.997(0.013) &  14.142(0.014) & 14.102(0.034)&0.009&-0.003&0.034
   \\
 14/06/05& 3536.21   &  2.55  & 14.076(0.014) &  14.154(0.034) & 14.093(0.019)&0.008&-0.006&0.038
    \\
 17/06/05& 3539.21   &  5.55  & 14.346(0.023) &  14.169(0.056) & 14.184(0.029)&0.006&-0.010&0.046
   \\
 29/06/05& 3551.16   &  17.50 & 15.365(0.025) &  13.974(0.064) & 14.288(0.019)&0.022&-0.021&0.037
   \\
 01/07/05& 3553.17   &  19.51 & 15.339(0.028) &  13.943(0.063) & 14.196(0.017)&0.025&-0.021&0.032
  \\
 02/07/05& 3554.17   &  20.51 & 15.333(0.023) &  14.060(0.033) &  \nodata     &0.027&-0.022&\nodata
   \\
 03/07/05 & 3555.28   &  21.62 & 15.294(0.018) &  13.979(0.060)& 14.181(0.017)&0.029&-0.023&0.026
    \\
 04/07/05 & 3556.17   &  22.51 & 15.261(0.023) &  14.040(0.017)& 14.167(0.021)&0.030&-0.024&0.023
    \\
 05/07/05 & 3557.15   &  23.49 & 15.174(0.023) &  13.984(0.081)& 14.101(0.011)&0.033&-0.022&0.019
    \\
 06/07/05 & 3558.15   &  24.49 & 15.191(0.036) &  13.983(0.089)& 14.105(0.023)&0.036&-0.021&0.016
    \\
 08/07/05 & 3560.16   &  26.50 & 14.995(0.026) &  14.017(0.018)& 14.123(0.021)&0.041&-0.017&0.008
    \\
 11/07/05 & 3563.15   &  29.49 & 14.846(0.024) &  14.087(0.021)& 14.219(0.019)&0.041&-0.008&0.005
    \\
\enddata
\tablenotetext{a}{Relative to the epoch of $B$-band maximum (JD =
2,453,533.66).} \tablenotetext{}{Note: The $K$ corrections listed in columns (7)-(9)
were added to the $JHKs$ magnitudes.}
\end{deluxetable*}

\begin{deluxetable}{ccccccccr}
\tablecolumns{9} \tablewidth{0pc} \tabletypesize{\scriptsize}
\tablecaption{Journal of spectroscopic observations of SN 2005cf}
\tablehead{ \colhead{UT Date} & \colhead{JD$-$2,450,000} &
\colhead{Phase\tablenotemark{a}}& \colhead{Range(\AA)} &
\colhead{Res.\tablenotemark{b}(\AA)} & \colhead{Airmass}  &
\colhead{Exposure(s)}& \colhead{Inst.\tablenotemark{c}} &
\colhead{Observers} } \startdata
31/05/2005&3521.7& $-$12.0&3480--7500&7    &1.3&1200&Fast &PB \\
01/06/2005&3522.7& $-$11.0&3480--7500&7    &1.3&960 &Fast &PB  \\
01/06/2005&3522.9& $-$10.8&3300--10400&5--12&1.8&600&Kast&MG;DW;BS \\
02/06/2005&3523.7& $-$10.0&3480--7500&7    &1.3&600&Fast&PB   \\
02/06/2005&3523.9& $-$9.8&3300--10400&5--12&1.3&300&Kast&DR \\
03/06/2005&3524.7& $-$9.0&3480--7500 &7    &1.3&900&Fast&MC   \\
03/06/2005&3524.9& $-$8.8&3300--10400&5--12&1.4&300&Kast&DR \\
04/06/2005&3525.9& $-$7.8&3300--10400&5--12&1.4&600&Kast&DR \\
05/06/2005&3526.9& $-$6.8&3300--10400&5--12&1.4&600&Kast&DR \\
06/06/2005&3527.9& $-$5.8&3300--10400&5--12&1.4&600&Kast&MAM\\
07/06/2005&3528.7& $-$5.0&3480--7500 &  7  &1.3&900&Fast&PB \\
08/06/2005&3529.7& $-$4.0&3480--7500 &  7  &1.3&900&Fast&PB \\
09/06/2005&3530.7& $-$3.0&3480--7500 &  7  &1.3&900&Fast&PB \\
10/06/2005&3531.8& $-$1.9&3480--7500 &  7  &1.7&900&Fast&PB \\
10/06/2005&3531.8& $-$1.9&3300--10400&5-12&1.4&300 &Kast&MG;FS\\
11/06/2005&3532.7& $-$1.0&3300--10400&5--12&1.5&600&Kast&FS \\
11/06/2005&3532.8& $-$0.9&3480--7500 &7    &1.5&600&Fast&MC \\
12/06/2005&3533.8& 0.2   &3480--7500 &7    &2.1&600&Fast&MC \\
13/06/2005&3534.7& 1.1   &3480--7500 &7    &1.4&600&Fast&PB \\
14/06/2005&3535.7& 2.0   &3300--10400&5--12&1.4&600&Kast&AF;MG;BS\\
14/06/2005&3535.8& 2.1   &3480--7500 &7    &1.5&600&Fast&PB \\
15/06/2005&3536.7& 3.0   &3480--7500 &7    &1.3&600&Fast&PB \\
16/06/2005&3537.6& 3.9   &3480--7500 &7    &1.3&780&Fast&MC \\
17/06/2005&3538.7& 5.0   &3480--7500 &7    &1.3&660&Fast&MC \\
29/06/2005&3550.8& 17.1  &3480--7500 &7    &2.0&900&Fast&RH \\
01/07/2005&3552.7& 19.0  &3300--10400&5--12&1.4&300&Kast&MG;DW\\
04/07/2005&3555.7& 22.0  &3480--7500 &7    &1.4&600&Fast&JG \\
06/07/2005&3557.7& 24.0  &3480--7500 &7    &1.4&900&Fast&JG \\
07/07/2005&3558.6& 24.9  &3480--7500 &7    &1.3&900&Fast&EF \\
08/07/2005&3559.6& 25.9  &3480--7500 &7    &1.3&900&Fast&PB \\
09/07/2005&3560.7& 27.0  &3480--7500 &7    &1.5&900&Fast&PB \\
10/07/2005&3561.7& 28.0  &3480--7500 &7    &1.4&900&Fast&PB \\
10/07/2005&3561.7& 28.0  &3300--10400&5--12&1.5&500&Kast&AF;FS\\
11/07/2005&3562.7& 29.0  &3480--7500 &7    &1.4&900&Fast&MC \\
12/07/2005&3563.9& 30.2  &3480--7500 &7    &1.3&1800&Fast&MC \\
26/07/2005&3577.6& 43.9  &3480--7500 &7    &1.4&1800&Fast&PB \\
28/07/2005&3579.6& 45.9  &3480--7500 &7    &1.5&1200&Fast&PB \\
03/09/2005&3616.6& 82.9  &3480--7500 &7    &2.2&1200&Fast&MC \\
26/04/2006&3853.1& 319.4&3250--9250&6&1.5&1200& LRISB&AF;RF\\
16/02/2007&4148.1& 614.4&4585--7230&1.3&1.3&6300&DEIMOS&AF;JS;RF;RC\\
\enddata
\tablenotetext{a}{Relative to the $B$ maximum (JD=2,453,533.66).}
\tablenotetext{b}{Approximate spectral resolution.}
\tablenotetext{*}{Fast = FLWO 1.5~m FAST;Kast= Lick Shane 3~m KAST;
LIRSB = Keck I 10~m LRISBLUE; DEIMOS = Keck II 10~m DEIMOS.}
\tablenotetext{*}{AF = Alex Filippenko; RH = Robert Hutchins; EF =
Emilio Falco; JG = Joseph Gallagher; PB = Perry Berlind; MC = Mike
Calkins; MG = Mohan Ganeshalingam; DW = Diane Wong; BS = Brandont
Swift; DR = David Reitzel; JS = Jeffrey Silverman; MAM =
Matthew.~A.~Malkan; FS = Frank Serduke; RC = Ryan Chronock; RF =
Ryan Floey}
\end{deluxetable}


\begin{deluxetable}{llrrrr}
\hspace{-1.0cm}
\tablecolumns{6} \tablewidth{0pc} \tabletypesize{\scriptsize}
\tablecaption{{\it HST} ACS Ultraviolet Photometry of SN 2005cf}
\tablehead{ \colhead{UT Date} & \colhead{JD} &
\colhead{Phase\tablenotemark{a}} & \colhead{F220W} & \colhead{F250W}
& \colhead{F330W} } \startdata
03/06/05 &3524.99   & -8.67   & 19.978(070) & 16.624(015) & 14.675(005) \\
05/06/05 &3527.26   & -6.40   & 19.024(030) & 15.638(005) & 13.766(004) \\
07/06/05 &3529.36   & -4.30   & 18.557(019) & 15.341(005) & 13.439(005) \\
11/06/05 &3532.96   & -0.70   & 18.189(038) & 14.972(004) & 13.272(004) \\
14/06/05 &3535.95   &  2.28   & 18.160(051) & 15.026(004) & 13.395(004) \\
16/06/05 &3537.95   &  4.29   & 18.131(027) & 15.243(005) & 13.719(006) \\
21/06/05 &3542.61   &  9.02   & 18.240(009) & 15.941(005) & 14.424(004) \\
25/06/05 &3547.28   & 13.62   & 18.665(011) & 16.589(005) & 15.051(004) \\
26/06/05 &3547.98   & 14.32   & 18.745(007) & 16.745(008) & 15.259(005) \\
29/06/05 &3551.28   & 17.68   & 19.060(013) & 17.068(005) & 15.677(004) \\
30/06/05 &3551.77   & 18.11   & 19.158(009) & 17.162(005) & 15.797(005) \\
05/07/05 &3557.07   & 23.41   & 19.524(023) & 17.766(008) & 16.369(005) \\
\enddata
\tablenotetext{a}{Relative to the epoch of $B$-band maximum (JD =
2,453,533.66)} \tablenotetext{}{Note: Uncertainties, in units of
0.001 mag, are $1\sigma$.}
\end{deluxetable}

\begin{deluxetable*}{lcrcccrcr}
\tablecolumns{9} \tablewidth{0pc}
\tabletypesize{\scriptsize}
\singlespace
\tablecaption{{\it HST} NICMOS3 NIR Photometry of SN 2005cf}
\tablehead{
\colhead{UT Date} &
\colhead{JD} &
\colhead{Phase\tablenotemark{a}} &
\colhead{$J$\tablenotemark{b}}&
\colhead{$H$\tablenotemark{b}}&
\colhead{Color} &
\colhead{Correction} &
\colhead{S} &
\colhead{Correction} \\
\colhead{ } &
\colhead{$-$2,450,000 } &
\colhead{ } &
\colhead{} &
\colhead{} &
\colhead{F110W$\rightarrow J$} &
\colhead{F160W$\rightarrow H$} &
\colhead{F110W$\rightarrow J$} &
\colhead{F160W$\rightarrow H$}
}
\startdata
2005 June 3 & 3525.08 & $-$8.58  & 14.246(0.030) & 14.185(0.025) & 0.126  & 0 & $-$0.162 & $-$0.120 \\
2005 June 7 & 3527.15 & $-$6.51  & 13.932(0.029) & 13.905(0.023) & 0.116  & 0 & $-$0.190 & $-$0.110 \\
2005 June 10& 3529.81 & $-$3.85  & 13.803(0.029) & 13.885(0.023) & 0.207  &-0.002 & $-$0.220 &$-$0.095\\
2005 June 13& 3533.14 & $-$0.52  & 13.821(0.029) & 13.945(0.023) & 0.222  &-0.002 & $-$0.235 &$-$0.078\\
2005 June 16& 3536.74 &  3.08    & 14.222(0.030) & 14.229(0.025) & 0.168  &-0.001 & $-$0.143 &$-$0.056 \\
2005 June 18 &3538.87 &  5.21    & 14.370(0.030) & 14.285(0.025) & 0.175  &-0.001 & $-$0.040 &$-$0.047  \\
2005 June 23 &3544.01 & 10.35    & 15.093(0.032) & 14.357(0.027) & 0.119  & 0     & 0.410    &$-$0.195\\
2005 June 27 &3548.08 & 14.42    & 15.389(0.032) & 14.197(0.026) & 0.021  & 0.002 & 0.730    &$-$0.275  \\
2005 July 1  &3552.08 & 18.42    & 15.350(0.032) & 14.132(0.025) &$-$0.045& 0.003 & 0.710    &$-$0.250\\
2005 July 7  &3558.00 & 24.34    & 15.271(0.032) & 14.063(0.025) &$-$0.042& 0.003 & 0.720    &$-$0.210 \\
\enddata
\tablenotetext{a}{Relative to the epoch of $B$-band maximum (JD =
2,453,533.66)}
\tablenotetext{b}{The $J$- and $H$-band magnitudes
were converted, respectively, from the F110W- and F160W-band
magnitudes using the color- and S-corrections listed in columns
6--9.}
\end{deluxetable*}

\clearpage


\begin{deluxetable}{lcrlllccccccc}
\tablewidth{0pt} \tabletypesize{\scriptsize} \tablecaption{Swift
UVOT Ultraviolet/Optical Photometry of SN 2005cf.}
\tablehead{
 \colhead{UT Date} &
 \colhead{JD\tablenotemark{a}} &
 \colhead{Phase\tablenotemark{b}}&
 \colhead{$uvw2$} &
 \colhead{$uvm2$} &
 \colhead{$uvw1$} &
 \colhead{$U$} &
 \colhead{$B$} &
 \colhead{$V$} &
 \colhead{$Sc_{U}$} &
 \colhead{$Sc_{B}$} &
 \colhead{$Sc_{V}$}
  }
\startdata
\multicolumn{12}{c}{}\\
\noalign{\smallskip}
2005 June 4 & 3525.55     & -8.11   & 17.79(0.09) & $>$19.69     & 16.34(0.07) & 14.31(0.05) & 14.26(0.05) & 14.36(0.05) & -0.17    & -0.02    & 0.01    \\
2005 June 5 & 3526.55     & -7.11   & 17.53(0.08) & $>$20.00     & 15.91(0.06) & 13.99(0.05) & 14.17(0.09) & 14.16(0.05) & -0.13    & -0.01    & 0.01 \\
2005 June 6 & 3527.55     & -6.11   & 17.33(0.08) & 19.45(0.31) & 15.58(0.06) & 13.76(0.05) & 13.92(0.06) & 14.02(0.04) & -0.11    & -0.01    & 0.01 \\
2005 June 8 & 3530.43     & -3.23   & 16.91(0.07) & 18.76(0.31) & 15.18(0.05) & 13.37(0.06) & 13.66(0.06) & 13.73(0.05) & -0.12    & -0.01    & 0.01 \\
2005 June 9 & 3530.77     & -2.89   & 16.94(0.06) & 19.26(0.26) & 15.10(0.05) & 13.40(0.06) & 13.65(0.07) & 13.70(0.05) & -0.12    & -0.01    & 0.01 \\
2005 June 10& 3531.97     & -1.69   & 16.85(0.06) & 18.36(0.16) & 15.10(0.05) & 13.34(0.06) & 13.60(0.07) & 13.66(0.05) & -0.12    & -0.01    & 0.01 \\
2005 June 11& 3533.05     & -0.61   & 16.83(0.06) & 18.50(0.16) & 15.13(0.05) &\nodata      &  \nodata    & 13.61(0.06) & \nodata  & \nodata  & 0.01 \\
2005 June 16& 3538.26     & 4.60    & 17.15(0.07) & 18.32(0.15) & 15.39(0.05) &\nodata      &   \nodata   & 13.58(0.06) &  \nodata & \nodata  & 0.02 \\
2005 June 17& 3538.74     & 5.08    & 17.07(0.07) & 18.31(0.15) & 15.44(0.05) &\nodata      &  \nodata    & 13.63(0.06) & \nodata  & \nodata  & 0.02 \\
2005 June 20& 3542.15     & 8.49    & 17.39(0.07) & 18.47(0.15) & 15.83(0.06) &\nodata      &  \nodata    & 13.70(0.06) & \nodata  & \nodata  & 0.03 \\
2005 June 22& 3543.70     & 10.04   & 17.47(0.08) & 18.58(0.20) & 16.03(0.09) &\nodata      &   \nodata   & 13.75(0.07) & \nodata  & \nodata  & 0.03 \\
2005 June 26& 3548.26     & 14.60   & 18.06(0.10) & 18.62(0.18) & 16.63(0.07) & 14.82(0.05) & 14.65(0.05) & 14.09(0.05) & -0.21    & 0    & 0.04 \\
2005 June 29& 3550.51     & 16.85   & 18.21(0.11) & 18.92(0.24) & 16.96(0.09) & 15.14(0.05) & 14.91(0.05) & 14.25(0.05) & -0.22    & 0    & 0.04 \\
2005 July 12& 3564.32     & 30.66   & 19.36(0.29) &  $>$19.76    & 18.17(0.16) & 16.46(0.08) & 16.07(0.06) & 14.94(0.05) & -0.17    & 0    & 0.06 \\
2005 July 23& 3575.16     & 41.50   & \nodata     &  \nodata    & \nodata     & 17.05(0.06) & 16.65(0.05) & 15.60(0.05) & -0.18    & 0    & 0.05 \\
2005 July 24& 3575.97     & 42.31   & 19.77(0.19) &  $>$20.29 & 18.60(0.11) &\nodata      & \nodata &   \nodata     & \nodata  & \nodata & \nodata \\
\enddata
\tablenotetext{a}{Julian Date $-$2,450,000}
\tablenotetext{b}{Relative to the epoch of $B$-band maximum (JD =
2,453,533.66).}
\end{deluxetable}

\begin{deluxetable}{lcccc}
\tablecolumns{5} \tablewidth{0pc} \tabletypesize{\scriptsize}
\tablecaption{Light-curve parameters of SN 2005cf}
\tablehead{
\colhead{Band} &
\colhead{$\lambda_{\rm central}$} &
\colhead{$t_{\rm max}$} &
\colhead{$m_{\rm peak}$} &
\colhead{$\Delta m_{15}\tablenotemark{a}$} \\
\colhead{ }    &
\colhead{(\AA)}&
\colhead{$-$2,450,000} &
\colhead{(mag)}&
\colhead{(mag)}
}
\startdata
$uvw2$&1928  &3533.05$\pm$0.50 &16.84$\pm$0.05 &1.11$\pm$0.06 \\
$uvm2$&2246  &3537.86$\pm$0.71 &18.30$\pm$0.13 &0.75$\pm$0.08  \\
$uvw1$&2600  &3532.35$\pm$0.44 &15.10$\pm$0.04 &1.44$\pm$0.05 \\
F220W&2228 &3537.17$\pm$0.48 &18.14$\pm$0.05 &1.00$\pm$0.06 \\
F250W&2696 &3532.49$\pm$0.42 &15.13$\pm$0.04 &1.54$\pm$0.05 \\
F330W&3354 &3532.30$\pm$0.40 &13.31$\pm$0.04 &1.91$\pm$0.05 \\
$U$    &3650 &3532.42$\pm$0.30  &13.40$\pm$0.03&1.26$\pm$0.04 \\
$B$    &4450 &3533.66$\pm$0.28  &13.63$\pm$0.02&1.05$\pm$0.03\\
$V$    &5500 &3535.54$\pm$0.33  &13.55$\pm$0.02&0.62$\pm$0.03\\
$R$    &6450 &3534.80$\pm$0.26  &13.53$\pm$0.03&0.67$\pm$0.03\\
$I$    &7870 &3532.56$\pm$0.34  &13.76$\pm$0.04&0.59$\pm$0.03\\
$J$    &12700&3530.54$\pm$0.44  &13.78$\pm$0.05&1.45$\pm$0.05\\
$H$    &16700&3529.48$\pm$0.42  &13.84$\pm$0.04&0.42$\pm$0.05 \\
$K$    &22200&3530.32$\pm$0.59  &13.94$\pm$0.05&0.40$\pm$0.06\\
\enddata
\tablenotetext{a}{The magnitude decline in 15 days after the initial
maximum of the light curve.}
\end{deluxetable}

\clearpage

\begin{deluxetable}{rccccc}
\tablecolumns{6} \tablewidth{0pc}
\tabletypesize{\scriptsize}
\singlespace
\tablecaption{The $UBVRI$ Template Light Curves of SN
2005cf\tablenotemark{a}}
\tablehead{ \colhead{Days} & \colhead{$U$}
& \colhead{$B$ } & \colhead{$V$} & \colhead{$R$} & \colhead{$I$}}
\startdata
-12 & 2.58    & 2.00    & 1.78    & 1.75    & 1.61 \\
-11 & 1.97    & 1.55    & 1.45    & 1.37    & 1.28 \\
-10 & 1.48    & 1.20    & 1.17    & 1.05    & 0.97 \\
-9  & 1.08    & 0.91    & 0.93    & 0.80    & 0.70 \\
-8  & 0.76    & 0.68    & 0.72    & 0.60    & 0.48 \\
-7  & 0.51    & 0.49    & 0.56    & 0.44    & 0.31 \\
-6  & 0.33    & 0.35    & 0.41    & 0.32    & 0.18 \\
-5  & 0.19    & 0.23    & 0.30    & 0.22    & 0.09 \\
-4  & 0.09    & 0.14    & 0.21    & 0.14    & 0.04 \\
-3  & 0.04    & 0.08    & 0.14    & 0.09    & 0.01 \\
-2  & 0.01    & 0.03    & 0.08    & 0.05    & 0   \\
-1  & 0       & 0.01    & 0.04    & 0.02    & 0.01 \\
0   & 0.02    & 0       & 0.02    & 0.01    & 0.02 \\
1   & 0.05    & 0.01    & 0       & 0       & 0.04 \\
2   & 0.10    & 0.03    & 0       & 0       & 0.05 \\
3   & 0.16    & 0.06    & 0.01    & 0.02    & 0.08 \\
4   & 0.22    & 0.11    & 0.02    & 0.04    & 0.10 \\
5   & 0.30    & 0.16    & 0.04    & 0.07    & 0.13 \\
6   & 0.39    & 0.23    & 0.07    & 0.11    & 0.16 \\
7   & 0.48    & 0.30    & 0.11    & 0.16    & 0.21 \\
8   & 0.58    & 0.37    & 0.15    & 0.21    & 0.27 \\
9   & 0.69    & 0.46    & 0.19    & 0.27    & 0.33 \\
10  & 0.80    & 0.55    & 0.24    & 0.34    & 0.41 \\
11  & 0.92    & 0.64    & 0.29    & 0.40    & 0.48 \\
12  & 1.04    & 0.73    & 0.34    & 0.47    & 0.54 \\
13  & 1.17    & 0.84    & 0.39    & 0.53    & 0.57 \\
14  & 1.31    & 0.94    & 0.45    & 0.59    & 0.59 \\
15  & 1.45    & 1.05    & 0.50    & 0.63    & 0.59 \\
16  & 1.59    & 1.15    & 0.56    & 0.66    & 0.59 \\
17  & 1.74    & 1.27    & 0.61    & 0.68    & 0.59 \\
18  & 1.89    & 1.38    & 0.67    & 0.70    & 0.57 \\
19  & 2.02    & 1.49    & 0.72    & 0.72    & 0.56 \\
20  & 2.15    & 1.60    & 0.78    & 0.74    & 0.54 \\
21  & 2.27    & 1.71    & 0.83    & 0.75    & 0.52 \\
22  & 2.38    & 1.81    & 0.88    & 0.76    & 0.50 \\
23  & 2.49    & 1.91    & 0.93    & 0.78    & 0.48 \\
24  & 2.59    & 2.00    & 0.98    & 0.80    & 0.47 \\
25  & 2.68    & 2.10    & 1.04    & 0.82    & 0.46 \\
26  & 2.77    & 2.19    & 1.09    & 0.85    & 0.45 \\
27  & 2.85    & 2.27    & 1.14    & 0.89    & 0.45 \\
28  & 2.92    & 2.36    & 1.21    & 0.93    & 0.46 \\
29  & 2.99    & 2.44    & 1.29    & 0.97    & 0.47 \\
30  & 3.06    & 2.51    & 1.35    & 1.02    & 0.50 \\
31  & 3.12    & 2.58    & 1.42    & 1.08    & 0.53 \\
34  & 3.27    & 2.77    & 1.60    & 1.27    & 0.66 \\
37  & 3.38    & 2.92    & 1.75    & 1.47    & 0.86 \\
40  & 3.46    & 3.03    & 1.88    & 1.64    & 1.08 \\
43  & 3.53    & 3.11    & 1.99    & 1.77    & 1.28 \\
46  & 3.58    & 3.16    & 2.09    & 1.89    & 1.40 \\
49  & 3.65    & 3.21    & 2.18    & 1.98    & 1.53 \\
52  & 3.72    & 3.26    & 2.26    & 2.08    & 1.66 \\
57  & 3.83    & 3.34    & 2.39    & 2.23    & 1.88 \\
62  & 3.94    & 3.42    & 2.51    & 2.39    & 2.10 \\
65  & 4.00    & 3.46    & 2.58    & 2.49    & 2.23 \\
70  & 4.11    & 3.54    & 2.71    & 2.64    & 2.45 \\
75  & 4.22    & 3.62    & 2.84    & 2.80    & 2.67 \\
80  & 4.33    & 3.70    & 2.98    & 2.96    & 2.89 \\
85  & 4.44    & 3.78    & 3.12    & 3.12    & 3.10 \\
90  & 4.55    & 3.86    & 3.25    & 3.28    & 3.32 \\
\enddata
\tablenotetext{a}{All of the light curves have been normalized to
the $B$-band maximum epoch and their peak values listed in Table
9.}
\end{deluxetable}

\clearpage
\begin{center}
\begin{deluxetable*}{lccccl}

\tablecolumns{6} \tablewidth{0pc} \tabletypesize{\scriptsize}
\tablecaption{The Intrinsic color $-$ $\Delta m_{15}(B)$ Relation}
\tablehead{
\colhead{Color} &
\colhead{$a$ } &
\colhead{$b_{1}$} &
\colhead{$b_{2}$} &
\colhead{$b_{3}$} &
\colhead{$\sigma$}
}
\startdata
$B_{\rm max}-V_{\rm max}$\tablenotemark{a} & $-$0.09(04)&0.15(03)&\nodata & \nodata & 0.03  \\
$B_{\rm max}-V_{\rm max}$\tablenotemark{b} & $-$1.61(35)& 2.60(47)&\nodata  & \nodata & 0.11 \\
$V_{\rm max}-I_{\rm max}$\tablenotemark{a} & $-$0.27(01)& 0.22(03)& \nodata & \nodata & 0.06   \\
$V_{\rm max}-I_{\rm max}$\tablenotemark{b} & $-$1.03(10)& 1.45(14)& \nodata & \nodata & 0.06   \\
$(B - V)_{12}$                             & 0.25(01)&0.34(05)&-0.38(21)&1.98(23)&  0.04 \\
$(B - V)_{35}$\tablenotemark{a}            & 1.02(01)& \nodata & \nodata & \nodata & 0.04\\
$(B - V)_{35}$\tablenotemark{b}            & 0.72(07)& 0.46(10)& \nodata & \nodata & 0.04 \\
\enddata
\tablenotetext{ }{$Color = a + b_{1}(\Delta m_{15} - 1.1) +
b_{2}(\Delta m_{15} - 1.1)^{2} + b_{3}(\Delta m_{15} - 1.1)^{3}$.}
\tablenotetext{a}{The correlation holds for SNe~Ia with $0.8 <
\Delta m_{15} < 1.7$.} \tablenotetext{b}{The correlation applies to
SNe~Ia with $\Delta m_{15} \gtrsim 1.7$.}

\end{deluxetable*}
\end{center}

\begin{deluxetable}{lcc}
\tablecolumns{4} \tablewidth{0pc} \tabletypesize{\scriptsize}
\tablecaption{Host-galaxy reddening of SN 2005cf}
\tablehead{
\colhead{Method} &
\colhead{E$(B - V)_{\rm host}$} &
\colhead{References }
}
\startdata
$B_{\rm max} - V_{\rm max}$ & $0.07 \pm 0.04$ & 1,2 \\
$V_{\rm max} - I_{\rm max}$ & $-0.06 \pm 0.06$ & 1,2 \\
$(B - V)_{12}$      & $0.12 \pm 0.04$ & 3,2 \\
$(B - V)_{35}$      & $0.13 \pm 0.05$ & 4,2 \\
$(B - V)_{\rm tail}$& $0.15 \pm 0.04$ & 1   \\
$V - J$             & $0.13 \pm 0.08$ & 5   \\
$V - H$             & $0.10 \pm 0.06$ & 5   \\
$V - K$             & $0.06 \pm 0.06$ & 5   \\
Mean                & $0.09 \pm 0.03$ & \nodata \\
\enddata
\tablerefs{(1) Phillips et~al. (1999); (2) this paper; (3) Wang
et~al. (2005); (4) Jha et~al. (2007); (5) Krisciunas et~al. (2004).}
\end{deluxetable}

\begin{table}[htb]
\begin{center}
\caption{Relevant parameters for SN 2005cf and its host.}
{\scriptsize
\begin{tabular}{llc}
\tableline\tableline
Parameter & Value & Source \\
\tableline
& Photometric parameters & \\
\tableline
Discovery date & 28.26 May 2005 & 1\\
Epoch of $B$ maximum & $2453533.66 \pm 0.28$ & 2\\
$B_{\rm max}$  & $13.63 \pm 0.02$ mag & 2 \\
$B_{\rm max} - V_{\rm max}$ & $0.08 \pm 0.03$ mag & 2\\
$E(B - V)_{\rm host}$ & $0.10 \pm 0.03$ mag & 2\\
$\Delta m_{15}$(true) & $1.07 \pm 0.03$ mag & 2\\
$\Delta C_{12}$(true) & $0.27 \pm 0.03$ & 2 \\
Late-time $B$ decline rate & $1.62 \pm 0.05$ mag (100~d)$^{-1}$ & 2\\
L$^{max}_{bol}$ & $(1.54 \pm 0.20) \times 10^{43}$ erg s$^{-1}$ & 2\\
$t_{r}$ & $18.4 \pm 0.5$~d & 2 \\
$^{56}$Ni & $0.77 \pm 0.11$~M$_{\odot}$ & 2\\
\tableline\tableline
& Spectroscopic parameters& \\
\tableline
$v_{\rm max}$(Si~II $\lambda$6355) & $\sim$10,100 km s$^{-1}$ & 2\\
$v_{\rm max}$(S~II $\lambda$5460) & $\sim$9600 km s$^{-1}$ &2 \\
$\dot{v}$(Si~II $\lambda$6355)& $30 \pm 5$ km s$^{-1}$ &2 \\
$R$(Si~II)$_{\rm max}$  & $0.28 \pm 0.04$ & 2\\
%
%
 \tableline
 \tableline
 & Parameters for MCG-01-39-003 & \\
 \tableline
 Galaxy type & S0 pec & 3\\
 E$(B - V)_{Gal}$ & 0.097& 3\\
 ($m - M$) & 32.31$\pm$0.11 & 2 \\
 v$_{hel}$ & 1937 km s$^{-1}$ & 3\\
 \tableline
\end{tabular}}

{References: (1) Hugh \& Li 2005; (2) this paper; (3) NASA
Extragalactic Database.}
\end{center}
\end{table}

\clearpage
\appendix
\section{The $S$- and $K$-corrections}

It is always tricky to transform photometric observations from one
filter system to another, requiring many response parameters
characterizing the instruments. According to the description in
Stritzinger et~al. (2002), the instrumental response $S(\lambda)$
can be simply defined as
\begin{equation}
S(\lambda) = F(\lambda) \times A(\lambda) \times QE(\lambda),
\end{equation}
where $F(\lambda)$ is the filter transmission function, $A(\lambda)$
is the transparency of the Earth's atmosphere, and $QE(\lambda)$ is
the detector quantum efficiency. The atmospheric transmission at the
sites where it is not directly available was obtained by modifying
the standard atmospheric model (Walker 1987) to be consistent with
the average broadband absorption coefficients.
Here we did not include the mirror reflectivities, dichroic
transmission, or dewar window transmissions due to the absence of
this information. Various instrumental responses, normalized to the
peak transmission, are shown in Figure 2.

To check whether the instrumental response curves match those
actually used at the telescopes, we computed the synthetic
magnitudes and hence the color terms by convolving the model curves
with a large sample of spectrophotometric standard stars from
Stritzinger et~al. (2005). The resulting synthetic color terms are
generally consistent with the values listed in Table 1 of the main
text, but small differences are present. The differences are
probably due to the mirror reflectivities, dichroic mirror
transmission, or other unknown transmissivity of the optical
elements. Following the method proposed by Stritzinger et~al.
(2002), we shifted the wavelength of the model response curves in
order to reproduce exactly the measured color terms. The wavelength
shifts of different instrumental responses are given in Table A1.
They are usually $\lesssim$100~\AA, except in the $R$ and $I$
filters at the CTIO 1.3~m telescope where the required shifts are
128~\AA\ and 377~\AA, respectively.

\begin{deluxetable}{lllllr}
 \tablecolumns{6} \tablewidth{0pc}
\tablecaption{Wavelength Shifts to Instrumental Response
Curves\tablenotemark{a}} \tablehead{ \colhead{Telescope} &
\colhead{$U$} & \colhead{$B$} & \colhead{$V$} & \colhead{$R$} &
\colhead{$I$} } \startdata
KAIT 0.8~m & 29~blue & 20~blue & 9~red  &  38~blue &  29~blue  \\
CfA 1.2~m  & 29~blue & 58~red & 9~red  &   0 &  0  \\
CTIO 1.3~m & \nodata  &70~blue & 58~blue  & 128~blue &  377~blue  \\
CTIO 0.9~m & 6~blue & 0 &  0 & 46~blue & 41~red  \\
Palomar 1.5~m &\nodata& 0 & 17~red   &  15~red  & 107~blue      \\
Lick 1.0~m & 38~blue& 38~red  &  46~red & 15~blue & 58~red  \\
Liverpool 2.0~m&\nodata& 35~blue & 9~red   &  0 &  29~red  \\
\enddata
\tablenotetext{a}{All values are measured in Angstrom units.}
\end{deluxetable}

With proper model response curves and better spectral coverage for
SN 2005cf, we are able to compute the $S$-corrections using
\begin{equation}
Sc_{\lambda 1} = M_{\lambda 1} - m_{\lambda 1} - CT_{\lambda
1}(m_{\lambda 1} - m_{\lambda 2}) - ZP_{\lambda 1},
\end{equation}
where $M_{\lambda 1}$ is the $\lambda 1$-band SN magnitude
synthesized with the Bessell function, and $m_{\lambda 1}$ and
$m_{\lambda 2}$ are (respectively) the $\lambda 1$-band and $\lambda
2$-band magnitudes synthesized with the instrumental response
function. The color term is $CT_{\lambda 1}$, and $ZP_{\lambda 1}$
is the zeropoint, which can be determined from the spectrophotomeric
standards with a precision close to 0.01 mag. However, since the
spectra of SN 2005cf taken 1--3 months after $B$-band maximum did
not have adequate wavelength coverage and were sparsely sampled, we
also used the spectra of SN 2003du (Stanishev et~al. 2007) to
compute the $S$-corrections during that phase.


In order to estimate the corresponding $S$-corrections at any epochs
without photometry, a polynomial function was used to fit the data
points shown in Figure 3. The resulting $S$-corrections are listed
in Table A2 (columns 4--8).

Owing to a redshift effect on the spectral energy distribution, we
further computed the $K$-corrections for SN 2005cf in the optical
bands.  The $K$-corrections, based on the response curves of the
Bessell filter band and the observed spectra of SNe 2005cf and
2003du, are listed in Table A2 (columns 9--13). Except in the $U$
band, the $K$-corrections are generally small, 0.02--0.03 mag around
maximum brightness, and they depend on the supernova phase.

\clearpage
\LongTables
\begin{deluxetable}{lcrrrrrrrrrrcc}
\tablewidth{0pt} \tabletypesize{\scriptsize} \tablecaption{The $K$-
and $S$-corrections added to the $UBVRI$ magnitudes of SN 2005cf.}
\tablehead{
 \colhead{UT Date} &
 \colhead{JD} &
 \colhead{Phase\tablenotemark{a}}&
 \colhead{$Sc_{U}$} &
 \colhead{$Sc_{B}$} &
 \colhead{$Sc_{V}$} &
 \colhead{$Sc_{R}$} &
 \colhead{$Sc_{I}$} &
 \colhead{$K_{U}$} &
 \colhead{$K_{B}$} &
 \colhead{$K_{V}$} &
 \colhead{$K_{R}$} &
 \colhead{$K_{I}$} &
 \colhead{Inst.\tablenotemark{b}}
 }
\startdata
\multicolumn{14}{c}{}\\
\noalign{\smallskip}
2005    May 31 & 3521.75 & -11.91 & 0.038  & -0.042 & -0.012 & -0.022  &0.015   & -0.087 & 0.004  & 0.014  & 0.016  & 0.017  & 2    \\
2005    May 31 & 3521.77 & -11.89 & 0.075  & -0.036 & 0.009  & 0       & -0.020 & -0.087 & 0.004  & 0.014  & 0.016  & 0.017  & 1    \\
2005    Jun 1  & 3522.74 & -10.92 & 0.032  & -0.043 & -0.014 & -0.025  & 0.018  & -0.073 & 0.006  & 0.013  & 0.016  & 0.018  & 2    \\
2005    Jun 1  & 3522.87 & -10.79 & 0.071  & -0.034 & 0.005  & 1E-3    & -0.015 & -0.071 & 0.007  & 0.013  & 0.016  & 0.019  & 1    \\
2005    Jun 1  & 3523.15 & -10.51 &\nodata & \nodata & \nodata & \nodata & \nodata&-0.069& 0.007  & 0.013  & 0.016  & 0.019  & 3    \\
2005    Jun 2  & 3523.77 & -9.89  & 0.021  & -0.040 & -0.015 & -0.027 & 0.017  & -0.060  & 0.008  & 0.013  & 0.016  & 0.019  & 2    \\
2005    Jun 2  & 3523.87 & -9.79  & 0.063  & -0.033 & 0.003  & 0.002  & -0.011 & -0.059  & 0.009  & 0.013  & 0.016  & 0.019  & 1    \\
2005    Jun 2  & 3524.13 & -9.53  & \nodata & \nodata & \nodata &\nodata&\nodata&-0.057  & 0.009  & 0.013  & 0.016  & 0.019  & 3    \\
2005    Jun 3  & 3524.63 & -9.03  & \nodata & -0.033 & -0.028 & 0.025  & 0.037  & \nodata & 0.010  & 0.013  & 0.016  & 0.020  & 4    \\
2005    Jun 3  & 3524.68 & -8.98  & \nodata & -0.037 & -0.015 & -0.029 & 0.017  & \nodata & 0.010  & 0.013  & 0.016  & 0.020  & 2    \\
2005    Jun 3  & 3524.79 & -8.87  & \nodata & -0.034 & 0.002  & 0.002  & 0.004  & \nodata & 0.010  & 0.013  & 0.016  & 0.020  & 5    \\
2005    Jun 3  & 3524.85 & -8.81  & 0.047  & -0.031 & 0  & 0.003  & -0.010 & -0.049 & 0.010  & 0.013  & 0.016  & 0.020  & 1    \\
2005    Jun 3  & 3525.42 & -8.24  & \nodata & -0.024 & -0.013 & 0.026  & 0.027  & \nodata & 0.011  & 0.013  & 0.016  & 0.020  & 6    \\
2005    Jun 4  & 3525.69 & -7.97  & 0.007  & -0.035 & -0.016 & -0.03  & 0.018  & -0.043 & 0.011  & 0.012  & 0.016  & 0.020  & 2    \\
2005    Jun 4  & 3525.76 & -7.90  & \nodata & -0.032 & -0.003 & 0.006  & -0.003 & \nodata & 0.011  & 0.012  & 0.016  & 0.020  & 5    \\
2005    Jun 4  & 3525.87 & -7.79  & \nodata & -0.031 & -0.009 & 0.007  & 0.026  & \nodata & 0.011  & 0.012  & 0.016  & 0.020  & 7    \\
2005    Jun 5  & 3526.68 & -6.98  & \nodata & -0.032 & -0.017 & -0.027 & 0.020  & \nodata & 0.012  & 0.012  & 0.017  & 0.019  & 2    \\
2005    Jun 5  & 3526.75 & -6.91  & 0.042  & -0.031 & -0.004 & \nodata & \nodata & -0.036 & 0.012  & 0.012  &\nodata & \nodata& 5    \\
2005    Jun 5  & 3527.44 & -6.22  & \nodata & -0.023 & -0.015 & 0.032  & 0.010  & \nodata & 0.012  & 0.012  & 0.018  & 0.019  & 6    \\
2005    Jun 6  & 3527.64 & -6.02  & \nodata & -0.028 & -0.021 & 0.020  & 0.018  & \nodata & 0.012  & 0.012  & 0.018  & 0.019  & 4    \\
2005    Jun 6  & 3527.69 & -5.97  & 0.003  & -0.031 & -0.018 & -0.026 & 0.012  & -0.032 & 0.012  & 0.012  & 0.018  & 0.019  & 2    \\
2005    Jun 6  & 3527.85 & -5.81  & 0.018  & -0.028 & -0.004 & 0.007  & -0.002 & -0.031 & 0.012  & 0.012  & 0.018  & 0.018  & 1    \\
2005    Jun 6  & 3528.43 & -5.23  & \nodata & -0.023 & -0.016 & 0.036  & 1E-3   & \nodata & 0.012  & 0.012  & 0.018  & 0.018  & 6    \\
2005    Jun 7  & 3528.75 & -4.91  & \nodata & -0.029 & -0.008 & 0.012  & 0.013  & \nodata & 0.012  & 0.011  & 0.019  & 0.018  & 7    \\
2005    Jun 7  & 3528.84 & -4.82  & 0.012  & -0.028 & -0.004 & 0.009  & -0.002 & -0.028 & 0.012  & 0.011  & 0.019  & 0.017  & 1    \\
2005    Jun 8  & 3529.43 & -4.23  & \nodata & -0.022 & -0.016 & 0.038  & -0.016 & \nodata & 0.012  & 0.011  & 0.019  & 0.017  & 6    \\
2005    Jun 8  & 3529.71 & -3.95  & 0.022  & -0.017 & 0.007  & 0.003  & -0.025 & -0.026 & 0.012  & 0.011  & 0.020  & 0.016  & 8    \\
2005    Jun 8  & 3530.42 & -3.24  & \nodata & -0.021 & -0.017 & 0.041  & -0.021 & \nodata & 0.012  & 0.011  & 0.020  & 0.015  & 6    \\
2005    Jun 8  & 3530.59 & -3.07  & \nodata & -0.028 & -0.020 & 0.017  & -0.012 & \nodata & 0.012  & 0.011  & 0.021  & 0.015  & 4    \\
2005    Jun 9  & 3530.68 & -2.98  & -0.011 & -0.03  & -0.019 & -0.022 & -0.024 & -0.025 & 0.012  & 0.011  & 0.021  & 0.015  & 2    \\
2005    Jun 10 & 3531.67 & -1.99  & -0.012 & -0.03  & -0.019 & -0.021 & -0.04  & -0.024 & 0.011  & 0.010  & 0.022  & 0.013  & 2    \\
2005    Jun 10 & 3531.79 & -1.87  & \nodata & -0.028 & -0.007 & 0.017  & 0  & \nodata & 0.011  & 0.010  & 0.022  & 0.013  & 7    \\
2005    Jun 10 & 3531.83 & -1.83  & -0.005  & -0.027 & -0.004 & 0.014  & -0.006 & -0.024 & 0.011  & 0.010  & 0.022  & 0.013  & 1    \\
2005    Jun 10 & 3532.42 & -1.24  & \nodata & -0.020 & -0.017 & 0.045  & -0.042 & \nodata & 0.011  & 0.010  & 0.022  & 0.012  & 6    \\
2005    Jun 11 & 3532.87 & -0.79  & 0  & -0.027 & -0.004 & 0.016  & -0.007 & -0.025 & 0.011  & 0.010  & 0.023  & 0.011  & 1    \\
2005    Jun 11 & 3533.42 & -0.24  & \nodata & -0.019 & -0.017 & 0.054  & -0.050 & \nodata & 0.010  & 0.010  & 0.024  & 0.010  & 6    \\
2005    Jun 12 & 3533.66 & 0.00   & \nodata & -0.028 & -0.018 & 0.011  & -0.044 & \nodata & 0.010  & 0.010  & 0.024  & 0.009  & 4    \\
2005    Jun 12 & 3533.72 & 0.06   & \nodata & -0.028 & -0.007 & 0.019  & -0.008 & \nodata & 0.010  & 0.010  & 0.024  & 0.009  & 7    \\
2005    Jun 12 & 3533.84 & 0.18   & 0  & -0.027 & -0.004 & 0.016  & -0.007 & -0.025       & 0.010  & 0.010  & 0.024  & 0.009  & 1    \\
2005    Jun 13 & 3534.73 & 1.07   & -0.011  & -0.031 & -0.020 & -0.006 & -0.08  & -0.026 & 0.009  & 0.009  & 0.025  & 0.007  & 2    \\
2005    Jun 13 & 3534.84 & 1.18   & 0.005  & -0.027 & -0.004 & 0.020  & -0.010 & -0.027 & 0.009  & 0.009  & 0.025  & 0.007  & 1    \\
2005    Jun 13 & 3535.43 & 1.77   & \nodata & -0.018 & -0.017 & 0.067  & -0.078 & \nodata & 0.009  & 0.009  & 0.025  & 0.006  & 6    \\
2005    Jun 14 & 3535.72 & 2.06   & \nodata & -0.027 & -0.008 & 0.022  & -0.015 & \nodata & 0.009  & 0.009  & 0.026  & 0.005  & 7    \\
2005    Jun 14 & 3535.74 & 2.08   & -0.009 & -0.032 & -0.020 & 0.005  & -0.090  & -0.028 & 0.009  & 0.009  & 0.026  & 0.005  & 2    \\
2005    Jun 14 & 3535.83 & 2.17   & 0.015  & -0.027 & -0.004 & 0.024  & \nodata & -0.028 & 0.008  & 0.009  & 0.026  & \nodata  & 1    \\
2005    Jun 14 & 3536.44 & 2.78   & \nodata & -0.018 & -0.017 & 0.07   & -0.089 & \nodata & 0.008  & 0.008  & 0.026  & 0.004  & 6    \\
2005    Jun 15 & 3536.70 & 3.04   & -0.004 & -0.032 & -0.020 & 0.007  & -0.103 & -0.029 & 0.008  & 0.008  & 0.026  & 0.003  & 2    \\
2005    Jun 15 & 3536.83 & 3.17   & 0.017  & -0.027 & -0.004 & 0.024  & -0.012 & -0.030  & 0.008  & 0.008  & 0.027  & 0.003  & 1    \\
2005    Jun 15 & 3537.47 & 3.81   & \nodata & -0.017 & -0.016 & 0.071  & -0.101 & \nodata & 0.007  & 0.008  & 0.027  & 0.002  & 6    \\
2005    Jun 16 & 3537.82 & 4.16   & \nodata & \nodata & -0.004 & 0.025  & -0.013 & -0.031 & 0.007  & 0.008  & 0.027  & 1E-3   & 1    \\
2005    Jun 17 & 3538.68 & 5.02   & \nodata & -0.025 & -0.017 & 0.019  & -0.085 & \nodata & 0.006  & 0.007  & 0.028  & -1E-3  & 4    \\
2005    Jun 21 & 3542.61 & 8.95   & \nodata & -0.020 & -0.015 & 0.024  & -0.104 & \nodata & 0.002  & 0.005  & 0.028  & -0.007 & 4    \\
2005    Jun 21 & 3542.72 & 9.06   & 0.035  & -0.015 & -0.004 & 0.021  & -0.016  & -0.040  & 0.002  & 0.005  & 0.028  & -0.007 & 1    \\
2005    Jun 21 & 3542.76 & 9.10   & \nodata & -0.024 & -0.013 & 0.022  & -0.023 & \nodata & 0.002  & 0.005  & 0.028  & -0.007 & 7    \\
2005    Jun 21 & 3543.06 & 9.40   & \nodata & \nodata & \nodata & \nodata & \nodata & -0.040 & 0.002  & 0.005  & 0.028  & -0.007 & 3    \\
2005    Jun 22 & 3543.68 & 10.02  & 0.016  & -0.024 & -0.018 & 0.015  & -0.160  & -0.041 & 1E-3   & 0.004  & 0.027  & -0.008 & 2    \\
2005    Jun 22 & 3543.82 & 10.16  & 0.040   & -0.013 & -0.004 & 0.021  & -0.018 & -0.042 & 1E-3   & 0.004  & 0.027  & -0.008 & 1    \\
2005    Jun 23 & 3544.77 & 11.11  & \nodata & -0.024 & -0.014 & 0.023  & -0.020 & \nodata & 0  & 0.003  & 0.027  & -0.009 & 7    \\
2005    Jun 23 & 3544.80 & 11.14  & 0.045  & -0.011 & -0.004 & 0.018  & -0.019 & -0.043 & 0  & 0.003  & 0.027  & -0.009 & 1    \\
2005    Jun 24 & 3545.82 & 12.16  & 0.050  & -0.010 & -0.004 & 0.016  & -0.020 & -0.044 & -1E-3  & 0.003  & 0.026  & -0.009 & 1    \\
2005    Jun 25 & 3546.75 & 13.09  & 0.058  & -0.009 & -0.004 & 0.015  & -0.020 & -0.045 & -0.002 & 0.002  & 0.024  & -0.009 & 1    \\
2005    Jun 26 & 3547.67 & 14.01  & \nodata & -0.014 & -0.012 & 0.026  & -0.108 & \nodata & -0.003 & 1E-3   & 0.023  & -0.009 & 4    \\
2005    Jun 26 & 3547.82 & 14.16  & 0.062  & -0.009 & -0.004 & 0.015  & -0.019 & -0.046 & -0.003 & 1E-3   & 0.023  & -0.009 & 1    \\
2005    Jun 27 & 3548.66 & 15.00  & 0.029  & -0.023 & -0.016 & 0.015  & -0.194 & -0.046 & -0.004 & 0  & 0.022  & -0.009 & 2    \\
2005    Jun 27 & 3548.79 & 15.13  & 0.064  & -0.009 & -0.005 & 0.014  & -0.016 & -0.047 & -0.004 & 0  & 0.021  & -0.009 & 1    \\
2005    Jun 28 & 3549.71 & 16.05  & 0.028  & -0.026 & -0.016 & 0.012  & -0.196 & -0.047 & -0.005 & -1E-3  & 0.020  & -0.008 & 2    \\
2005    Jun 28 & 3549.74 & 16.08  & \nodata & -0.026 & -0.013 & 0.021  & -0.007 & \nodata & -0.005 & -1E-3  & 0.019  & -0.008 & 7    \\
2005    Jun 28 & 3549.79 & 16.13  & 0.067  & -0.011 & -0.005 & 0.013  & -0.016 & -0.047 & -0.005 & -1E-3  & 0.019  & -0.008 & 1    \\
2005    Jun 29 & 3550.66 & 17.00  & 0.031  & -0.027 & -0.015 & 0.010  & -0.196 & -0.048 & -0.006 & -1E-3  & 0.017  & -0.008 & 2    \\
2005    Jun 29 & 3550.67 & 17.01  & \nodata & -0.012 & -0.011 & 0.043  & -0.104 & \nodata & -0.006 & -1E-3  & 0.017  & -0.008 & 4    \\
2005    Jun 29 & 3550.79 & 17.13  & 0.068  & -0.012 & -0.005 & 0.012  & -0.015 & -0.048 & -0.006 & -0.002 & 0.017  & -0.008 & 1    \\
2005    Jun 29 & 3551.49 & 17.83  & \nodata & -0.015 & 0  & 0.044  & -0.172 & \nodata & -0.007 & -0.002 & 0.015  & -0.008 & 6    \\
2005    Jun 30 & 3551.73 & 18.07  & 0.069  & -0.011 & -0.005 & 0.013  & -0.014 & -0.048 & -0.007 & -0.002 & 0.015  & -0.007 & 1    \\
2005    Jun 30 & 3552.07 & 18.41  & \nodata & \nodata & \nodata & \nodata & \nodata & \nodata & -0.007 & -0.003 & 0.015  & -0.007 & 3    \\
2005    Jul 1  & 3553.44 & 19.78  & \nodata & -0.016 & 0.0015 & 0.053  & -0.163 & \nodata & -0.008 & -0.004 & 0.015  & -0.007 & 6    \\
2005    Jul 2  & 3553.73 & 20.07  & 0.072  & -0.010 & -0.005 & 0.013  & -0.012 & -0.048 & -0.009 & -0.005 & 0.015  & -0.006 & 1    \\
2005    Jul 2  & 3553.83 & 20.17  & \nodata & -0.032 & -0.010 & 0.018  & 0.006  & \nodata & -0.009 & -0.005 & 0.015  & -0.006 & 7    \\
2005    Jul 2  & 3554.45 & 20.79  & \nodata & -0.017 & 0.003  & 0.057  & -0.157 & \nodata & -0.009 & -0.005 & 0.016  & -0.006 & 6    \\
2005    Jul 3  & 3554.66 & 21.00 & \nodata & -0.012 & -0.006 & 0.043  & -0.091 & \nodata & -0.010 & -0.006 & 0.016  & -0.006 & 4    \\
2005    Jul 4  & 3555.77 & 22.11  & 0.075  & -0.013 & -0.003 & 0.012  & -0.010 & -0.049 & -0.010 & -0.007 & 0.016  & -0.005 & 1    \\
2005    Jul 6  & 3557.68 & 24.02  & 0.033  & -0.04  & -0.011 & 0.020  & -0.153 & -0.049 & -0.012 & -0.009 & 0.016  & -0.005 & 2    \\
2005    Jul 6  & 3557.73 & 24.07  & 0.077  & -0.016 & -0.003 & 0.012  & -0.008 & -0.049 & -0.012 & -0.009 & 0.016  & -0.004 & 1    \\
2005    Jul 7  & 3559.48 & 25.82  & \nodata & -0.020 & 0.008  & 0.08   & -0.126 & \nodata & -0.013 & -0.012 & 0.017  & -0.004 & 6    \\
2005    Jul 8  & 3559.59 & 25.93  & \nodata & 0.035  & -0.016 & 0.006  & -0.002 & \nodata & -0.013 & -0.012 & 0.017  & -0.004 & 8    \\
2005    Jul 8  & 3559.61 & 25.95  & \nodata & -0.022 & 0.003  & 0.023  & -0.074 & \nodata & -0.013 & -0.012 & 0.017  & -0.003 & 4    \\
2005    Jul 8  & 3559.70 & 26.04  & 0.073  & -0.025 & 1E-3   & 0.012  & -0.006 & -0.049 & -0.013 & -0.012 & 0.017  & -0.003 & 1    \\
2005    Jul 8  & 3559.73 & 26.07  & \nodata & -0.046 & -0.002 & 0.011  & 0.019  & \nodata & -0.013 & -0.012 & 0.017  & -0.003 & 7    \\
2005    Jul 8  & 3559.76 & 26.10  & \nodata & -0.046 & 0.010  & 0.023  & 0.045  & \nodata & -0.013 & -0.012 & 0.017  & -0.003 & 5    \\
2005    Jul 10 & 3561.73 & 28.07  & 0.067  & -0.026 & 0.004  & 0.011  & -0.003 & -0.049 & -0.013 & -0.015 & 0.017  & -0.002 & 1    \\
2005    Jul 10 & 3562.42 & 28.76  & \nodata & -0.023 & 0.010  & 0.081  & -0.118 & \nodata & -0.013 & -0.016 & 0.017  & -0.002 & 6    \\
2005    Jul 11 & 3562.63 & 28.97  & \nodata & -0.023 & 0.004  & 0.010  & -0.063 & \nodata & -0.013 & -0.017 & 0.017  & -0.002 & 4    \\
2005    Jul 11 & 3562.72 & 29.06  & 0.078  & -0.048 & 0.009  & 0.019  & \nodata & -0.05  & -0.013 & -0.017 & 0.017  & \nodata & 5    \\
2005    Jul 11 & 3563.03 & 29.37  & \nodata & \nodata & \nodata & \nodata & \nodata & \nodata & -0.013 & -0.017 & 0.017  & -0.002 & 3    \\
2005    Jul 11 & 3563.42 & 29.76  & \nodata & -0.023 & 0.010  & 0.08   & -0.116 & \nodata & -0.013 & -0.018 & 0.018  & -1E-3  & 6    \\
2005    Jul 12 & 3563.70 & 30.04  & \nodata & -0.055 & 0.003  & 0.009  & 0.022  & \nodata & -0.013 & -0.018 & 0.018  & -1E-3  & 7    \\
2005    Jul 12 & 3563.73 & 30.07  & 0.063   & -0.024 & 0.004  & 0.011  & -1E-3  & -0.050  & -0.013 & -0.018 & 0.018  & -1E-3  & 1    \\
2005    Jul 13 & 3565.39 & 31.73  & \nodata & -0.025 & 0.009  & 0.079  & -0.1   & \nodata & -0.013 & -0.021 & 0.018  & 0  & 6    \\
2005    Jul 14 & 3565.71 & 32.05  & 0.058  & -0.026 & 1E-3   & 0.010   & 1E-3   & -0.051  & -0.013 & -0.022 & 0.018  & 0  & 1    \\
2005    Jul 17 & 3569.40 & 35.74  & \nodata & -0.027 & 0.006  & 0.077  & -0.089 & \nodata & -0.013 & -0.020 & 0.019  & 0.002  & 6    \\
2005    Jul 18 & 3569.54 & 35.88  & \nodata & -0.027 & 0.003  & 0  & -0.043 & \nodata     & -0.013 & -0.020 & 0.019  & 0.002  & 4    \\
2005    Jul 19 & 3570.70 & 37.04  & 0.047  & -0.025 & 0.003  & 0.011  & 0.004   & -0.050  & -0.013 & -0.019 & 0.019  & 0.003  & 1    \\
2005    Jul 19 & 3570.80 & 37.14  & \nodata & -0.058 & 0.003  & 0.009  & 0.021  & \nodata & -0.013 & -0.019 & 0.019  & 0.003  & 7    \\
2005    Jul 21 & 3572.70 & 39.04  & 0.046   & -0.026 & 0.003  & 0.012  & 0.005  & -0.052  & -0.013 & -0.019 & 0.020  & 0.004  & 1    \\
2005    Jul 22 & 3573.60 & 39.94  & \nodata & -0.029 & 0.002  & -0.005 & -0.036 & \nodata & -0.013 & -0.018 & 0.020  & 0.005  & 4    \\
2005    Jul 23 & 3574.70 & 41.04  & 0.044   & -0.026 & 0.004  & 0.012  & 0.005  & -0.051  & -0.013 & -0.018 & 0.020  & 0.006  & 1    \\
2005    Jul 23 & 3574.76 & 41.10  & \nodata & -0.058 & 0.002  & 0.010  & 0.025  & \nodata & -0.013 & -0.018 & 0.020  & 0.006  & 7    \\
2005    Jul 24 & 3576.40 & 42.74  & \nodata & -0.027 & 0.006  & \nodata & -0.077 & \nodata& -0.013& -0.017& \nodata & 0.007  & 6    \\
2005    Jul 25 & 3576.61 & 42.95  & \nodata & -0.029 & 1E-3   & -0.008 & -0.032 & \nodata & -0.013 & -0.017 & 0.021  & 0.007  & 4    \\
2005    Jul 25 & 3576.73 & 43.07  & 0.044  & -0.027 & 0.004  & 0.013  & 0.005  & -0.050  & -0.013 & -0.017 & 0.021  & 0.007  & 1    \\
2005    Jul 27 & 3578.69 & 45.03  & 0.045  & -0.028 & 0.005  & 0.015  & 0.007  & -0.050  & -0.013 & -0.016 & 0.021  & 0.008  & 1    \\
2005    Jul 28 & 3579.55 & 45.89  & \nodata & -0.031 & 0  & -0.004 & -0.029 & \nodata & -0.013 & -0.016 & 0.021  & 0.009  & 4    \\
2005    Jul 29 & 3580.69 & 47.03  & 0.046  & -0.029 & 0.004  & 0.017  & 0.007  & -0.049 & -0.013 & -0.015 & 0.021  & 0.010  & 1    \\
2005    Jul 29 & 3581.00 & 47.34  & \nodata & \nodata & \nodata & \nodata & \nodata & \nodata & -0.013 & -0.015 & 0.021  & 0.010  & 3    \\
2005    Jul 31 & 3582.69 & 49.03  & 0.048  & -0.027 & 0.003  & 0.017  & 0.006  & -0.049 & -0.013 & -0.015 & 0.021  & 0.011  & 1    \\
2005    Aug 2  & 3584.69 & 51.03  & 0.049  & -0.025 & 0.002  & 0.016  & 0.007  & -0.048 & -0.013 & -0.014 & 0.021  & 0.013  & 1    \\
2005    Aug 4  & 3586.69 & 53.03  & 0.05   & -0.023 & 0.002  & 0.016  & 0.007  & -0.048 & -0.013 & -0.013 & 0.021  & 0.014  & 1    \\
2005    Aug 6  & 3588.69 & 55.03  & 0.051  & -0.021 & 0  & 0.017  & 0.007  & -0.047 & -0.013 & -0.012 & 0.021  & 0.016  & 1    \\
2005    Aug 6  & 3588.71 & 55.05  & \nodata & -0.046 & -0.002 & 0.018  & 0.071  & \nodata & -0.013 & -0.012 & 0.021  & 0.016  & 7    \\
2005    Aug 8  & 3590.69 & 57.03  & 0.05   & -0.019 & 0  & 0.017  & 0.006  & -0.047 & -0.013 & -0.011 & 0.021  & 0.017  & 1    \\
2005    Aug 10 & 3592.69 & 59.03  & 0.05   & -0.017 & -1E-3  & 0.016  & 0.006  & -0.046 & -0.013 & -0.011 & 0.021  & 0.019  & 1    \\
2005    Aug 14 & 3596.68 & 63.02  & \nodata & -0.013 & -0.002 & 0.016  & \nodata & \nodata & -0.014 & -0.009 & 0.020  & \nodata & 1    \\
2005    Aug 17 & 3599.68 & 66.02  & \nodata & -0.008 & -0.002 & 0.017  & 0.006  & \nodata & -0.014 & -0.008 & 0.019  & 0.025  & 1    \\
2005    Aug 18 & 3601.01 & 67.35  & \nodata & \nodata & \nodata & \nodata & \nodata & \nodata & -0.014 & -0.008 & 0.019  & 0.026  & 3    \\
2005    Aug 19 & 3602.67 & 69.01  & \nodata & -0.005 & -0.003 & 0.016  & 0.006  & \nodata & -0.014 & -0.007 & 0.018  & 0.027  & 1    \\
2005    Aug 20 & 3604.01 & 70.35  & \nodata & \nodata & \nodata & \nodata & \nodata & \nodata & -0.014 & -0.007 & 0.018  & 0.027  & 3    \\
2005    Aug 21 & 3605.11 & 71.35  & \nodata & \nodata & \nodata & \nodata & \nodata & \nodata & -0.014 & -0.006 & \nodata  & \nodata   & 3    \\
2005    Aug 22 & 3605.67 & 72.01  & \nodata & -0.002 & -0.004 & 0.017  & 0.006  & \nodata & -0.014 & -0.005 & 0.017  & 0.03   & 1    \\
2005    Aug 25 & 3608.67 & 75.01  & \nodata & 0.003  & -0.003 & 0.016  & 0.006  & \nodata & -0.014 & -0.004 & 0.015  & 0.033  & 1    \\
2005    Aug 28 & 3611.67 & 78.01  & \nodata & 0.007  & -0.003 & 0.016  & 0.006  & \nodata & -0.014 & -0.003 & 0.013  & 0.036  & 1    \\
2005    Aug 31 & 3614.66 & 81.00  & \nodata & 0.011  & -0.003 & 0.016  & 0.006  & \nodata & -0.014 & -0.002 & 0.011  & 0.039  & 1    \\
2005    Sept 3 & 3617.66 & 84.00  & \nodata & 0.015  & -0.003 & 0.016  & 0.007  & \nodata & -0.014 & -1E-3  & 0.008  & 0.042  & 1    \\
2005    Sept 4 & 3618.65 & 84.99  & \nodata & 0.018  & -0.003 & 0.015  & 0.008  & \nodata & -0.014 &  0     & 0.007  & 0.043  & 1    \\
2005    Sept 7 & 3621.66 & 88.00  & \nodata & 0.022  & -1E-3  & 0.012  & 0.009  & \nodata & -0.014 & 1E-3   & 0.005  & 0.046  & 1    \\
2005    Sept 10& 3624.65 & 90.99  & \nodata & 0.026  & 0      & 0.010  & 0.010  & \nodata & -0.014 & 0.002  & 0.002  & 0.049  & 1    \\
\enddata
\tablenotetext{a}{Relative to the epoch of $B$-band maximum (JD =
2,453,533.66).} \tablenotetext{b}{1 = KAIT 0.76~m; 2 = FLWO 1.2~m; 3
= TNT 0.8~m; 4 = CTIO 1.3~m; 5 = Lick 1.0~m; 6 = Liverpool 2.0~m; 7
= Palomar 1.5~m; 8 = CTIO 0.9~m}
\end{deluxetable}

\end{document}